\documentclass{emulateapj}
\usepackage{apjfonts}
\usepackage{natbib}
\def\chandra{{\it Chandra\/}}
\def\xray{\hbox{X-ray}}

\def\asca{{\it ASCA\/}}
\def\hst{{\it HST\/}}
\def\lum{erg~s$^{-1}$}
\def\flux{erg~cm$^{-2}$~s$^{-1}$}
\def\etal{{et\,al.\,}}
\def\lsim{\mathrel{\rlap{\lower4pt\hbox{\hskip1pt$\sim$}}
    \raise1pt\hbox{$<$}}}                % less than or approx. symbol
\def\gsim{\mathrel{\rlap{\lower4pt\hbox{\hskip1pt$\sim$}}
    \raise1pt\hbox{$>$}}}                % greater than or approx. symbol

\def\cdfs{\hbox{CDF-S}}
\def\cdfn{\hbox{CDF-N}}
\def\ecdfs{\hbox{E-CDF-S}}
\def\goodsn{\hbox{GOODS-N}}
\def\goodss{\hbox{GOODS-S}}
\def\hhdfn{\hbox{H-HDF-N}}
\def\hdfn{\hbox{HDF-N}}
\def\zs{$z_{\rm spec}$}
\def\zp{$z_{\rm phot}$}
\def\zf{$z_{\rm final}$}
\def\nmad{$\sigma_{\rm NMAD}$}

\begin{document}
%\slugcomment{Data and images available at http://www.astro.psu.edu/users/niel/cdfs/cdfs-chandra.html}

\title{The 2~Ms Chandra Deep Field-North Survey and the 250~ks Extended Chandra Deep Field-South Survey: Improved Point-Source Catalogs}

\author{
Y.~Q.~Xue,\altaffilmark{1}
B.~Luo,\altaffilmark{2,3}
W.~N.~Brandt,\altaffilmark{2,3,4}
D.~M.~Alexander,\altaffilmark{5}
F.~E.~Bauer,\altaffilmark{6,7,8}
B.~D.~Lehmer,\altaffilmark{9,10,11}
and G.~Yang\altaffilmark{2,3}
}
\altaffiltext{1}{CAS Key Laboratory for Researches in Galaxies and Cosmology, Center for Astrophysics, Department of Astronomy, University of Science and Technology of China, Chinese Academy of Sciences, Hefei, Anhui 230026, China; xuey@ustc.edu.cn}
\altaffiltext{2}{Department of Astronomy and Astrophysics, Pennsylvania State University, University Park, PA 16802, USA}
\altaffiltext{3}{Institute for Gravitation and the Cosmos, Pennsylvania State University, University Park, PA 16802, USA}
\altaffiltext{4}{Department of Physics, Pennsylvania State University, University Park, PA 16802, USA}
\altaffiltext{5}{Centre for Extragalactic Astronomy, Department of Physics, Durham University, Durham, DH1 3LE, UK}
\altaffiltext{6}{Instituto de Astrof\'{\i}sica, Facultad de F\'{i}sica, Pontificia Universidad Cat\'{o}lica de Chile, Casilla 306, Santiago 22, Chile}
\altaffiltext{7}{Millennium Institute of Astrophysics}
\altaffiltext{8}{Space Science Institute, 4750 Walnut Street, Suite 205, Boulder, Colorado 80301}
\altaffiltext{9}{The Johns Hopkins University, Homewood Campus, Baltimore, MD 21218, USA}
\altaffiltext{10}{NASA Goddard Space Flight Centre, Code 662, Greenbelt, MD 20771, USA}
\altaffiltext{11}{Department of Physics, University of Arkansas, 226 Physics Building, 835 West Dickson Street, Fayetteville, AR 72701, USA}

\begin{abstract}

We present improved point-source catalogs
for the 2~Ms \chandra\ Deep Field-North (\cdfn) that covers
an area of 447.5~arcmin$^{2}$ and the 250~ks
Extended Chandra Deep Field-South (\ecdfs) that covers
an area of 1128.6~arcmin$^{2}$,
implementing a number of recent improvements
in \chandra\ source-cataloging methodology.
For the \cdfn, we provide a main catalog that contains
683 \hbox{X-ray} sources detected with {\sc wavdetect} at a false-positive probability
threshold of $10^{-5}$ in at least one of three standard \xray\ bands
(\hbox{0.5--7~keV}, full band; \hbox{0.5--2~keV}, soft band;
and \hbox{2--7~keV}, hard band) that also satisfy a
binomial-probability source-selection criterion of \mbox{$P<0.004$}.
Such an approach maximizes the number of reliable sources detected:
a total of 196 \cdfn\ main-catalog sources are new compared to the
Alexander \etal (2003) 2~Ms \cdfn\ main catalog.
We also provide a \cdfn\ supplementary catalog that consists of 72 sources
detected at the same {\sc wavdetect} threshold and having $0.004<P<0.1$
and $K_s\le 22.9$~mag counterparts.
For the \ecdfs, we provide likewise a main catalog containing 1003 sources
and a supplementary catalog consisting of 56 sources,
with the only differences lying in the corresponding adopted threshold values of
$P<0.002$ and $K_s\le 22.3$~mag.
A total of 275 \ecdfs\
main-catalog sources are new compared to the Lehmer \etal (2005) \ecdfs\ main catalog.
For all \mbox{$\approx 1800$} \cdfn\ and \ecdfs\ sources, including the
$\approx 500$ newly detected ones (these being generally fainter and more obscured),
we determine \hbox{X-ray} source positions utilizing centroid
and matched-filter techniques and achieve median positional uncertainties
of $0\farcs47$ for \cdfn\ and $0\farcs63$ for \ecdfs.
We provide multiwavelength identifications (with a 98.1\% identification rate for \cdfn\
and 95.5\% for \ecdfs), apparent magnitudes of counterparts, and
spectroscopic and/or photometric redshifts (with a 95.2\% redshift success rate for \cdfn\
and 84.6\% for \ecdfs).
Finally, by analyzing \hbox{X-ray} and multiwavelength properties of the sources, we find that
86.5\%/90.6\% of the \cdfn/\ecdfs\ main-catalog sources are likely AGNs
and the galaxy fraction among the new \cdfn/\ecdfs\ main-catalog sources
is larger than that among the corresponding old sources,
reflecting the rise of normal and starburst galaxies when probing fainter fluxes.
In the areas within respective off-axis angles of~$3\arcmin$ of
the \cdfn\ average aim point and the four \ecdfs\ aim points,
the observed AGN and galaxy source densities reach
$12400_{-1300}^{+1400}$~deg$^{-2}$ and
$4200_{-700}^{+900}$~deg$^{-2}$ for \cdfn, and
$5200_{-800}^{+1000}$~deg$^{-2}$ and
$500_{-200}^{+400}$~deg$^{-2}$ for \ecdfs, respectively.
Simulations show that both the \cdfn\ and \ecdfs\ main catalogs are highly reliable and reasonably complete.
The mean soft- and hard-band backgrounds are
0.055 and 0.108 count~Ms$^{-1}$~pixel$^{-1}$ for \cdfn, and
0.048 and 0.109 count~Ms$^{-1}$~pixel$^{-1}$ for \ecdfs, respectively;
$\gsim 92$\%/$\gsim 97$\% of the pixels have zero background counts
in any of the three standard bands for \cdfn/\ecdfs.
The soft- and hard-band on-axis mean flux limits reached
are $\approx 1.2\times 10^{-17}$ and $5.9\times 10^{-17}$ \flux\
for the 2~Ms \cdfn\ (i.e., a factor of $\approx 2$ improvement over the previous
\cdfn\ limits), and
$\approx 7.6\times 10^{-17}$ and $3.0\times 10^{-16}$ \flux\
for the 250~ks \ecdfs\ (i.e., a factor of \mbox{$\approx 1.5$--2.0} improvement over the previous
\ecdfs\ limits), respectively.
We make our data products publicly available.

\end{abstract}

\keywords{catalogs --- cosmology: observations --- diffuse radiation --- galaxies: active ---
surveys --- \hbox{X-rays}: galaxies}

\section{INTRODUCTION}\label{sec:intro}

Deep and wide cosmic \hbox{X-ray} surveys of active galactic nuclei (AGNs)
over the past few decades,
and their critical complementary multiwavelength observations,
have dramatically improved our understanding of many aspects of 
growing supermassive black holes in the distant universe, e.g.,
the AGN population and its evolution (``demographics''),
the physical processes operating in AGNs (``physics''),
and the interactions between AGNs and their environments (``ecology'')
(see Brandt \& Alexander 2015 for a review).  
The \chandra\ Deep Fields (CDFs) have critically contributed to
the characterization of the \hbox{0.5--8 keV} cosmic \xray\ 
background sources, the majority of which are AGNs.
The CDF-North (\cdfn; 1~Ms \cdfn, Brandt \etal 2001;
2~Ms \cdfn, Alexander \etal 2003, hereafter A03) and
the CDF-South (\cdfs; 1~Ms \cdfs, Giacconi \etal 2002; 
2~Ms \cdfs, Luo \etal 2008;
4~Ms \cdfs, Xue \etal 2011, hereafter X11)
are the two deepest \chandra\ surveys,
and the latter is complemented by
the 1~Ms Extended-\cdfs\ (\ecdfs, which consists of four flanking,
contiguous 250~ks \chandra\ observations; 
Lehmer \etal 2005, hereafter L05).
The CDFs have enormous supporting multiwavelength investments
that are key to source identification and characterization,
and will remain a crucial resource in interpreting the nature of
extragalactic populations identified using superb
multiwavelength surveys (e.g., {\it JWST}, ALMA, and EVLA) 
over the coming decades, thereby continuing 
the lasting legacy value.

Over the last $\approx 10$~yr there have been major improvements 
in the methodology of producing \chandra\ source catalogs,
as evidenced by, e.g., the 4~Ms \cdfs\ point-source catalogs (X11).
Similar applications of a two-stage source-detection approach,
which is a key ingredient of such an improved methodology, 
have also been presented in, e.g., Getman \etal (2005); Nandra \etal (2005, 2015);
Elvis \etal (2009); Laird \etal (2009); Lehmer \etal (2009); 
Puccetti \etal (2009); and Ehlert \etal (2013) (see Sections~\ref{sec:cdfn-cand} and 
\ref{sec:cdfn-main-select} for details).
Given the parallel importance of the \cdfn\ and \ecdfs\ to the \cdfs,
it is imperative to create improved
2~Ms \cdfn\ and 250~ks \ecdfs\ source catalogs implementing
such improvements in methodology, thereby contributing to
the most effective exploitation of the large investments in 
the CDF surveys.
The \ecdfs, though not as deep as the \cdfn\ and \cdfs,
is also a premiere deep-survey field, and its data help significantly
with measurements of sources located at large off-axis angles 
in the \cdfs\ proper.

We present in this paper the improved \chandra\ point-source catalogs and
associated data products, together with observation details,
data reduction, and technical analyses, 
for the 2~Ms \cdfn\ and the 250~ks \ecdfs.
Table~\ref{tab:impro} gives a list of major improvements implemented 
in X11 and here in the production of the improved source catalogs
over the existing 2~Ms \cdfn\ (A03) and 250~ks \ecdfs\ (L05) catalogs.
The key improvements include  
(1) adoption of the flexible and reliable two-stage source-detection approach (leading to a significant number of new sources with high confidence in their validity without new \chandra\ observational investment),
(2) optimal extractions of \xray\ photometry (enabling the best possible \xray\ characterization of detected sources), and
(3) secure identification of multiwavelength counterparts of detected
\xray\ sources (allowing for detailed follow-up studies).
The details of the improvements are given in the relevant sections
as indicated in Table~\ref{tab:impro}.
To implement the improved methodology, we make extensive use of 
the ACIS Extract (AE; %version released on 2010 February 26;
Broos \etal 2010)\footnote{Details on AE can be found at
http://www.astro.psu.edu/xray/docs/TARA/ae\_users\_guide.html.\label{ft:ae}} 
point-source analysis software that accurately computes source 
\xray\ properties (most importantly point spread function; PSF), when combining
multiple observations that have different roll angles and/or aim points.
The improved 2~Ms \cdfn\ and 250~ks \ecdfs\ point-source catalogs 
presented here supersede those presented in A03 and L05, respectively.

\begin{table*}[ht]
\caption{Improvements over Existing 2~Ms \cdfn\ (A03) and 250~ks \ecdfs\ (L05) Catalogs}
\centering
%\small
%\resizebox{1.00\textwidth}{!}{%
\begin{tabular}{lllr}\hline\hline
 & A03 and L05 & Improved Catalogs & Example Section(s) \\\hline
Astrometric alignment & Using merged observations & Frame by frame (i.e., observation by observation) & \ref{sec:cdfn-img} \\
Source detection & {\sc wavdetect}-only & {\sc wavdetect} + ACIS Extract (AE)$^{\ref{ft:ae}}$ no-source probability & \ref{sec:cdfn-cand} \& \ref{sec:cdfn-main-select} \\
Extraction region & Circular aperture & AE polygonal region that approximates the PSF shape & \ref{sec:cdfn-cand} \\
Crowded sources & Manual extraction & AE extraction by automatically shrinking regions & \ref{sec:cdfn-cand} \\
Background estimate & Source-masking approach & AE BETTER\_BACKGROUNDS algorithm & \ref{sec:cdfn-cand} \\
\xray\ photometry & Cumulative images & AE merging of extractions on individual images & \ref{sec:cdfn-cand} \\\hline
Comprehensive source identification & Not provided & Provided & \ref{sec:cdfn-id} \\
Redshift compilation & Not provided & Provided & \ref{sec:cdfn-maincat} \\
Source classification & Not provided & Provided & \ref{sec:cdfn-maincat} \\\hline
\end{tabular}%}
%\vspace{-3.7mm}
\label{tab:impro}
\end{table*}

This paper is structured as follows.
Section~2 is dedicated to the production of the improved 2~Ms \cdfn\ 
source catalogs, which covers a range of contents organized into 
subsections and sub-subsections as appropriate, including
observations and data reduction (Section~\ref{sec:cdfn-obs}),
creation of the images, exposure maps, and candidate-list catalog (Section~\ref{sec:cdfn-img-cand}),
production of the main and supplementary catalogs (Sections~\ref{sec:cdfn-main} and \ref{sec:cdfn-supp}),
completeness and reliability analyses (Section~\ref{sec:cdfn-comp}), and
background and sensitivity analyses (Section~\ref{sec:cdfn-bkg}).
Section~3 is parallel to Section~2,
but dedicated to the production of the improved 250~ks \ecdfs\ 
source catalogs, in basically the same manner as Section~2.
Section~4 summarizes the results of this work.

Throughout this paper, Galactic column densities
of $N_{\rm H}=1.6\times 10^{20}$~cm$^{-2}$ and
$N_{\rm H}=8.8\times 10^{19}$~cm$^{-2}$
along the lines of sight to the \cdfn\ and \ecdfs\ are adopted, respectively 
(e.g., Stark \etal 1992). 
The J2000.0 coordinate system, the AB magnitude system, and a cosmology with
$H_0=69.7$ km~s$^{-1}$~Mpc$^{-1}$,
$\Omega_{\rm M}=0.282$, and
$\Omega_\Lambda=0.718$ (Hinshaw \etal 2013) are used.

\section{Production of the improved 2~Ms \cdfn\ point-source catalogs}\label{sec:cdfn}

\begin{figure}
\centerline{\includegraphics[width=8.5cm]{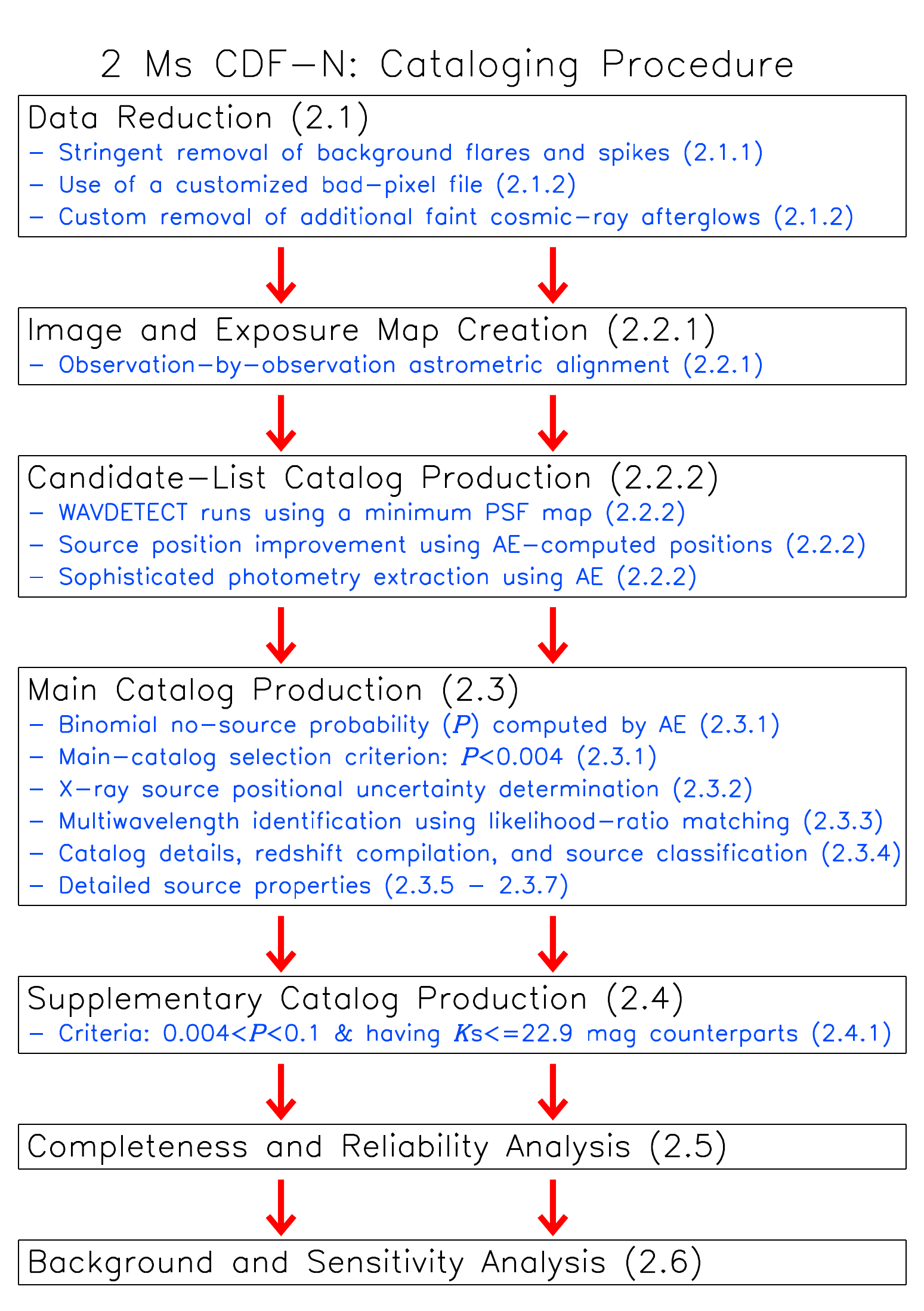}}
\figcaption{Flow chart of the overall 2~Ms \cdfn\ cataloging procedure.
The black texts describe the major cataloging steps, while
the blue texts highlight some key points in the corresponding steps.
The numbers in parentheses indicate the relevant 
subsections/sub-subsections.
\label{fig:cataloging_flow}}
\end{figure}

The overall production procedure,
as illustrated in Fig.~\ref{fig:cataloging_flow},
 is similar to that described in X11.
For ease of reading, we provide here only essential details 
and refer readers to X11 for full details.
In addition, we make our 2~Ms \cdfn\ data products publicly 
available.\footnote{The data products, including the final event files, raw images, 
effective exposure maps, background maps, sensitivity maps, and solid-angle vs. flux-limit
curves for the 2~Ms \cdfn\ and 250~ks
\ecdfs\ are available at 
http://www2.astro.psu.edu/users/niel/hdf/hdf-chandra.html and
http://www2.astro.psu.edu/users/niel/ecdfs/ecdfs-chandra.html, respectively.\label{ft:datalink}}

\subsection{Observations and Data Reduction}\label{sec:cdfn-obs}

\subsubsection{Observations and Observing Conditions}\label{sec:cdfn-obs-info}

The 2~Ms \cdfn\ consists of a total of 20 separate observations
taken between 1999 November 13 and 2002 February 22 (see Table~1 of A03
for the journal of these 20 \cdfn\ observations).
The 20 \cdfn\ observations were made with  
the Advanced CCD Imaging Spectrometer  
(ACIS; Garmire \etal 2003) onboard \chandra\ 
that consists of an imaging array (\mbox{ACIS-I}; 
with an overall field of view of 
$16\farcm9\times 16\farcm9 = 285.6$ arcmin$^2$)
and a spectroscopic array (\mbox{ACIS-S}). 
The four \mbox{ACIS-I} CCDs were in operation throughout the 20
\cdfn\ observations, while the \mbox{ACIS-S} CCD S2 was operated for 
the first 12 observations.
We do not use the data taken with the \mbox{ACIS-S} CCD S2 in this work
due to its large off-axis angle and consequently its low sensitivity.
The focal-plane temperature was $-110\degr$C for the first three
\cdfn\ observations (ObsIDs=580, 967, and 966) 
and $-120\degr$C for the remaining ones.
The first 12 \cdfn\ observations were carried out in Faint mode, 
while the later 8 observations were carried out in Very Faint mode
to help screen background events and thus improve
the ACIS sensitivity for detecting faint \xray\ sources (Vikhlinin 2001).

The background light curves for all the 20 \cdfn\ observations
were examined utilizing the \chandra\ Imaging and Plotting System 
(ChIPS).\footnote{The ChIPS analysis threads can be found at 
http://cxc.harvard.edu/chips/.}
During observation 2344,
there are two significant flares in the background, with
each lasting $\approx 1.5$~ks and being $\gsim 2$ times higher than nominal;
a time span of $\approx 18.0$~ks between these two flares was
affected moderately. 
The background increased significantly 
(up to $\approx 4$ times higher than nominal)  
toward the end of observation 3389,
affecting an exposure of $\approx 17.0$~ks.
All the other observations are free from strong flaring
and are stable within $\approx 20$\% of typical quiescent \chandra\ values,
except for a number of short moderate ``spikes'' (up to $\approx 1.5$ times
higher than nominal).
To remove these significant flares and moderate spikes, 
we utilize an iterative sigma-clipping tool 
{\sc lc\_sigma\_clip}, which is part of the \chandra\ Interactive Analysis of Observations ({\sc ciao}; we use {\sc ciao} 4.5 and {\sc caldb} 4.5.9 
in this work) package.
We adopt 2.6-, 2.0-, and 3.5-sigma clippings for observation 2344,
observation 3389, and the other observations, respectively.
After filtering the data on good-time intervals,
we obtain a total effective exposure time of 
1.896~Ms for the 20 \cdfn\ observations (see Section~\ref{sec:cdfn-img}),
which is smaller than the value of 1.945~Ms reported in A03
due to our more stringent filtering process.

For the majority of the 20 observations,
the \mbox{ACIS-I} aim point was placed near 
the \hdfn\ (Williams \etal 1996) center 
and the roll angles varied around two main values of 
$\approx 40$ and $\approx 140$ degrees.
Such a pointing scheme and roll constraints not only lead to 
a total region of 447.5 arcmin$^2$ covered by these 20 \cdfn\ observations
that is considerably larger than the \mbox{ACIS-I} field of view,
but also result in all the individual pointings being separated from
the average aim point by $>1\arcmin$.
The average aim point is
$\alpha_{\rm J2000.0}=12^{\rm h}36^{\rm m}45.^{\rm s}7$,
$\delta_{\rm J2000.0}=+62\degr13\arcmin58\farcs 0$,
weighted by the 20 individual exposures 
that typically range from $\approx 50$ to $\approx 170$~ks.

\subsubsection{Data Reduction}\label{sec:cdfn-obs-data}

We make use of {\sc ciao} tools and custom software for data reduction.
We utilize {\sc acis\_process\_events} to
reprocess each level~1 observation,
which takes into account the radiation damage sustained by the CCDs 
during the beginning of
\chandra\ operations by implementing a Charge Transfer Inefficiency (CTI)
correction procedure presented in Townsley \etal (2000, 2002; 
this procedure
is only applicable to $-120\degr$C observations, but not to $-110\degr$C ones)
and applies a modified bad-pixel file instead of the standard CXC one. 
Our customized bad-pixel file retrieves several percent of 
the \hbox{ACIS-I} pixels
on which numerous events are valid for
source detection, photometry extraction, and spectral analysis
that would be discarded otherwise
(see Section~2.2 of Luo \etal 2008 for reasoning). 
We set {\sc check\_vf\_pha=yes} in {\sc acis\_process\_events} 
for observations carried out in Very Faint mode for better cleaning of
background events, which utilizes a $5\times 5$ pixel event island 
to identify imposter cosmic-ray background events.
We then use {\sc acis\_detect\_afterglow} to eliminate cosmic-ray afterglows.
To reject further surviving faint afterglows,
we remove a number of additional faint afterglows with $\gsim 3$ counts 
arriving within a timespan of 20~s on a pixel that almost certainly signifies
an association with cosmic-ray afterglows
(see Footnote~27 of X11 for reasoning).

\subsection{Images, Exposure Maps, and Candidate-List Catalog}\label{sec:cdfn-img-cand}

\subsubsection{Image and Exposure Map Creation}\label{sec:cdfn-img}

\begin{figure}
\centerline{\includegraphics[width=8.5cm]{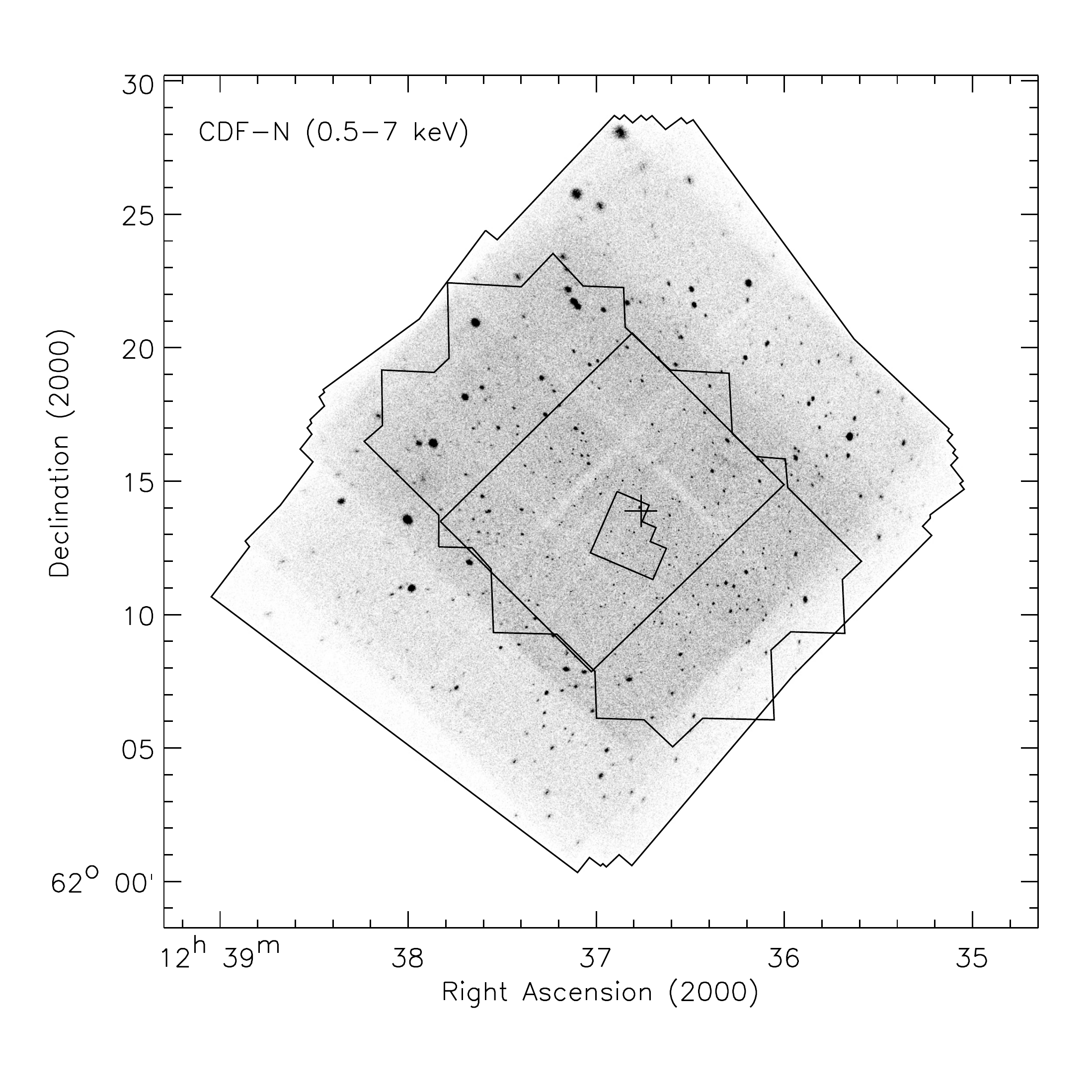}}
\figcaption{Full-band (0.5--7.0~keV) raw image of the 2~Ms \cdfn\ rendered using
linear gray scales.
The outermost segmented boundary indicates the coverage of the entire \cdfn.
Toward the direction of the exposure-weighted average aim point (denoted as a plus sign)
sitting roughly at the field center,
the large polygon, the rectangle, and the small polygon denote the regions for the
\goodsn\ (Giavalisco \etal 2004),
the CANDELS \goodsn\ deep (Grogin \etal 2011; Koekemoer \etal 2011),
and the \mbox{HDF-N} (Williams \etal 1996), respectively.
The light grooves running through the image are caused by the \mbox{ACIS-I} CCD gaps,
thereby having lower effective exposures than the nearby non-gap areas
(clearly revealed in Fig.~\ref{fig:cdfn-fb-exp}).
The apparent trend of sources having larger sizes off the center is due to
the PSF degradation toward larger off-axis angles (also see Fig.~\ref{fig:cdfn-smooth-false}).
\label{fig:cdfn-fb-img}}
\end{figure}

For astrometric alignment purposes,
we first run {\sc wavdetect} (Freeman \etal 2002) 
with the option of ``psffile=none''\footnote{This option means
that no PSF map file is provided to {\sc wavdetect} such that
{\sc wavdetect} will not compute PSF sizes for source detection.
With this option on, the source-detection results are still secure
although the source characteristics might not be reliable.
However, in Section~\ref{sec:cdfn-cand} 
where we perform formal source detections on merged images,
we do provide an appropriate PSF map to {\sc wavdetect}.}
at a false-positive probability threshold of
$10^{-6}$ on each of the individual cleaned \hbox{0.5--7}~keV\footnote{Throughout
this work we switch to an upper energy bound of 7~keV from
the ``traditional'' 8~keV adopted by our previous CDF catalogs (e.g., A03,
L05, X11) for the following reasons:
(1) the \chandra\ High Resolution Mirror Assembly (HRMA) 
effective area decreases
significantly toward high energies, e.g., $\approx 180$~cm$^2$ at 7~keV vs.
$\approx 80$~cm$^2$ at 8~keV 
(as opposed to $\approx 800$~cm$^2$ at 1~keV);  
(2) the gain of net counts at 7--8~keV is modest due to increasing
background toward high energies; and
(3) the upper energy bound of 7~keV has been adopted
for source detection and/or \xray\ photometry by a number of
other cataloging works (e.g., Elvis \etal 2009; 
Laird \etal 2009; Nandra \etal 2015).
(However, as noted in Footnote~32 of X11,
there appears to be no significant statistical difference between catalogs 
made with upper energy cuts of 7~keV and 8~keV for the case
of 4~Ms \cdfs.)
We note that the cataloging of the coming 7~Ms \cdfs\ (PI: W. N. Brandt; 
the \chandra\ observations are scheduled to be completed by March 2016)
will adopt the upper energy cut of 7~keV as well to ensure uniformity
among the latest CDF catalogs.\label{ft:7keV}}
images to construct initial source lists
and utilize AE to determine centroid positions of detected sources.
In order to register the observations to a common astrometric frame, 
we then match \mbox{X-ray} centroid positions to the $K_s\le 21.0$~mag
sources in the \goodsn\ WIRCam $K_s$-band catalog (Wang \etal 2010)
rather than the Very Large Array (VLA) 1.4~GHz \goodsn\ radio
sources used by A03 (Morrison \etal 2010; note that an earlier
VLA 1.4~GHz \goodsn\ radio catalog presented in Richards 2000 
was adopted in A03),
because we find the astrometric frame of the $K_s$-band catalog
in better agreement with that of other multiwavelength
catalogs that are used for our \xray\ source identifications
in Section~\ref{sec:cdfn-id}.
We carry out \xray/$K_s$-band matching and astrometric reprojection utilizing 
{\sc reproject\_aspect} and {\sc wcs\_update} with a matching radius of 2$\arcsec$ 
and a $0\farcs6$ residual rejection limit,
resulting in typical false-match rates of
$\lsim 8\%$ that are estimated using the simple shifting-and-recorrelating
approach (this approach of estimating false-match rates 
is adopted throughout this paper except for 
Sections~\ref{sec:cdfn-id} and \ref{sec:ecdfs-id} 
where we perform multiwavelength 
identifications for the detected \xray\ sources).
We then reproject all the observations to the frame of observation 3293
that is among the observations with longest exposures and has raw coordinates
closely matched to the $K_s$-band astrometric frame.
Subsequently we combine the individual event files into
a merged event file using {\sc dmmerge},
from which we construct images using the standard
\asca\ grade set for three standard bands:
\hbox{0.5--7.0~keV} (full band; FB), \hbox{0.5--2.0~keV} (soft band; SB), 
and \hbox{2--7~keV} (hard band; HB).$^{\ref{ft:7keV}}$
Figure~\ref{fig:cdfn-fb-img} shows the full-band raw image.

\begin{figure}
\centerline{\includegraphics[width=8.5cm]{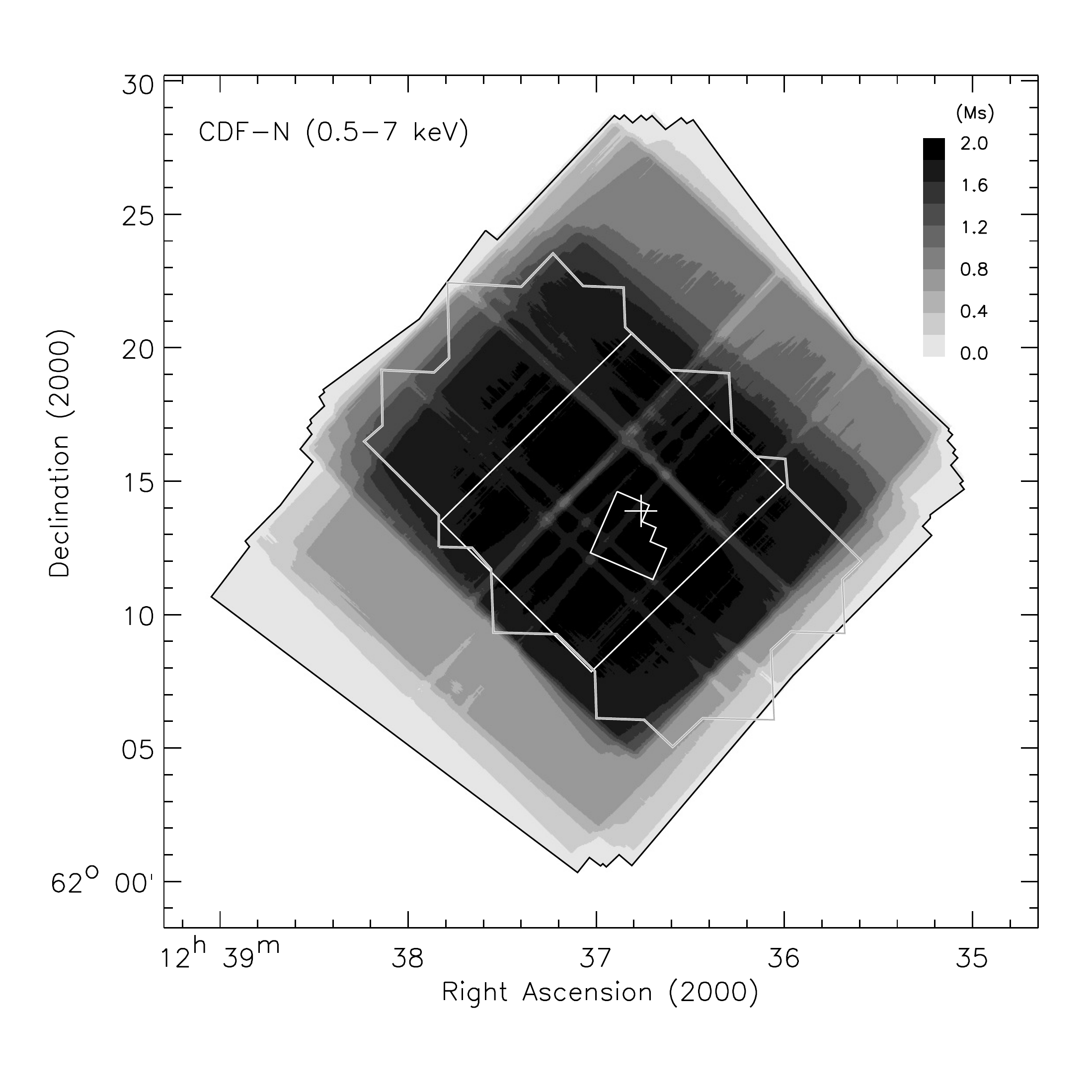}}
\figcaption{Full-band (0.5--7.0~keV) effective-exposure map of the 2~Ms \cdfn\ rendered using
linear gray scales (indicated by the inset scale bar).
The darkest areas indicate the highest effective exposure times, reaching a maximum of 1.896~Ms.
The \hbox{ACIS-I} CCD gaps can be clearly identified as the light grooves.
The regions and the plus sign have the same meanings as those in Fig.~\ref{fig:cdfn-fb-img}.
\label{fig:cdfn-fb-exp}}
\end{figure}

\begin{figure}
\centerline{\includegraphics[width=8.5cm]{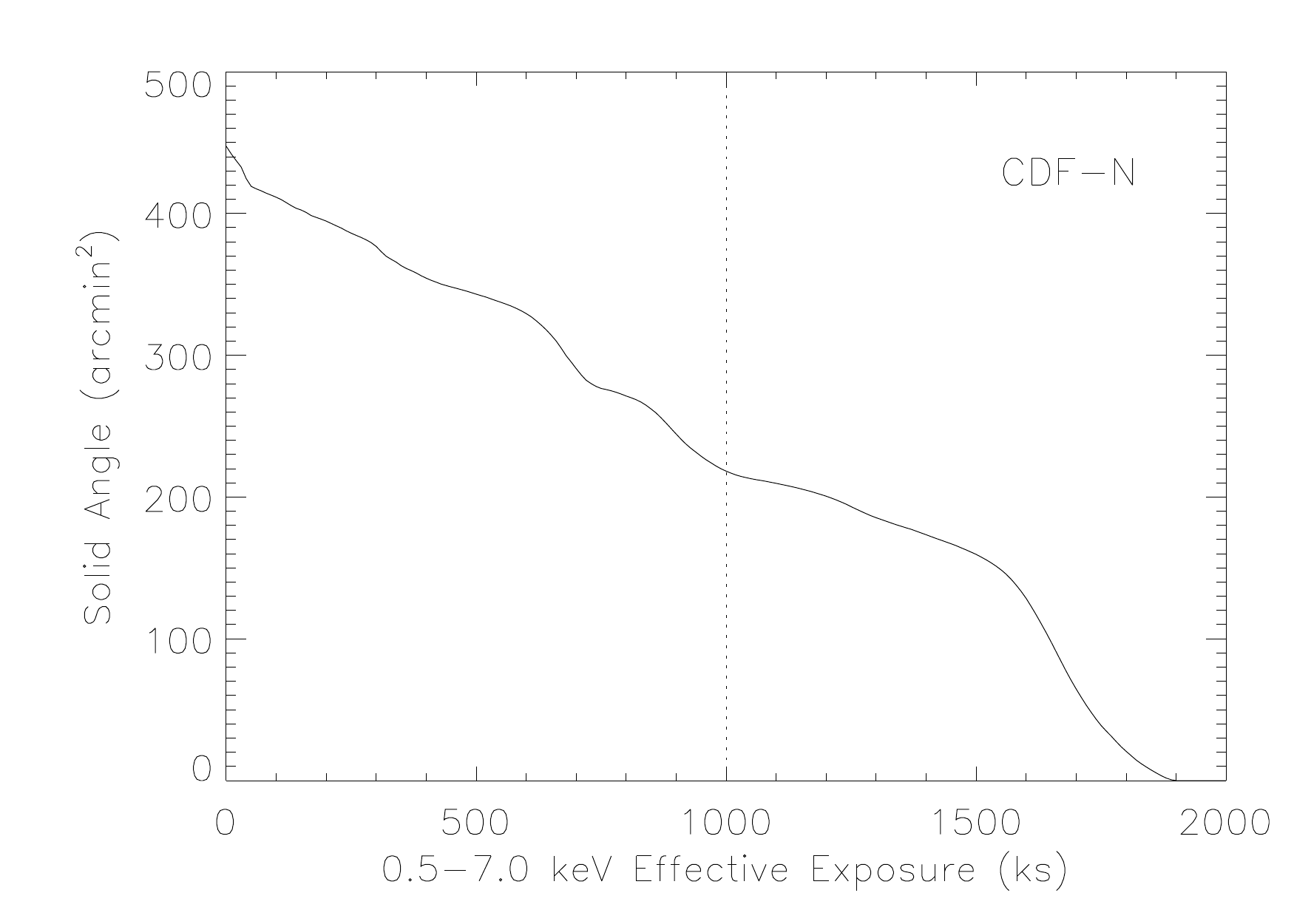}}
\figcaption{Survey solid angle as a function of minimum full-band (\mbox{0.5--7.0~keV}) effective exposure for the 2~Ms \cdfn.
The 2~Ms \cdfn\ covers a total area of 447.5 arcmin$^2$ and has a maximum exposure of 1.896~Ms.
The vertical dotted line denotes an effective exposure of 1~Ms.
218.3 arcmin$^2$ ($48.8\%$) of the \cdfn\ survey area has $>1$~Ms effective exposure.
\label{fig:cdfn-solid-exp}}
\end{figure}

Following the basic procedure
detailed in Section~3.2 of Hornschemeier \etal (2001),
we produce effective-exposure maps for the three standard bands
and normalize them to the effective exposures of a pixel lying 
at the average aim point, assuming a photon index of
$\Gamma=1.4$ that is the slope of the cosmic \hbox{2--10~keV} 
\mbox{X-ray} background (e.g., Marshall \etal 1980; Gendreau \etal 1995; 
Hasinger \etal 1998; Hickox \& Markevitch 2006).
This procedure accounts for the effects of vignetting, 
CCD gaps, bad-column and bad-pixel filtering, as well as the spatial and time
dependent degradation in quantum efficiency caused by
contamination on the ACIS optical-blocking filters.
Figure~\ref{fig:cdfn-fb-exp} presents the full-band effective-exposure map
and Figure~\ref{fig:cdfn-solid-exp} shows the survey solid angle as a
function of the minimum full-band effective exposure.

\begin{figure}
\centerline{\includegraphics[width=8.5cm]{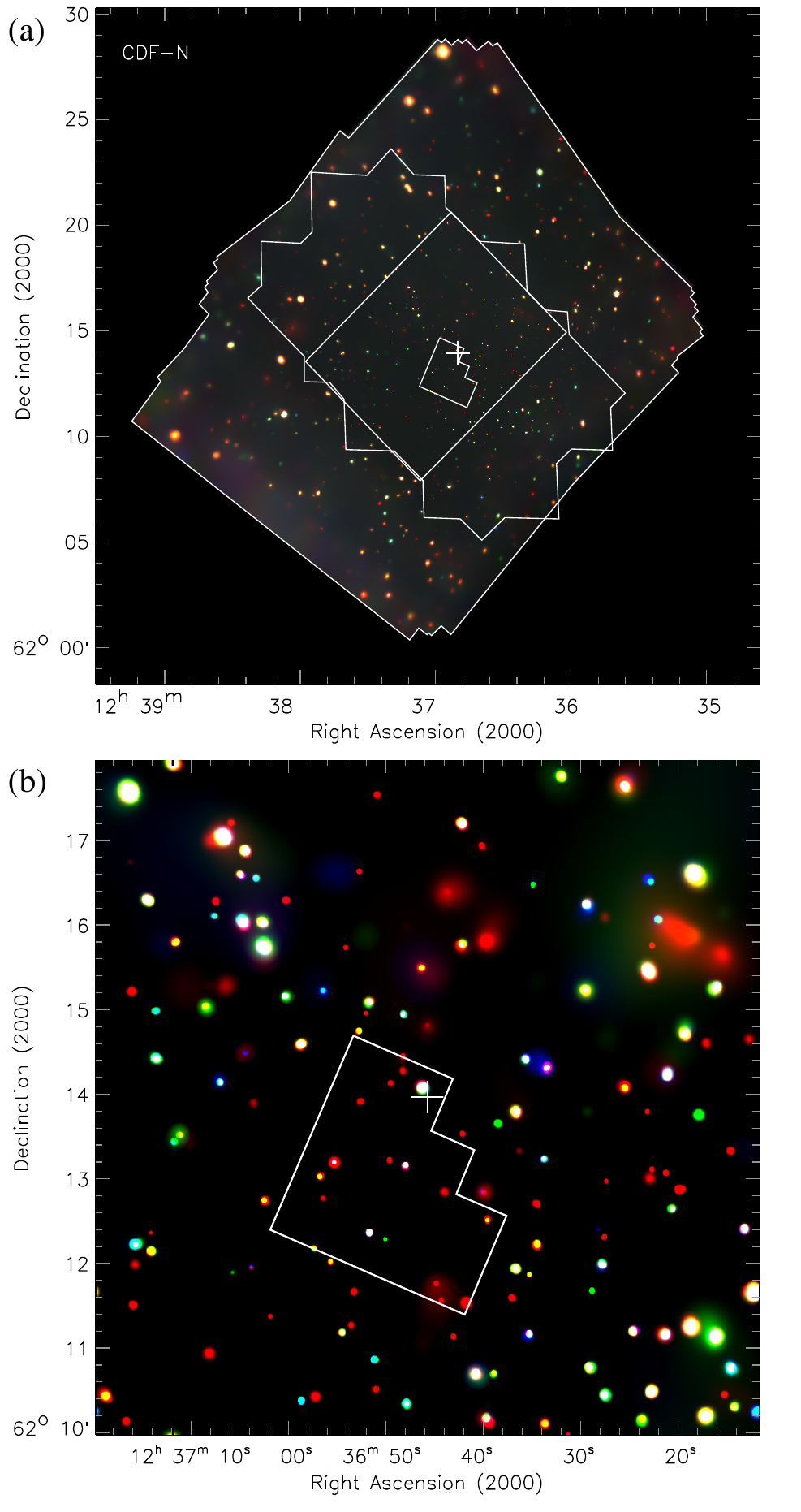}}
\figcaption{(a) False-color image of the 2~Ms \cdfn\ that
is a color composite of the exposure-weighted and adaptively smoothed images
in the 0.5--2.0 keV ({\it red}), 2--4 keV ({\it green}), and 4--7 keV ({\it blue}) bands. 
(b) A zoomed-in view of the false-color image of the
central $8\arcmin \times 8\arcmin$ region.
Near the field center, the seemingly smaller sizes and lower brightnesses of sources
are due to the smaller PSF size on-axis.
The regions and the plus sign have the same meanings as those in Fig.~\ref{fig:cdfn-fb-img}. 
\label{fig:cdfn-smooth-false}}
\end{figure}

We create exposure-weighted smoothed images 
following Section~3.3 of Baganoff \etal (2003).
We first generate the raw images and effective-exposure maps 
in the \hbox{0.5--2},
\hbox{2--4}, and \hbox{4--7~keV} bands.
We then utilize {\sc csmooth} (Ebeling, White, \& Rangarajan 2006) to
adaptively smooth the raw images and effective-exposure maps.
We finally divide the smoothed images by the corresponding smoothed 
effective-exposure maps and combine the exposure-weighted smoothed 
images into a false-color composite, which is shown in 
Figure~\ref{fig:cdfn-smooth-false}.

\subsubsection{Candidate-List Catalog Production}\label{sec:cdfn-cand}

To perform a blind search of potential sources, we
run {\sc wavdetect} on each combined raw image 
in the three standard bands to search for likely sources
and to generate a candidate-list catalog,
utilizing a ``$\sqrt{2}$~sequence'' of wavelet scales 
(i.e., from 1, $\sqrt{2}$, 2, $\ldots$ to 16 pixels),
a false-positive probability threshold of $10^{-5}$ (sigthresh=$10^{-5}$),
and an appropriate merged PSF map.
We obtain the PSF map in the following way.
We first utilize {\sc mkpsfmap} to produce a soft-band PSF map
pixel by pixel for each individual observation, 
setting the ``energy'' parameter to 1.497~keV and the parameter of
``encircled counts fraction (ECF)'' to 0.393.
We then use {\sc dmimgfilt} to combine the individual PSF maps
into a merged one with the option of adopting the minimum PSF map
size at each pixel location rather than the average.
In the above {\sc wavdetect} runs,
our choices of parameters (i.e.,  
energy=1.497~keV, ECF=0.393, and minimum PSF map size) 
provide the best sensitivity to point-like 
sources across the entire field,
thus being able to detect as many candidate-list sources as possible.
However, these parameter choices in combination with sigthresh=$10^{-5}$
would inevitably introduce a non-negligible number of 
spurious sources that have \mbox{$\lsim 2$--3} source counts.
In Section~\ref{sec:cdfn-main-select},
we therefore construct a more conservative main catalog 
by determining additional detection significances of each candidate-list 
source in the three standard
bands and discarding sources with significances below 
an adopted threshold value.

Our candidate-list catalog contains 1003 \cdfn\ source candidates,
with each being detected in at least one of the three standard bands.
For these candidate sources, we adopt source positions  
in a prioritized order, i.e., the full-, soft-, or hard-band position.
We adopt a $2\farcs5$ matching radius to carry out cross-band matching for
sources lying within $6\arcmin$ of the average aim point 
(i.e., \mbox{$\theta<6\arcmin$}) and a $4\farcs0$ matching radius for sources
with $\theta \ge 6\arcmin$, with the mismatch probability
being $\approx 1$\% across the entire field.
We then make use of the AE-computed centroid and matched-filter positions
to improve the above {\sc wavdetect} source positions.
The {\sc wavdetect}, centroid, and matched-filter positions
are comparably accurate on-axis, while the matched-filter positions
are of better accuracy off-axis.
As such, we adopt centroid positions for sources located inside $\theta =8\arcmin$
and matched-filter positions for sources lying outside $\theta =8\arcmin$.

Utilizing AE, we compute photometry for the candidate-list catalog sources.
AE calculates the PSF model by simulating the \chandra\ HRMA with
the MARX\footnote{See http://space.mit.edu/CXC/MARX/index.html for the MARX manual.\label{ft:marx}}
ray-tracing simulator (version 4.4).
It then creates a polygonal extraction region,
rather than the ``traditional'' circular aperture (e.g., A03, L05), 
to approximate the $\approx 90$\% encircled-energy fraction (EEF) 
contour of a local PSF that is measured at 1.497~keV. 
For crowded sources with overlapping polygonal extraction regions,
AE automatically shrinks extraction regions (\mbox{$\approx$40--75\%} EEFs)
that are not overlapping and chosen to be as large as possible.
We utilize the AE ``BETTER\_BACKGROUNDS'' 
algorithm for background extraction, which models the spatial distributions of flux for the source 
of interest and its adjacent sources making use of unmasked data,
and then calculates local background counts
inside background regions removing
contributions from the source and its adjacent sources.
This algorithm generates accurate background extractions,
being particularly critical for crowded sources.
For each source, AE analyzes individual observations independently and
merges the data to produce photometry
with appropriate energy-dependent aperture corrections applied.

\subsection{Main \chandra\ Source Catalog}\label{sec:cdfn-main}

\subsubsection{Selection of Main-Catalog Sources}\label{sec:cdfn-main-select}

\begin{figure*}
\centerline{\includegraphics[width=13cm]{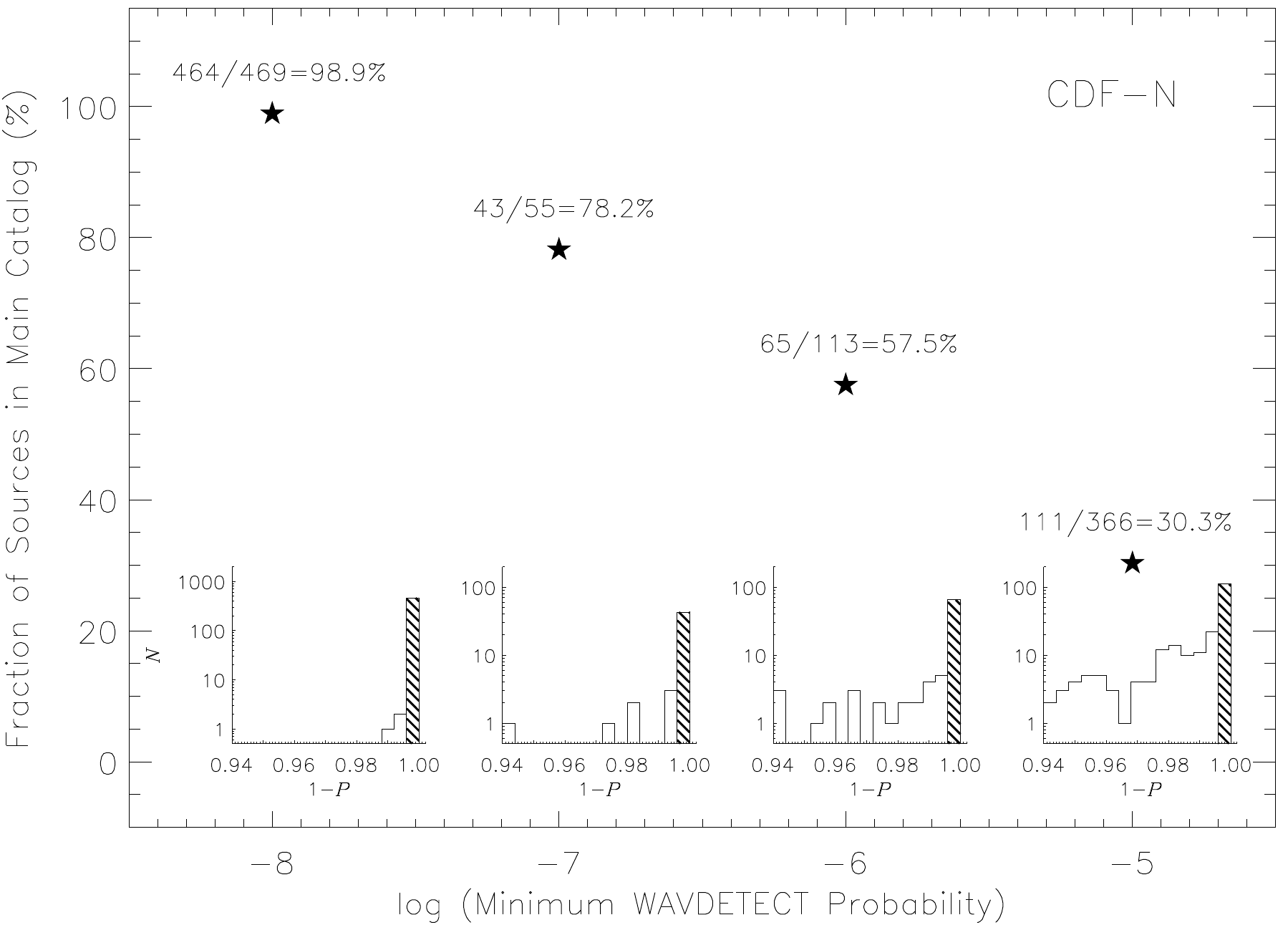}}
\figcaption{Fraction of the candidate-list catalog sources having an AE
binomial no-source probability $P<0.004$ that are included in
the 2~Ms \cdfn\ main catalog, as a function of minimum {\sc wavdetect} probability$^{\ref{ft:siglev}}$  
(denoted as five-pointed stars).
The number of sources having $P<0.004$ versus the number of
candidate-list catalog
sources detected at each minimum {\sc wavdetect} probability are
displayed (note that there are 
464+43+65+111=683 main-catalog sources and 469+55+113+366=1003 candidate-list catalog sources).
The fraction of candidate-list catalog sources included in the
main catalog falls monotonically from 98.9\% to 30.3\%
between minimum {\sc wavdetect} probabilities of $10^{-8}$
and $10^{-5}$.
The insets present the $1-P$ distributions for the
candidate-list catalog sources at each minimum {\sc wavdetect}
probability, and the shaded areas highlight those included in
the main catalog (i.e., satisfying $1-P>0.996$).
\label{fig:cdfn-prob-siglev}}
\end{figure*}

To cull spurious candidate-list catalog sources and thus
produce a reliable main \chandra\ source catalog,
we calculate for each candidate source 
the binomial no-source probability $P$ 
that no source exists
given the source and local background measurements,
which can be calculated as
\begin{equation}
P(X\ge S)=\sum_{X=S}^N \frac{N!}{X!(N-X)!} p^X (1-p)^{N-X},
\label{equ:bi}
\end{equation}
\noindent where $S$
is the total number of counts in the source-extraction region 
without subtracting the background
counts $B_{\rm src}$ therein;
$N=S+B_{\rm ext}$, with $B_{\rm ext}$ being the total background counts extracted
within a background-extraction region; and
$p=1/(1+BACKSCAL)$ with $BACKSCAL=B_{\rm ext}/B_{\rm src}$, 
being the probability that a photon is located inside
the source-extraction region.
AE computes $P$ in each of the three standard bands.
We include a candidate source in the main catalog only if
it has $P < 0.004$ in at least one of the three standard 
bands.\footnote{We note that our $P<0.004$ source-detection procedure associated 
with Eq.~\ref{equ:bi} can also be discussed in terms of False Discovery Rate (FDR) and 
Type I/II errors (i.e., false positives/negatives). We refer interested readers to
Benjamini \& Hochberg (1995) for a discussion of FDR.}$^{\rm ,}$\footnote{The adopted source-detection criterion of $P < (P_0=0.004)$ does not
straightforwardly indicate that, for a source with
$P < P_0$ in each of the three standard bands,
its final probability of being fake is $1-(1-P_0)^3\approx 3P_0$.
This is because only the SB and HB are truly distinct, while the FB, 
being the sum of the SB and HB, is dependent both on the SB and HB.
Furthermore, $P < P_0$ is only the second stage of the overall two-stage source-detection
approach (i.e., {\sc wavdetect} plus $P < P_0$), which implies that
the probability of a source being fake is not strictly $P_0$ even as far as 
only one single band is concerned.
Therefore, we rely on simulations (see Section~\ref{sec:cdfn-comp}) to obtain a realistic estimate of the reliability of
our main-catalog sources.}
The criterion of $P < 0.004$ results from a balance between
keeping the fraction of spurious sources small and 
recovering the maximum possible number of real sources,
primarily based on joint maximization of the total number of sources and
minimization of the fraction of sources without significant
multiwavelength counterparts (see Section~\ref{sec:cdfn-id}).
Our main catalog consists of a total of 683 sources given this $P < 0.004$
criterion.
Figure~\ref{fig:cdfn-prob-siglev} presents the fraction of 
candidate-list sources that satisfy the $P<0.004$ main-catalog source-selection criterion
and the $1-P$ distribution of candidate-list sources 
as a function of the minimum {\sc wavdetect} 
probability.\footnote{We also run {\sc wavdetect} with sigthresh=$10^{-6}$,
$10^{-7}$, and $10^{-8}$ in order to provide a more detailed
{\sc wavdetect}-based perspective on source significance.
The minimum {\sc wavdetect} probability
gives the {\sc wavdetect} significance with lower values
{representing} higher significances.
For instance, if a source was detected with {\sc wavdetect} in at least one of the
three standard bands at sigthresh=$10^{-7}$ but was not
detected in any of the three standard bands at sigthresh=$10^{-8}$, then the minimum {\sc wavdetect} probability
is $10^{-7}$ for this source.\label{ft:siglev}}

The cataloging procedure adopted in this work
is characterized by a number of
advantages over a ``traditional'' {\sc wavdetect}-only approach
(e.g., A03, L05; see Table~\ref{tab:impro}), including, e.g.,:
(1) the better PSF approximation (i.e., using MARX-simulated 
polygonal source-extraction regions rather than circular apertures)
that lays the foundation of accurate \xray\ photometry,
(2) the more sophisticated background treatment that takes into account
effects of both adjacent sources and CCD gaps, and
(3) the more flexible and reliable two-stage source-detection approach
that provides an effective identification of
real \mbox{X-ray} sources including those falling below the 
traditional more stringent {\sc wavdetect} searching threshold 
(e.g., sigthresh=$10^{-6}$).
Note that such a two-stage source-detection approach
has been implemented in a similar way in a number of previous 
studies (e.g., Getman \etal 2005; Nandra \etal 2005, 2015; 
Elvis \etal 2009; Laird \etal 2009; Lehmer \etal 2009; Puccetti \etal 2009; X11;
Ehlert \etal 2013).

\subsubsection{X-ray Source Positional Uncertainty}\label{sec:cdfn-dpos}

\begin{figure}
\centerline{\includegraphics[scale=0.6]{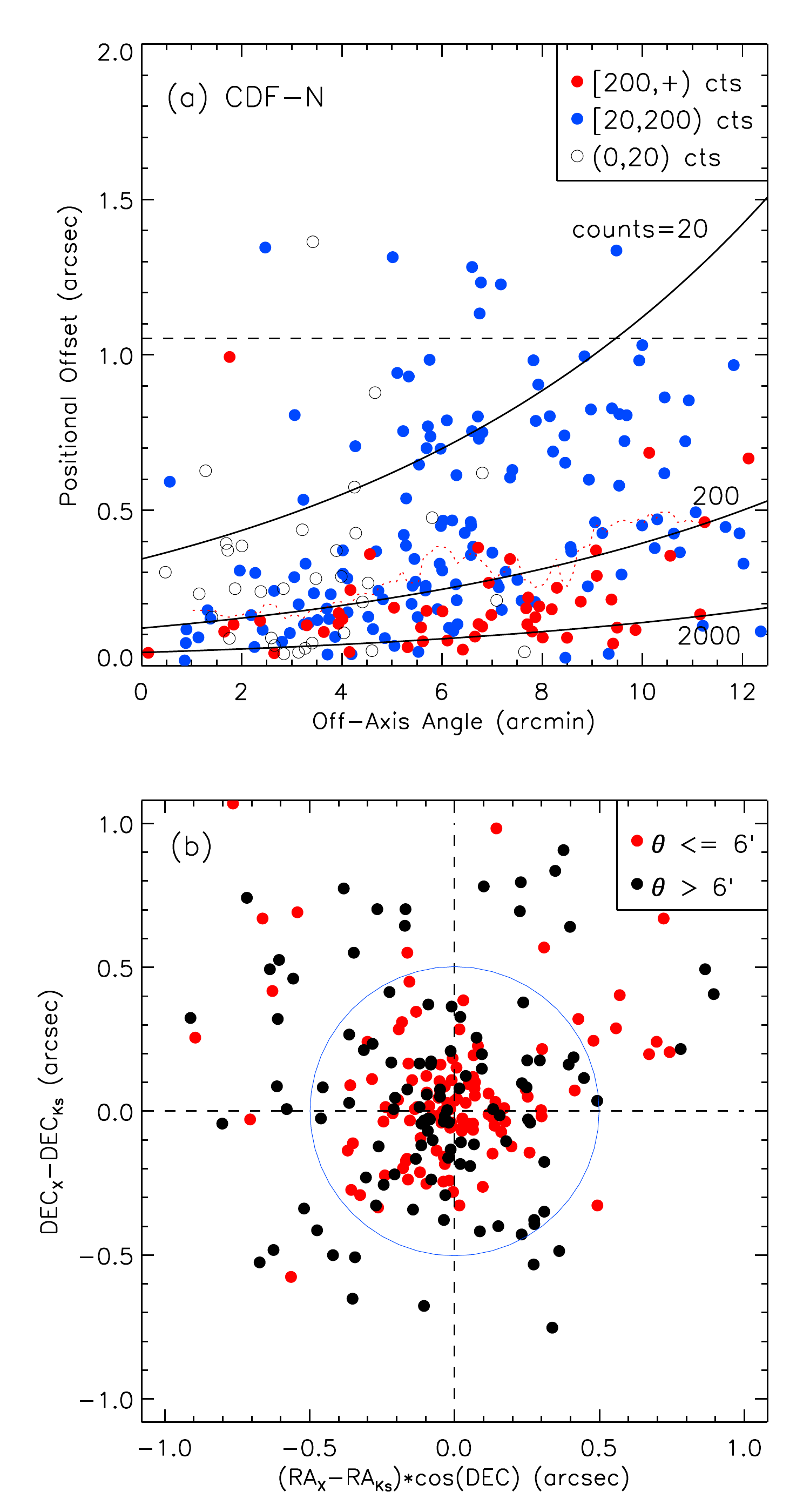}}
\figcaption{(a) Positional offset vs. off-axis angle for
the 230 2~Ms \cdfn\ main-catalog sources that have $K_s\le 20.0$~mag counterparts in the
\goodsn\ WIRCam $K_s$-band catalog (Wang \etal 2010) utilizing a matching radius of $1\farcs5$ (see Section~\ref{sec:cdfn-dpos}
for the description of an apparent outlier,
i.e., the red filled circle located around the top-left corner,
that deviates significantly from the relation defined as
Equation~\ref{equ:dpos}).
Red filled, blue filled, and black open circles indicate
\mbox{X-ray} sources having $\ge200$, $\ge20$, and $<20$
counts in the energy band that is used to determine the source position, respectively.
The red dotted curve denotes the running median of positional offset in
bins of $2\arcmin$.
The horizontal dashed line represents the median offset ($1\farcs05$)
of the false matches expected.
The three solid curves correspond to the $\approx68\%$ confidence-level \xray\ positional
uncertainties (derived according to Equation~\ref{equ:dpos}) for sources with 20, 200 and 2000 counts.
(b) Positional residuals between the \xray\ and $K_s$-band positions for 
the remaining 229 \xray-$K_s$-band matches.
Red and black filled circles represent sources with an off-axis angle
of $\le 6\arcmin$ and $>6\arcmin$, respectively.
A blue circle with a $0\farcs5$ radius is drawn at the center as visual guide.
\label{fig:cdfn-dposfit}}
\end{figure}

\begin{figure}
\centerline{\includegraphics[width=8.5cm]{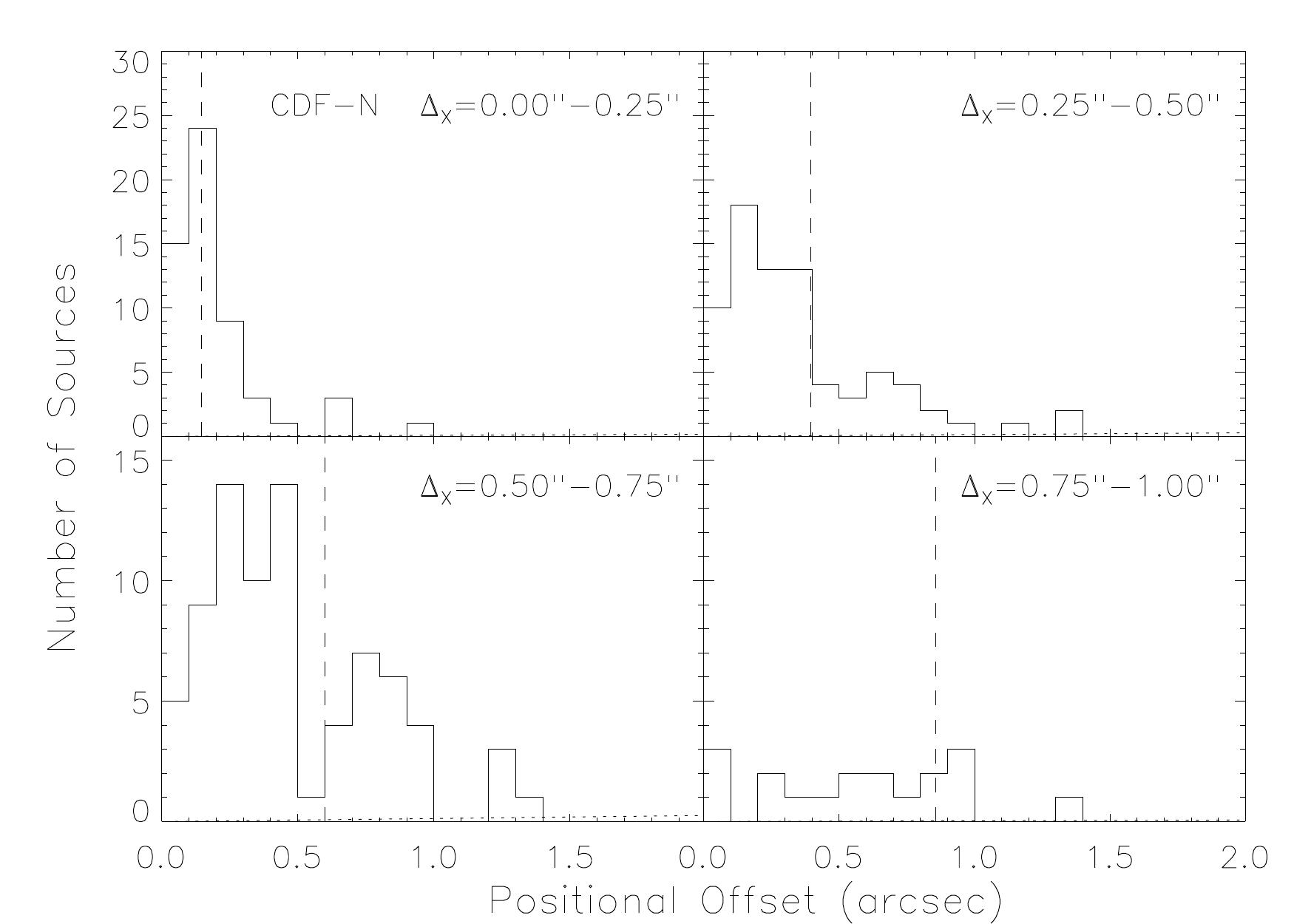}}
\figcaption{Histograms of \xray-$K_s$-band positional offsets for the 229 2~Ms \cdfn\ main-catalog
sources that are matched to the \goodsn\ WIRCam $K_s\le 20.0$~mag sources (Wang \etal 2010)
utilizing a matching radius of $1\farcs5$.
Based on their \xray\ positional
uncertainties estimated with Equation~(\ref{equ:dpos}),
these matched sources are divided into four bins of $0\arcsec$--$0\farcs25$, $0\farcs25$--$0\farcs50$,
$0\farcs50$--$0\farcs75$, and $0\farcs75$--$1\arcsec$.
In each panel (bin), the vertical dashed line denotes the median \xray\ positional uncertainty;
the dotted line (almost indistinguishable from the bottom $x$-axis) displays
the expected numbers of false matches
assuming a uniform spatial distribution of $K_s$-band sources
as a function of \xray-$K_s$-band positional offset.
\label{fig:cdfn-poshist}}
\end{figure}

We find 230 matches between the 683 main-catalog sources and
the $K_s\le 20.0$~mag sources in the \goodsn\ WIRCam $K_s$-band catalog 
using a matching radius of $1\farcs5$.
On average $\approx 5.1$ (2.2\%) false matches are expected, with
a median offset of $1\farcs05$ for these false matches.
Figure~\ref{fig:cdfn-dposfit}(a) presents the positional offset between
these 230 \hbox{X-ray}-$K_s$-band matches
(the median offset is $0\farcs28$) as a function of off-axis angle.
We find that 
the source indicated as a red filled circle around the top-left corner
is likely an off-nuclear source 
based on inspecting its \xray\ and {\it HST} images and
therefore do not include it in the following analysis of
\xray\ positional uncertainty.
Figure~\ref{fig:cdfn-dposfit}(b) presents 
the positional residuals between the \xray\
and $K_s$-band positions for the remaining 229 sources,  
which appear roughly symmetric.
Figure~\ref{fig:cdfn-dposfit}(a) reveals clear off-axis angle and 
source-count dependencies for these sources,
with the former caused by the degrading \chandra\ PSF
toward large off-axis angles
and the latter caused by statistical difficulties in 
identifying the centroid of a faint
\hbox{X-ray} source.
We adopt the Kim \etal (2007) functional form and
obtain an empirical relation for the positional uncertainty of our
main-catalog \mbox{X-ray} sources by fitting to the 229 \mbox{X-ray} sources 
with $K_s$-band counterparts, which is given as
\begin{equation}
\log \Delta_{\rm X}=0.0514 \theta-0.4538\log C+0.1262,\label{equ:dpos}
\end{equation}
\noindent
where $\Delta_{\rm X}$ is the \mbox{X-ray} positional uncertainty in units of arcseconds at the 68\% confidence level,
$\theta$ denotes the off-axis angle in units of arcminutes, and $C$ represents the source counts quoted in the energy
band that is used to determine the source position
(note that our Equation~\ref{equ:dpos} is very similar to
Equation~2 of X11). 
Figure~\ref{fig:cdfn-poshist} shows the distributions of
\xray-$K_s$-band positional
offsets in four bins of X-ray positional uncertainty.
When deriving Equation~\ref{equ:dpos} and presenting 
\hbox{X-ray}-$K_s$-band positional offsets 
in Figures~\ref{fig:cdfn-dposfit} and \ref{fig:cdfn-poshist},
we allow for positional uncertainties arising from
the $K_s$-band sources that are typically $\lsim$0\farcs1.

\subsubsection{Multiwavelength Identifications}\label{sec:cdfn-id}

We implement the Luo \etal (2010) likelihood-ratio matching procedure 
to identify the primary optical/near-infrared/mid-infrared/radio (ONIR) counterparts
for our main-catalog \xray\ sources.
We adopt, in order of priority (given factors of positional accuracy,
angular resolution, false-match rate, and catalog depth), 
six ONIR catalogs for identification purposes.
\begin{enumerate}
\item The VLA 1.4 GHz \goodsn\ radio catalog (denoted as ``VLA'';
Morrison \etal 2010), with a 5$\sigma$ detection threshold of
$\approx 20~\mu$Jy at the field center.
\item The \mbox{GOODS-N} {\it HST} version~2.0
F850LP ($z$)-band catalog 
(denoted as ``\mbox{GOODS-N}''; Giavalisco \etal 2004), 
with a $5\sigma$ limiting magnitude of $\approx 28.1$.
\item The CANDELS \goodsn\ WFC3 F160W-band catalog
(denoted as ``CANDELS''; Grogin \etal 2011; Koekemoer \etal 2011),
with a $5\sigma$ limiting magnitude of $\approx 27.4$.
\item The \goodsn\ WIRCam $K_s$-band catalog (denoted as ``Ks'';
Wang \etal 2010), with a $5\sigma$ limiting magnitude of $\approx 24.5$.
\item The \hhdfn\ Suprime-Cam $R$-band catalog (denoted as ``CapakR'';
Capak \etal 2004), with a $5\sigma$ limiting magnitude of $\approx 26.6$;
this catalog is complemented by the \hhdfn\ photometric-redshift catalog
(denoted as ``Yang14''; Yang \etal 2014).
\item The SEDS IRAC 3.6~$\mu$m-band catalog (denoted as ``IRAC'';
Ashby \etal 2013), with a $3\sigma$ limiting magnitude of $\approx 26.0$.
\end{enumerate}

We shift the above ONIR source positions appropriately to be consistent with the \goodsn\ WIRCam $K_s$-band 
astrometry (see Section~\ref{sec:cdfn-img}), by removing systematic positional offsets between
the ONIR and $K_s$-band coordinates of common sources that are matched using a matching radius of $0\farcs5$.
We identify primary ONIR counterparts for 670 (98.1\%) of 
the 683 main-catalog sources.
We estimate the false-match rates for the above six catalogs 
in the listed order to be
0.2\%, 3.7\%, 1.5\%, 1.3\%, 4.5\%, and 1.3\%, respectively, 
utilizing the Monte Carlo approach described in Broos \etal (2007, 2011)
rather than the simple shifting-and-recorrelating approach,
given that the Monte Carlo approach provides more realistic and reliable
estimates of false-match rates by taking into account 
different levels of susceptibilities to false matching 
associated with different \xray\ source populations
(also see Section~4.3 of X11 for more details). 
We derive the average false-match rate as 1.9\% by means of weighting the false-match
rates of individual ONIR catalogs with the number of 
counterparts in each catalog.
The high identification rate in conjunction with 
the small false-match rate serves as independent evidence 
that the vast majority of our main-catalog sources are robust.

We visually examine the \hbox{X-ray} images of
the 13 main-catalog sources without highly significant multiwavelength
counterparts, and find that
the majority of them have apparent or even strong \hbox{X-ray} signatures.
Of these 13 sources,
two are relatively bright sources (with 47.4 and 25.0 full-band counts) 
that are free of associations with 
any background flares or cosmic-ray afterglows.
These two sources are located near a very bright optical source
(their counterparts might thus be hidden by light of the
bright sources) and are also present
in the A03 main catalog. 
The other 11 sources are all fainter with $<20$ full-band counts
(some of them are thus likely false detections), 
none of which is present in the A03 main or supplementary catalog.

\subsubsection{Main-Catalog Details}\label{sec:cdfn-maincat}

For easy use of our main catalog,
we provide in Table~\ref{tab:cdfn-cols} a list of
a total of 72~columns in our main \chandra\ \hbox{X-ray} source catalog
(note that the contents of Table~\ref{tab:cdfn-cols} are very similar
to those of Table~2 of X11).
We present the main catalog itself in Table~\ref{tab:cdfn-main}.
Below we give the details of these 72~columns.

\begin{table*}
\caption{2~Ms \cdfn\ Main Catalog: Overview of Columns}
\begin{tabular}{ll}\hline\hline
Column(s) & \hspace{6.6cm}Description\\ \hline
1 & Source sequence number (i.e., XID) in this work \\
2, 3 & J2000.0 right ascension and declination of the \xray\ source \\
4 & Minimum value of $\log P$ among the three standard bands ($P$ is the binomial no-source probability calculated by AE) \\
5 & Logarithm of the minimum {\sc wavdetect} false-positive probability detection threshold \\
6 & \mbox{X-ray} positional uncertainty (in units of arcseconds) at the $\approx68\%$ confidence level \\
7 & Off-axis angle (in units of arcminutes) of the \hbox{X-ray} source \\
8--16 & Aperture-corrected net (i.e., background-subtracted) source counts and the associated errors for the three standard bands \\
17 & Flag of whether a source has a radial profile consistent with that of the local PSF \\
18, 19 & J2000.0 right ascension and declination of the primary optical/near-infrared/mid-infrared/radio (ONIR) counterpart \\
20 & Offset (in units of arcseconds) between the \mbox{X-ray} source and its primary ONIR counterpart \\
21 & AB magnitude of the primary ONIR counterpart \\
22 & Catalog name of the primary ONIR counterpart \\
23--40 & J2000.0 right ascension, declination, and AB magnitude of the counterpart in the six ONIR catalogs\\
41, 42 & Secure spectroscopic redshift and its reference \\
43--53 & Photometric-redshift information compiled from the literature \\
54 & Preferred redshift adopted in this work \\
55--57 & Corresponding XID, J2000.0 right ascension, and declination of the A03 main- and supplementary-catalog sources \\
58--60 & Effective exposure times (in units of seconds) derived from the exposure maps for the three standard bands\\
61--63 & Band ratio and the associated errors \\
64--66 & Effective photon index and the associated errors \\
67--69 & Observed-frame fluxes (in units of \flux) for the three standard bands\\
70 & Absorption-corrected, rest-frame \hbox{0.5--7~keV} luminosity (in units of \lum)$^a$ \\
71 & Estimate of likely source type (AGN, Galaxy, or Star)\\
72 & Note on the source (whether the source is in a close double or triple)\\ \hline
\end{tabular}
\\$^a$ Note that $L_{\rm 0.5-8\ keV}$=1.066$\times L_{\rm 0.5-7\ keV}$
and $L_{\rm 2-10\ keV}$=0.721$\times L_{\rm 0.5-7\ keV}$, given the assumed $\Gamma_{\rm int}=1.8$ (see the description of Column~70 in Section~\ref{sec:cdfn-maincat} for details).
\label{tab:cdfn-cols}
\end{table*}

\begin{table*}
%\tabletypesize{\scriptsize}
%\tablewidth{0pt}
\caption{2~Ms \cdfn\ Main {\it Chandra} Source Catalog}
\begin{tabular}{lllcccccccccc}\hline\hline
No. & $\alpha_{2000}$ & $\delta_{2000}$ & $\log P$ & {\sc wavdetect} & Pos Err & Off-axis & FB & FB Upp Err & FB Low Err & SB & SB Upp Err & SB Low Err \\
(1) & (2) & (3) & (4) & (5) & (6) & (7) & (8) & (9) & (10) & (11) & (12) & (13) \\ \hline
1 & 12 35 12.35 & +62 16 35.1 &     \phantom{0}$-$2.7 &  $-$5 &  1.3 &  11.18 &    \phantom{0}39.1 &  $-$1.0 &  $-$1.0 &    20.7 &   \phantom{0}9.2 &   \phantom{0}8.0 \\
2 & 12 35 15.09 & +62 14 06.9 &    $-$17.8 &  $-$8 &  0.8 &  10.56 &    \phantom{0}55.2 &  10.3 &   \phantom{0}9.1 &    43.9 &   \phantom{0}8.8 &   \phantom{0}7.5 \\
3 & 12 35 16.70 & +62 15 37.9 &     \phantom{0}$-$9.0 &  $-$7 &  0.7 &  10.50 &    \phantom{0}61.2 &  13.3 &  12.1 &    33.3 &   \phantom{0}9.1 &   \phantom{0}7.9 \\
4 & 12 35 18.77 & +62 15 51.9 &    $-$14.7 &  $-$8 &  0.6 &  10.30 &   101.4 &  16.6 &  15.4 &    61.6 &  11.8 &  10.6 \\
5 & 12 35 19.40 & +62 13 40.5 &    $-$11.3 &  $-$8 &  0.6 &  10.06 &    \phantom{0}73.8 &  14.2 &  13.0 &    44.4 &   \phantom{0}9.8 &   \phantom{0}8.6 \\ \hline
\end{tabular}
The full table contains 72~columns of information for the 683 \xray\ sources.\\
(This table is available in its entirety in a machine-readable form in the online journal. A portion is shown here for guidance regarding its form and
content.)
\label{tab:cdfn-main}
\end{table*}

\begin{enumerate}

\item Column 1 gives the source sequence number (i.e., XID) in this work. 
Sources are sorted in order of increasing right ascension.

\item Columns 2 and 3 give the J2000.0 right ascension and declination 
(determined in Section~\ref{sec:cdfn-cand}) of the \mbox{X-ray} source, respectively.

\item Columns 4 and 5 give the minimum value of $\log P$ among the three standard bands,
with $P$ being the binomial no-source probability computed by AE, 
and the logarithm of the minimum {\sc wavdetect} false-positive 
probability detection threshold, respectively.
More negative values of $\log P$ and
{\sc wavdetect} false-positive probability threshold 
correspond to a source detection of higher significance.
For sources with $P=0$, we set $\log P=-99.0$.
We find a median value of $\log P=-10.7$ for the main-catalog sources, being
much smaller than the main-catalog selection threshold value of 
$\log P<-2.4$ (i.e., $P<0.004$; see Section~\ref{sec:cdfn-main-select}).
We find that 464, 43, 65, and 111 sources have minimum {\sc wavdetect} 
probabilities$^{\ref{ft:siglev}}$ of $10^{-8}$, $10^{-7}$, $10^{-6}$, and $10^{-5}$, 
respectively (see Fig.~\ref{fig:cdfn-prob-siglev}).

\item Column 6 gives the \mbox{X-ray}
positional uncertainty in units of arcseconds at the $\approx68\%$ confidence level,
which is computed utilizing 
Equation~(\ref{equ:dpos}) that is dependent on both off-axis angle
and aperture-corrected net source counts.
For the main-catalog sources, the positional uncertainty ranges from
$0\farcs10$ to $2\farcs02$, with a median value of $0\farcs47$.

\item Column 7 gives the off-axis angle of each \hbox{X-ray} source in units of arcminutes
that is the angular separation between the \hbox{X-ray} source
and the average aim point given in Section~\ref{sec:cdfn-obs-info}.
For the main-catalog sources, the off-axis angle ranges from
$0\farcm 13$ to $14\farcm 63$, with a median value of $6\farcm 01$
(see Section~\ref{sec:cdfn-obs-info} for the observational pointing
scheme and roll constraints that lead to such a wide range 
of off-axis angles).

\item Columns 8--16 give the aperture-corrected net 
(i.e., background-subtracted) source counts and
the associated $1\sigma$ upper and lower statistical errors (Gehrels 1986)
for the three standard bands (computed in Section~\ref{sec:cdfn-cand}), respectively.
We treat a source as being ``detected'' for photometry purposes
in a given band only if it satisfies $P<0.004$ in that band.
We calculate upper limits for sources not detected in a given band,
according to the Bayesian method of
Kraft \etal (1991) for a 90\% confidence level,
and set the corresponding errors to $-1.00$.

\item Column 17 gives a flag indicating whether a source shows
a radial profile consistent with that of the local PSF.
This analysis is motivated by the fact that 
the use of 9 wavelet scales up to 16 pixels in the {\sc wavdetect} runs
in Section~\ref{sec:cdfn-cand}
potentially allows detection of extended sources on such scales 
compared to local PSFs.
From the merged PSF image, we initially derive a set of cumulative EEFs 
by means of extracting the PSF power within a series of circular apertures up to
a 90\% EEF radius.
From the merged source image, we subsequently derive another set of cumulative EEFs
by means of extracting source counts within a series of circular apertures 
up to the same 90\% EEF.
Finally, we make use of a Kolmogorov-Smirnov (K-S) test to calculate
the probability ($\rho_{\rm K-S}$) of the two sets of
cumulative EEFs being consistent with each other.
Of the 683 main-catalog sources, we find that
all but 15 have $\rho_{\rm K-S}>0.05$, i.e., 
these sources have radial profiles consistent with that of 
their corresponding PSFs above a 95\% confidence level
(thus being likely point-like sources), 
and set the value of this column to 1 for these sources.
We then set the value of this column to 0 for 
the 15 sources with $\rho_{\rm K-S}\le 0.05$, which
are located across the entire \cdfn\ field and 
show no pattern of spatial clustering.
Furthermore, we visually inspect these 15 sources and do not find any significant signature of extension.

\item Columns 18 and 19 give the right ascension and declination
of the primary ONIR counterpart (shifted accordingly to be
consistent with the $K_s$-band astrometric frame; 
see Section~\ref{sec:cdfn-id}).
Sources without ONIR counterparts have these two columns
set to \hbox{``00 00 00.00''} and \hbox{``$+$00 00 00.0''}.

\item Column 20 gives the offset between the \mbox{X-ray} source and
the primary ONIR counterpart in units of arcseconds. 
Sources without ONIR counterparts have this column set to $-1.00$.

\item Column 21 gives the AB magnitude of the primary ONIR counterpart
in the counterpart-detection band.\footnote{The radio AB
magnitudes are converted from the radio flux densities using $m({\rm AB}) =-2.5 \log(f_\nu)-48.60$.\label{ft:radio_mag}}
Sources without ONIR counterparts have this column set to $-1.00$.

\item Column 22 gives the name of the ONIR catalog (i.e., VLA,
GOODS-N, CANDELS, Ks, CapakR/Yang14, or IRAC; 
see Section~\ref{sec:cdfn-id}) where the primary 
counterpart is found.
Sources without ONIR counterparts have a value set to ``...''.

\item Columns 23--40 give the counterpart right ascension, declination, 
and AB magnitude$^{\ref{ft:radio_mag}}$ from the above six ONIR catalogs
(the coordinates have been shifted accordingly to be
consistent with the $K_s$-band astrometric frame; 
see Section~\ref{sec:cdfn-id}).
We match the position of the primary ONIR counterpart (i.e., 
Columns 18 and 19) with the six ONIR catalogs
using a matching radius of $0\farcs5$. 
We set values of right ascension
and declination to \hbox{``00 00 00.00''} and \hbox{``$+$00 00 00.0''}
and set AB magnitudes to $-1.00$ for sources without matches.
We find 31.3\%, 55.2\%, 57.2\%, 91.7\%, 68.5\%, and 87.0\%
of the main-catalog sources have
VLA, \mbox{GOODS-N}, CANDELS, Ks, CapakR/Yang14, and IRAC 
counterparts, respectively.

\item Columns 41 and 42 give the spectroscopic redshift ($z_{\rm spec}$)
and its corresponding reference.  
Only secure \zs's are collected and they are from
(1) Barger \etal (2008), 
(2) Cowie \etal (2004),
(3) Wirth \etal (2004),
(4) Cooper \etal (2011),
(5) Chapman \etal (2005),
(6) Barger \etal (2003), and
(7) Skelton \etal (2014).
The number preceding the corresponding reference is listed in Column~42.
We match the positions of primary ONIR counterparts
with the above \zs\ catalogs utilizing a $0\farcs5$ matching radius.
For the 670 main-catalog sources with ONIR counterparts, we find that
351 (52.4\%) have \zs\ measurements
(307/351=87.5\% have $R\le 24$~mag and 44/351=12.5\% have $R>24$~mag).
Sources without \zs\ have these two columns set to
$-1.000$ and $-1$, respectively.
The \zs\ histogram is shown in Fig.~\ref{fig:cdfn-zphot-comp}(a).

\begin{figure*}
\centerline{\includegraphics[width=17.5cm]{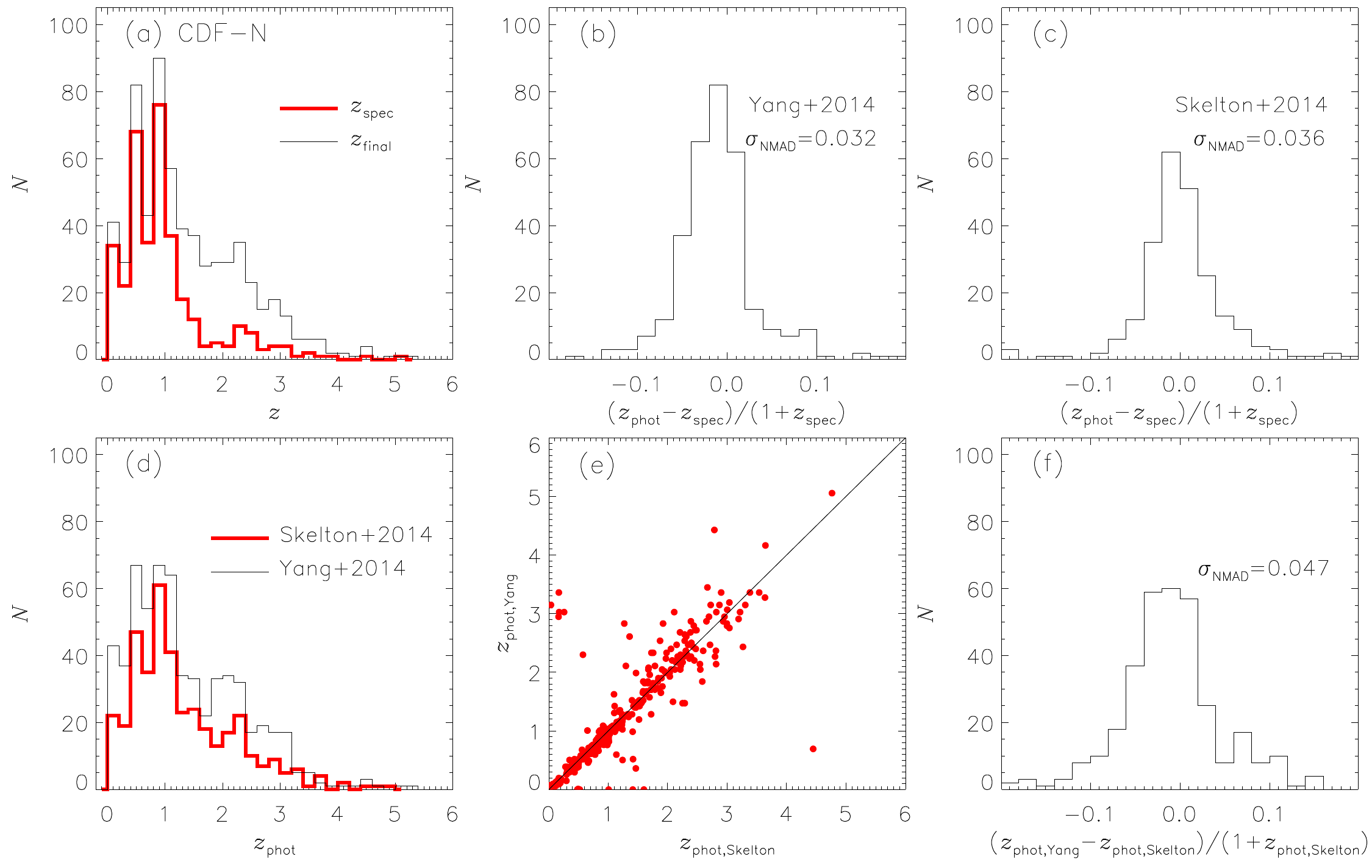}}
\figcaption{Redshift information for the 2~Ms \cdfn\ main-catalog sources.
(a) Histograms of \zs\ (351 sources; 351/683=51.4\%) and $z_{\rm final}$ (638 sources; 638/683=93.4\%). 
(b) Histogram of (\zp-\zs)/(1+\zs) from Yang \etal (2014; 347 sources) with \nmad\ annotated.  
(c) Histogram of (\zp-\zs)/(1+\zs) from Skelton \etal (2014; 264 sources) with \nmad\ annotated.  
(d) Histograms of $z_{\rm phot}$ from Yang \etal (2014; 612 sources) and Skelton \etal (2014; 389 sources).
(e) The Yang \etal\ \zp\ versus the Skelton \etal\ \zp\ for the 365 sources that have \zp\ estimates from both catalogs.
(f) Histogram of $(z_{\rm phot,Yang}-z_{\rm phot,Skelton})/(1+z_{\rm phot,Skelton})$ for these 365 sources with \nmad\ annotated.
\label{fig:cdfn-zphot-comp}}
\end{figure*}

\item Columns 43--53 give the photometric-redshift ($z_{\rm phot}$) 
information compiled from the literature.
Columns 43--48 give the \zp, the associated $1\sigma$
lower and upper bounds, the associated quality flag $Q_z$
(smaller $Q_z$ values denote better quality, with \hbox{$0<Q_z\lsim 1$}
indicating a reliable \zp\ estimate),
the alternative \zp\ (set to $-1.000$ if not available), and
the likely photometric classification 
(``Galaxy'', ``Star'', or ``Xray\_AGN'';
``...'' indicates lacking relevant information)
from the \hhdfn\ \zp\ catalog of Yang \etal (2014).
Columns 49--53 give the \zp, the associated $1\sigma$
lower and upper bounds, $Q_z$, and 
the likely photometric classification
(``Galaxy'' or ``Star'') from the CANDELS/3D-\hst\ \zp\ catalog
of Skelton \etal (2014). 
We match the positions of primary ONIR counterparts 
with the above \zp\ catalogs utilizing a $0\farcs5$ matching radius.
Of the 670 main-catalog sources with ONIR counterparts,
612 (91.3\%) and 389 (58.1\%) have \zp\ estimates from
Yang \etal (2014) and Skelton \etal (2014), respectively.
Sources without \zp's have all these columns set to $-1.000$ or ``...'' correspondingly.
Figures~\ref{fig:cdfn-zphot-comp}(b--d) show
the histograms of $(z_{\rm phot}-z_{\rm spec})/(1+z_{\rm spec})$ and
\zp\ for these two \zp\ catalogs.
The Yang \etal histogram of $(z_{\rm phot}-z_{\rm spec})/(1+z_{\rm spec})$ seems
skewed toward slightly negative values (i.e., by $\lsim 2\%$), which is likely caused
by some systematic errors of a small subset of the adopted templates in \zp\ estimation;
there appears no such skewness for the Skelton \etal histogram of $(z_{\rm phot}-z_{\rm spec})/(1+z_{\rm spec})$. 
Figures~\ref{fig:cdfn-zphot-comp}(e) and (f) show the Yang \etal\ \zp\ versus the Skelton \etal\
\zp\ and the histogram of $(z_{\rm phot,Yang}-z_{\rm phot,Skelton})/(1+z_{\rm phot,Skelton})$,
with both revealing general agreement between the two sets of \zp\ estimates and the latter again
indicating the above slightly negative skewness of the Yang \etal\ \zp.
We caution that the quoted \zp\ qualities, as indicated by values of \nmad\ 
annotated in Figures~\ref{fig:cdfn-zphot-comp}(b) and (c),
do not necessarily represent realistic estimates because
those \zp\ qualities are not derived using blind tests
(see, e.g., Section~3.4 of Luo \etal 2010 for relevant discussion)
and in some cases ``training biases'' are involved in \zp\ derivation
(e.g., the Skelton \etal \zp\ catalog makes use of template correction).

\item Column 54 gives the preferred redshift (\zf) adopted in this work.
We choose \zf\ for a source in the following order of preference:
secure \zs, the CANDELS/3D-\hst\ \zp, and the \hhdfn\ \zp.
Of the 670 main-catalog sources with ONIR counterparts,
638 (95.2\%) have \zs's or \zp's.

\item Column 55 gives the corresponding source ID number
in the A03 2~Ms \cdfn\ catalogs.
We match our \mbox{X-ray} source positions 
to the A03 source positions (shifted accordingly to be 
consistent with the $K_s$-band astrometric frame) 
using a $2\farcs5$ matching radius for sources
having $\theta <6\arcmin$ and a $4\farcs0$ matching radius
for sources having $\theta \ge 6\arcmin$.
Among the 683 main-catalog sources, we find that
(1) 487 have matches 
in the 503-source A03 main catalog
(the value of this column is that from Column~1 of Table~3a in A03), i.e.,
there are 196 (i.e., $683-487=196$) new main-catalog sources 
(see Section~\ref{sec:cdfn-new} for more details of these 196 new sources),
compared to the A03 main catalog;
(2) 45 have matches in the 78-source A03 supplementary
catalog (the value of this column is that from Column~1 of Table~7a in 
A03 added with a prefix of ``SP\_''); and 
(3) 151 have no match 
in either the A03 main or supplementary
catalog, which are detected now thanks to our two-stage source-detection
approach (the value of this column is set to $-1$).
We refer readers to Section~\ref{sec:cdfn-comp-old}
for the information of the 16 A03 main-catalog sources that
are not included in our main catalog.

\item Columns 56 and 57 give the right ascension and declination of
the corresponding A03 source (shifted accordingly to be         
consistent with the $K_s$-band astrometric frame).
Sources without an A03 match have these two columns
set to \hbox{``00 00 00.00''} and \hbox{``$+$00 00 00.0''}.

\item Columns 58--60 give the effective exposure times in units of seconds
derived from the exposure maps (see Section~\ref{sec:cdfn-img}) 
for the three standard bands.
Effective count rates that are corrected for effects of
vignetting, quantum-efficiency degradation, and exposure time variations
can be obtained by dividing the counts in Columns~8--16 by the corresponding effective
exposure times.

\item Columns 61--63 give the band ratio and the associated
upper and lower errors, respectively.
Band ratio is defined as the ratio of effective count rates 
between the hard and soft bands.
Band-ratio errors are computed according to
the numerical error-propagation method detailed in 
Section~1.7.3 of Lyons (1991),
which avoids the failure of the standard approximate variance
formula in the case of a small number of counts
(e.g., see Section~2.4.5 of Eadie \etal 1971).
Upper limits are computed for sources detected in the soft
band but not the hard band, while lower limits are computed for sources detected
in the hard band but not the soft band;
for these sources, the upper and lower
errors are set to the calculated band ratio.
Band ratios and associated errors are set to $-1.00$ for sources
with full-band detections only.

\item Columns 64--66 give the effective photon index ($\Gamma$) and
the associated upper and lower errors, respectively, 
assuming a power-law model with the Galactic
column density that is given in Section~\ref{sec:intro}.
$\Gamma$ is calculated based on the band ratio and
a conversion between $\Gamma$ and the band ratio.
This conversion is derived utilizing the band ratios and photon indices
computed by the AE-automated XSPEC-fitting procedure for sources
with $>200$ full-band counts.
Upper limits are computed for sources detected in the hard band 
but not the soft band, while lower limits are computed for sources detected in the soft band but not the hard band;
for these sources, the upper and lower errors are set to 
the calculated $\Gamma$.
A value of $\Gamma=1.4$ is assumed for low-count sources,
being a representative value for faint
sources that enables reasonable flux estimates,
and the associated upper and lower errors are set to 0.00.
Low-count sources are defined as those that were (1) detected in the
soft band having $<30$ counts and not detected in the hard band, 
(2) detected in the hard band having $<15$ counts and not detected 
in the soft band, (3) detected
in both the soft and hard bands, but having $<15$ counts in each, 
or (4) detected only in the full band.

\item Columns 67--69 give observed-frame fluxes in units of \flux\ 
for the three standard bands.
Fluxes are calculated making use of the net counts (Columns~\hbox{8--16}), 
the effective exposure times (Columns~58--60), and 
$\Gamma$ (Column~64), based on conversions
derived from XSPEC-fitting results.
Fluxes are not corrected for Galactic absorption or 
intrinsic absorption of the source.
Negative fluxes denote upper limits.

\item Column 70 gives a basic estimate of the absorption-corrected, 
rest-frame \hbox{0.5--7~keV} luminosity
($L_{\rm 0.5-7\ keV}$ or $L_{\rm X}$) in units of \hbox{erg s$^{-1}$}.
$L_{\rm 0.5-7\ keV}$ is computed utilizing the procedure presented in 
Section~3.4 of Xue \etal (2010).
First,  
this procedure adopts a power law with both Galactic and intrinsic absorption
to model the \hbox{X-ray} emission,
thus obtaining an estimate of the intrinsic column density 
that reproduces the observed band ratio under the assumption of
$\Gamma_{\rm int}=1.8$ for intrinsic AGN spectra;
subsequently, it derives the absorption-corrected flux by means of correcting for 
both Galactic and intrinsic absorption and
obtains $L_{\rm 0.5-7\ keV}$ using \zf. 
We note that $L_{\rm 0.5-8\ keV}$=1.066$\times L_{\rm 0.5-7\ keV}$
and $L_{\rm 2-10\ keV}$=0.721$\times L_{\rm 0.5-7\ keV}$, given the assumed $\Gamma_{\rm int}=1.8$.
In this procedure, 
the observed band ratio is set to a value that
corresponds to $\Gamma=1.4$ for sources with full-band detections only;
and for sources with upper or lower limits on the band ratio, 
their upper or lower limits are adopted.
For sources without full-band detections,
their observed-frame full-band fluxes are estimated
by extrapolating their soft- or hard-band fluxes 
assuming $\Gamma=1.4$ and Galactic absorption.
Crude luminosity estimates derived this way 
are in general agreement (i.e., within $\approx 30\%$) with those 
derived through direct and detailed spectral fitting.
Sources without \zf\ have this column set to $-1.000$.

\item Column 71 gives a basic estimate of likely source type: 
``AGN'', ``Galaxy'', or ``Star''.
A source is classified as an AGN once it
satisfies at least one of the following four criteria
(see Section~4.4 of X11 for reasoning and caveats):
$L_{\rm 0.5-7\ keV}\ge 3\times 10^{42}$ \lum\ (i.e., luminous AGNs),
$\Gamma \le 1.0$ (i.e., obscured AGNs),
$\log(f_{\rm X}/f_R)>-1$ ($f_{\rm X}=f_{\rm 0.5-7\ keV}$, $f_{\rm 0.5-2\ keV}$, or $f_{\rm 2-7\ keV}$; $f_R$ is the \mbox{$R$-band} flux), and
$L_{\rm 0.5-7\ keV}\gsim 3\times (8.9\times 10^{17}L_{\rm R})$ 
($L_{\rm R}$ is the rest-frame 1.4~GHz monochromatic luminosity 
in units of W~Hz$^{-1}$).
A source is classified as a star if
(1) it has \zs=0, 
(2) it is one of the old late-type \xray-detected \cdfn\ stars studied 
in Feigelson \etal (2004), or
(3) it has a photometric classification of ``Star'' 
(see Columns~48 and 53) and is further
confirmed through visual inspection of optical images.
The sources that are not classified as either AGNs or stars are then 
regarded as ``galaxies''.
There are 591 (86.5\%), 75 (11.0\%), and 17 (2.5\%) of the 683 main-catalog sources
identified as AGNs, galaxies, and stars, respectively.

\item Column 72 gives brief notes on the sources.
Sources in close doubles or triples are annotated with ``C'' 
(a total of 27 such sources, which
have overlapping polygonal extraction regions 
corresponding to \hbox{$\approx 40$--75\%} EEFs; 
see \S~\ref{sec:cdfn-cand}); otherwise,
sources are annotated with ``...''.

\end{enumerate}

\subsubsection{Comparison with the A03 Main-Catalog Sources}\label{sec:cdfn-comp-old}

Table~\ref{tab:cdfn-det} summarizes the source detections 
in the three standard bands for the main catalog.
Of the 683 main-catalog sources,
622, 584, and 411 are detected in the full, soft, and hard bands, 
respectively;
as a comparison (see Table~4 of A03),
of the 503 A03 main-catalog sources,
479, 451, and 332 are detected in the full, soft, and hard bands, 
respectively (note that A03 adopt an upper energy bound of 8~keV).
As stated in Section~\ref{sec:cdfn-maincat} 
(see the description of Column~55),
487 of the main-catalog sources have matches in the A03 main catalog.
For these 487 common sources, we find that
the \hbox{X-ray} photometry derived in this work is in 
general agreement with that in A03,
e.g., the median ratio between our and the A03
soft-band count rates for the soft-band detected common sources
is 1.03, with an interquartile range of 0.92--1.11.
The significant increase in the number of main-catalog sources,
i.e., an increase of $683-487=196$ new main-catalog sources, 
is mainly due to the improvements of our cataloging methodology
that are summarized in Table~\ref{tab:impro}, in particular,
due to our two-stage source-detection approach.
Indeed, we are able to detect fainter sources than A03 
that are yet reliable, 
with median detected counts (see Table~\ref{tab:cdfn-det}) 
in the three standard bands being $\approx 70\%$ of those of A03.

\begin{table}
%\tabletypesize{\small}
%\tablewidth{0pt}
\caption{2~Ms \cdfn\ Main Catalog: Summary of Source Detections}
\centering
\begin{tabular}{lccccc}\hline\hline
Band & Number of & Maximum & Minimum & Median & Mean \\
(keV) & Sources & Counts & Counts & Counts & Counts \\\hline
Full (0.5--7.0)   & 622 & 19748.4 & 8.1 & 66.2 & 342.8 \\
Soft (0.5--2.0)  & 584 & 14227.3 & 5.4 & 35.0 & 234.7  \\
Hard (2--7)  & 411 & \phantom{0}5540.6 & 7.7 & 57.5 & 181.4\\ \hline
\end{tabular}
\label{tab:cdfn-det}
\end{table}

Sixteen (i.e., $503-487=16$) of the A03 main-catalog sources 
are not recovered in our main catalog,
among which 7 are recovered in our supplementary catalog 
(see Section~\ref{sec:cdfn-supp}).
Among the 9 A03 main-catalog sources that are not recovered in our main or
supplementary catalogs,
6 sources not only have faint \xray\ signatures,
but also have multiwavelength counterparts with a few being bright,
which indicates that most of them are likely real \hbox{X-ray} sources
although they do not satisfy our main-catalog or supplementary-catalog 
source-selection criterion.
The remaining 3 sources have marginal \xray\ signatures
and have no multiwavelength counterparts, thus being likely false detections. 

Table~\ref{tab:cdfn-undet} summarizes 
the number of sources detected in one band but not another in the main catalog
(cf. Table~5 of A03).
There are 19, 53, and 8 sources detected only in the full, soft, and hard bands, 
in contrast to 5, 23, and 1 source(s) 
in the A03 main catalog, respectively.

\begin{table}
%\tabletypesize{\small}
%\tablewidth{0pt}
\caption{2~Ms \cdfn\ Main Catalog: Sources Detected in One Band but not Another}
\centering
\begin{tabular}{lccc}\hline\hline
Detection Band & Nondetection & Nondetection & Nondetection \\
(keV) & Full Band & Soft Band & Hard Band \\\hline
Full (0.5--7.0)  & \ldots & 91 & 219 \\
Soft (0.5--2.0)  & ~~~~~53~~~~ & \ldots & 253 \\
Hard (2--7)   & ~~~~~\phantom{0}8~~~~ & 80  & \ldots \\\hline
\end{tabular}
\label{tab:cdfn-undet}
\end{table}

\subsubsection{Properties of Main-Catalog Sources}\label{sec:cdfn-prop}

\begin{figure}
\centerline{\includegraphics[width=8.5cm]{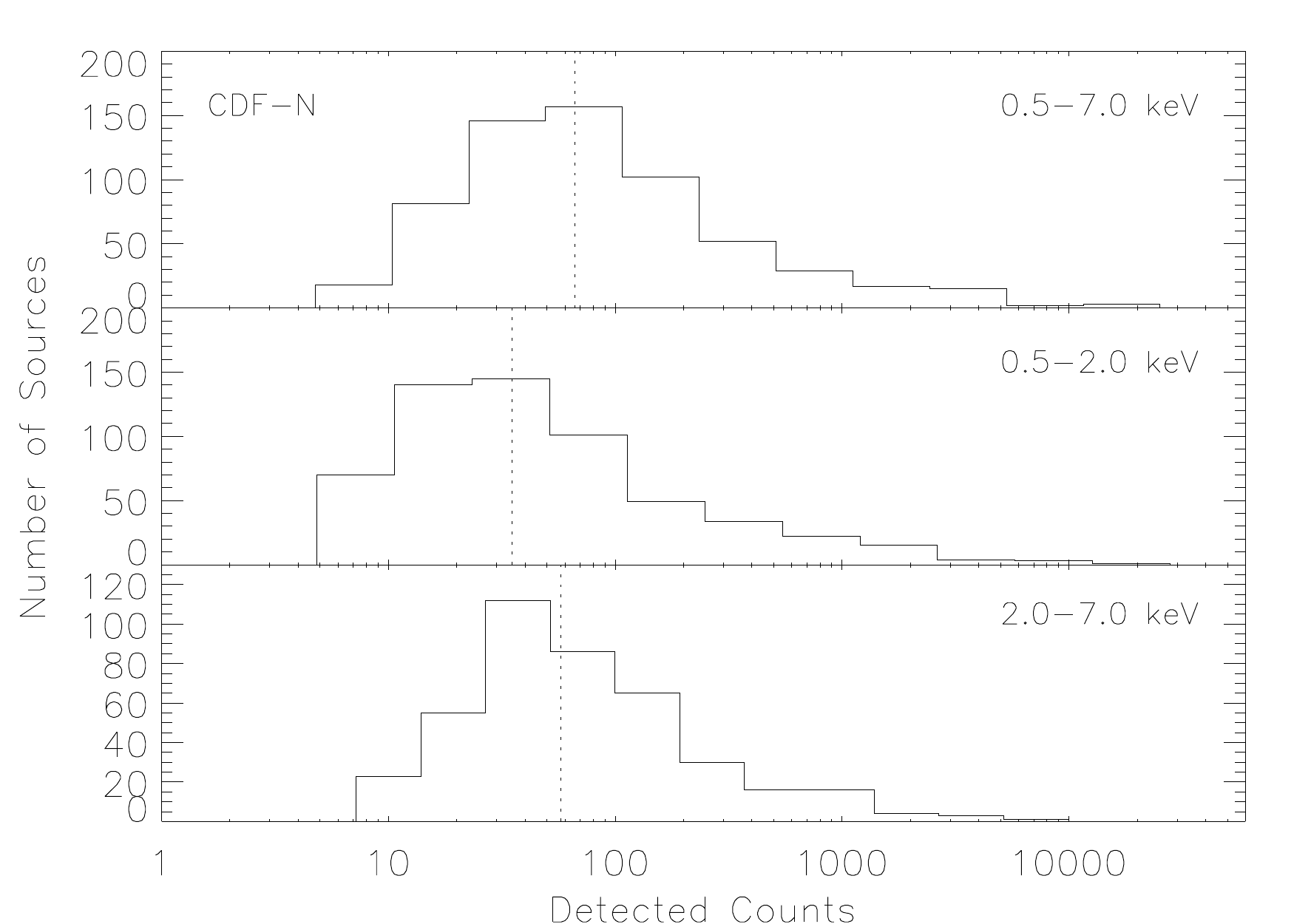}}
\figcaption{Distributions of detected source
counts for the 2~Ms \cdfn\ main-catalog sources
in the full, soft, and hard bands.
Sources with upper limits are not plotted.
The vertical dotted lines indicate the median detected counts
of 66.2, 35.0, and 57.5, for the full, soft, and hard bands, respectively
(detailed in Table~\ref{tab:cdfn-det}).
\label{fig:cdfn-cnthist}}
\end{figure}

Figure~\ref{fig:cdfn-cnthist} presents the histograms of detected
source counts in the three standard bands for the sources in the main catalog.
The median detected counts are 66.2, 35.0, and 57.5
for the full, soft, and hard bands, respectively;
and there are 232, 136, 67, and 41 sources having $>100$, $>200$, $>500$,
and $>1000$ full-band counts, respectively.

\begin{figure}
\centerline{\includegraphics[width=8.5cm]{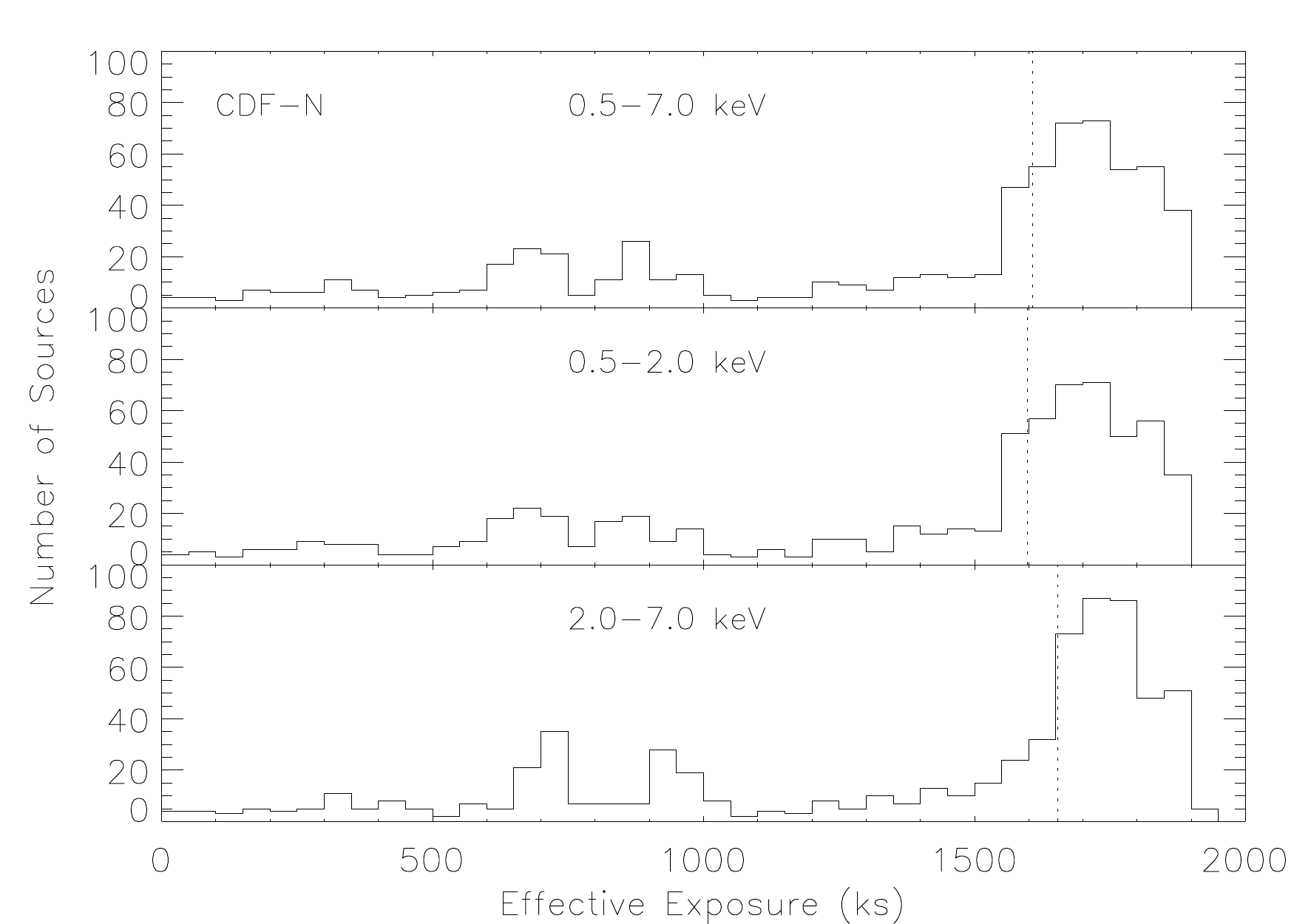}}
\figcaption{Distributions of effective exposure times
for all the 683 2~Ms \cdfn\ main-catalog sources
in the full, soft, and hard bands.
The vertical dotted lines indicate the median effective exposures
of 1607.5, 1597.0, and 1653.6~ks, for the full, soft, and hard bands, respectively.
\label{fig:cdfn-exphist}}
\end{figure}

Figure~\ref{fig:cdfn-exphist} presents the histograms of 
effective exposure times in the three standard bands for all the 683 main-catalog sources.
The median effective exposures are
1607.5, 1597.0, and 1653.6~ks for the full, soft, and hard bands, respectively.

\begin{figure}
\centerline{\includegraphics[width=8.5cm]{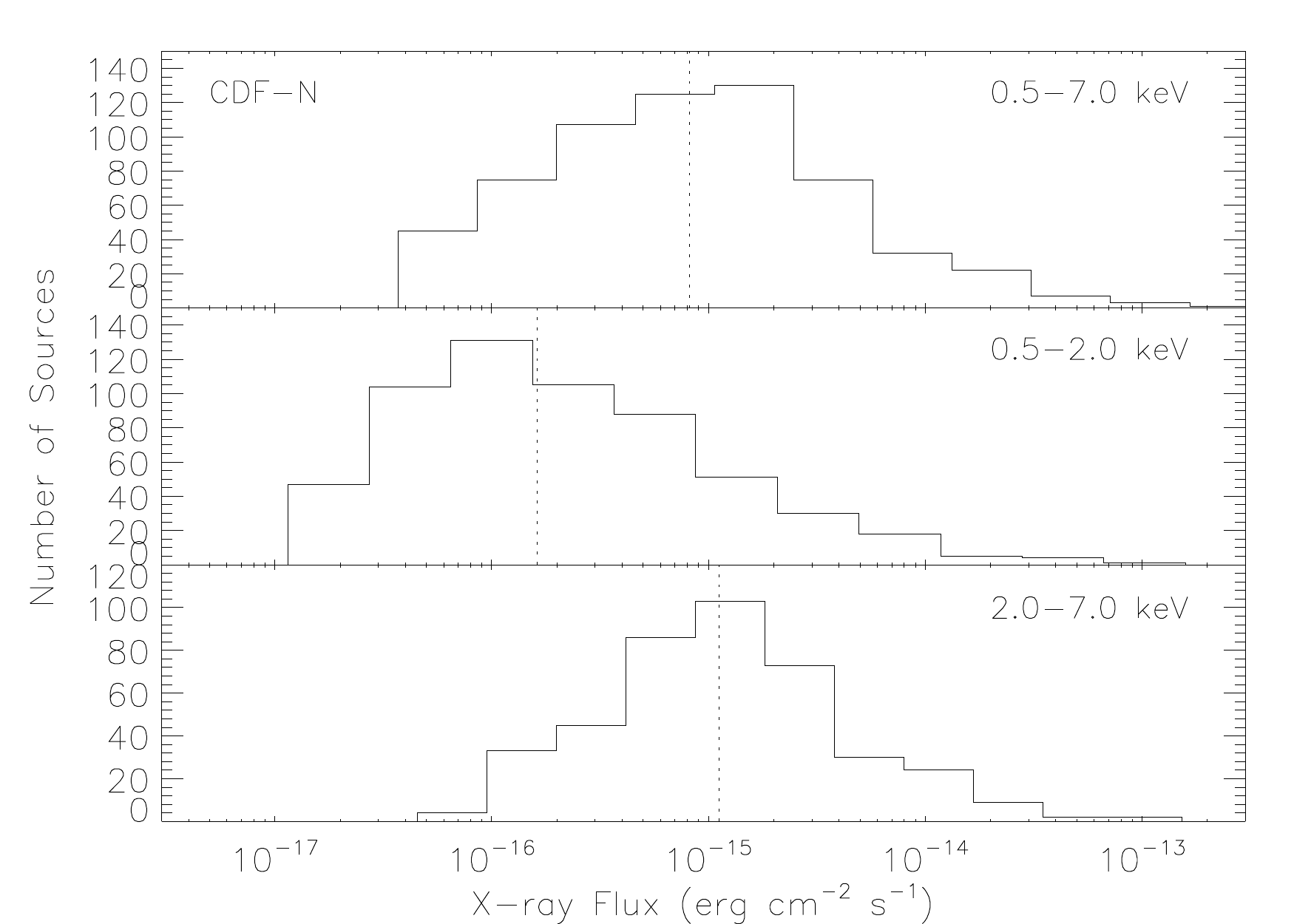}}
\figcaption{Distributions of observed-frame \hbox{X-ray} fluxes for the 2~Ms \cdfn\ main-catalog
sources in the full, soft, and hard bands.
Sources with upper limits are not plotted.
The vertical dotted lines denote the median fluxes of
$8.1\times10^{-16}$, $1.6\times10^{-16}$ and $1.1\times10^{-15}$ \flux\
for the full, soft, and hard bands, respectively.
\label{fig:cdfn-fluxhist}}
\end{figure}

Figure~\ref{fig:cdfn-fluxhist} presents the histograms of observed-frame \hbox{X-ray} fluxes
in the three standard bands for the sources in the main catalog.
The \hbox{X-ray} fluxes distribute within roughly four orders of
magnitude, with median values of
$8.1\times 10^{-16}$, $1.6\times 10^{-16}$, and $1.1\times 10^{-15}$ \flux\
for the full, soft, and hard bands, respectively.

\begin{figure}
\centerline{\includegraphics[width=8.5cm]{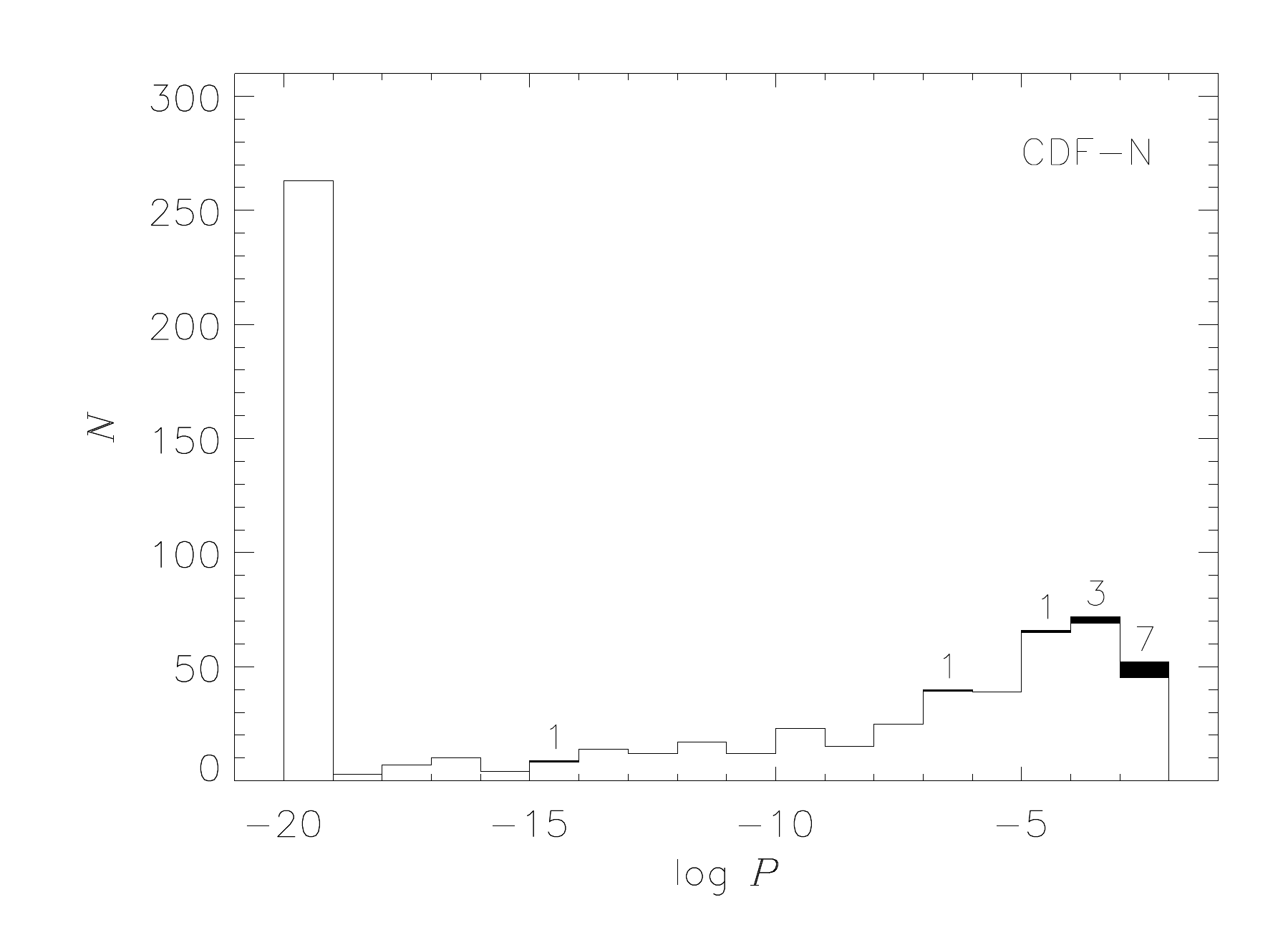}}
\figcaption{Distribution of the AE-computed binomial
no-source probability $P$ for the 2~Ms \cdfn\
main-catalog sources.
The values of $\log P<-20$ are set to $\log P=-20$ for easy illustration.
The shaded areas denote sources without multiwavelength counterparts,
with their corresponding numbers annotated.
\label{fig:cdfn-p-idrate}}
\end{figure}

Figure~\ref{fig:cdfn-p-idrate} presents the histogram of the AE-computed binomial
no-source probability $P$ for the sources in the main catalog,
with a total of 13 sources having no multiwavelength counterparts highlighted by shaded areas.
The majority of the main-catalog sources have low $P$ values that indicate significant detections,
with a median $P$ of $1.95\times 10^{-11}$ and an interquartile range of 
0.00 to $2.36\times 10^{-5}$.
We find that 0.4\% of the $\log P\le -5$ sources
have no ONIR counterparts, in contrast to 5.8\% of $\log P> -5$
sources lacking ONIR counterparts.
Given the small false-match rate estimated in Section~\ref{sec:cdfn-id},
a main-catalog source with a secure ONIR counterpart is almost certain to be real
(note that sources without ONIR counterparts are more likely
but not necessarily false detections).

\begin{figure*}
\centerline{\includegraphics[width=17.5cm]{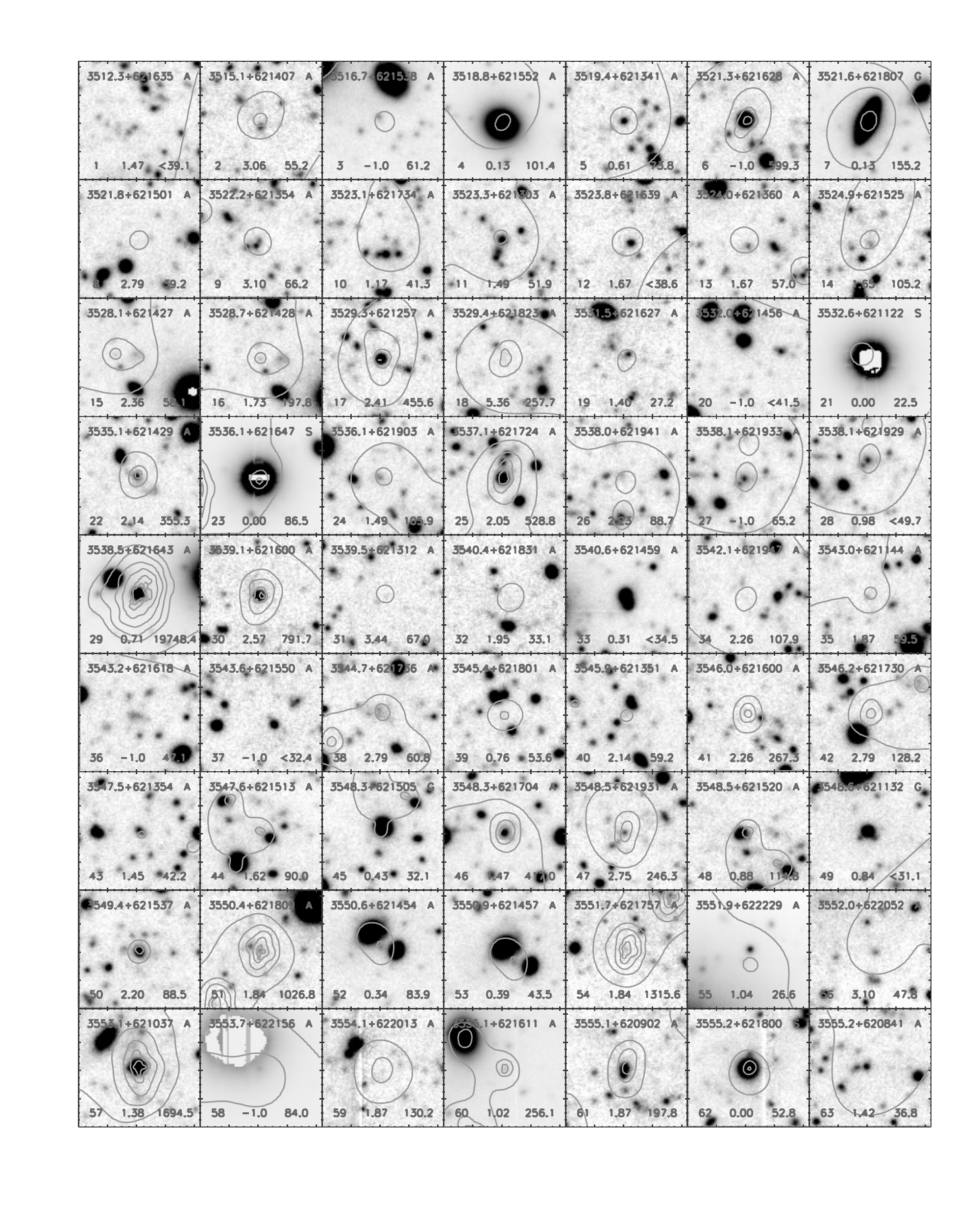}}
\figcaption{$25\arcsec\times 25\arcsec$ postage-stamp images from the \hhdfn\ Suprime-Cam $R$ band (Capak \etal 2004)
for the 2~Ms \cdfn\ main-catalog sources that are
centered on the \xray\ positions, overlaid
with full-band adaptively smoothed \hbox{X-ray} contours
that have a logarithmic scale and range from \hbox{$\approx$0.003\%--30\%}
of the maximum pixel value.
In each image, the labels at the top are the
source name (the hours ``12'' of right ascension are omitted for succinctness) 
and source type (A=AGN, G=Galaxy, and S=Star);
the bottom numbers
indicate the source \xray\ ID number, adopted redshift,
and full-band counts or upper limit (with a ``$<$'' sign).
There are cutouts (i.e., nearly plain white portions) in some images that
are caused by stellar light-induced saturation. 
In some cases there are
no \hbox{X-ray} contours present,
either due to these sources being not detected in the full band or
having low full-band counts leading to their observable emission in the adaptively
smoothed image being suppressed by {\sc csmooth}.
\\(An extended version of this figure is available in the online journal.)
\label{fig:cdfn-capakr-stamps-page1}}
\end{figure*}

\begin{figure*}
\centerline{\includegraphics[width=17.5cm]{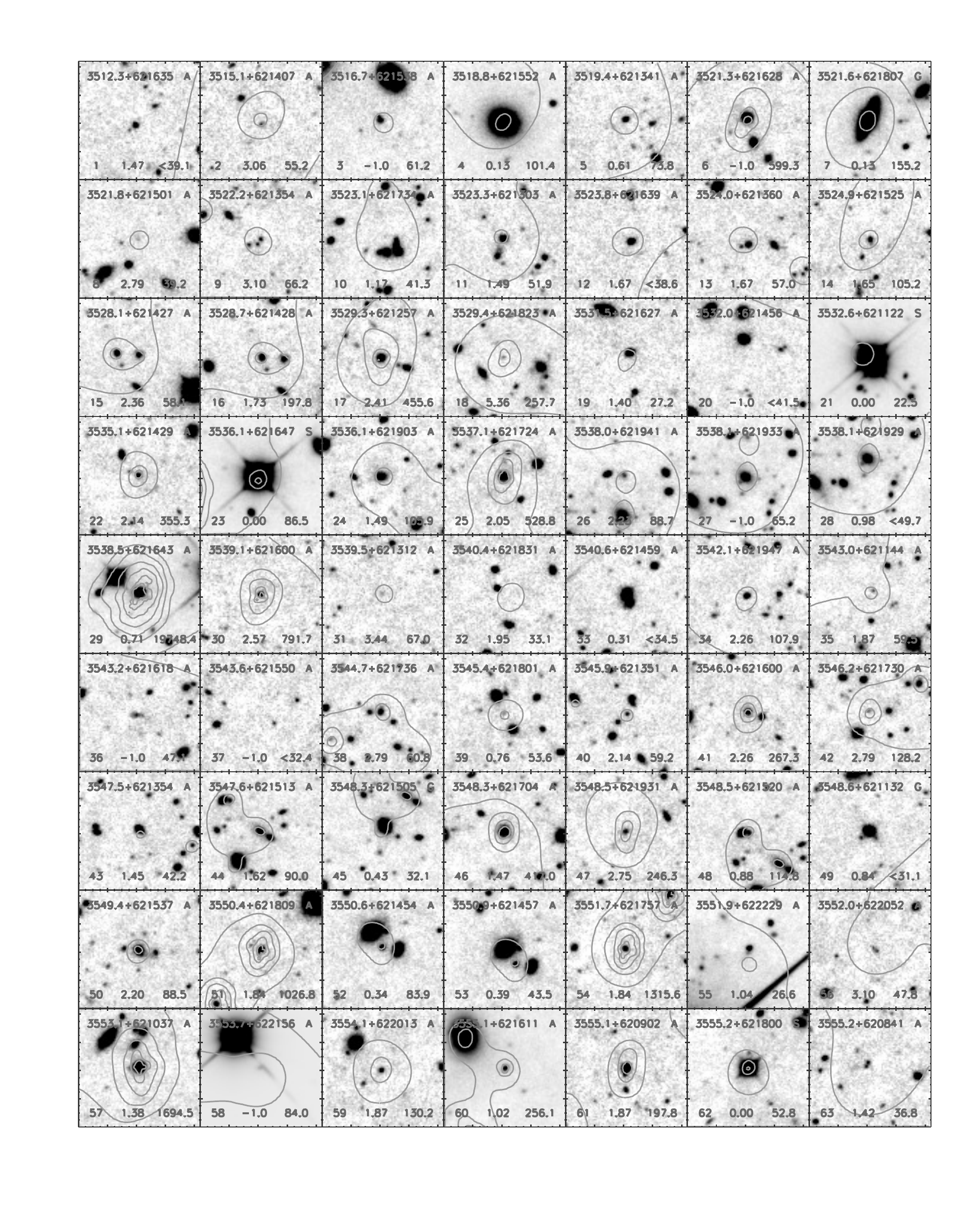}}
\figcaption{Same as Figure~\ref{fig:cdfn-capakr-stamps-page1}, but for the WIRCam $K_s$ band (Wang \etal 2010).
\\(An extended version of this figure is available in the online journal.)
\label{fig:cdfn-ks-stamps-page1}}
\end{figure*}

\begin{figure*}
\centerline{\includegraphics[width=17.5cm]{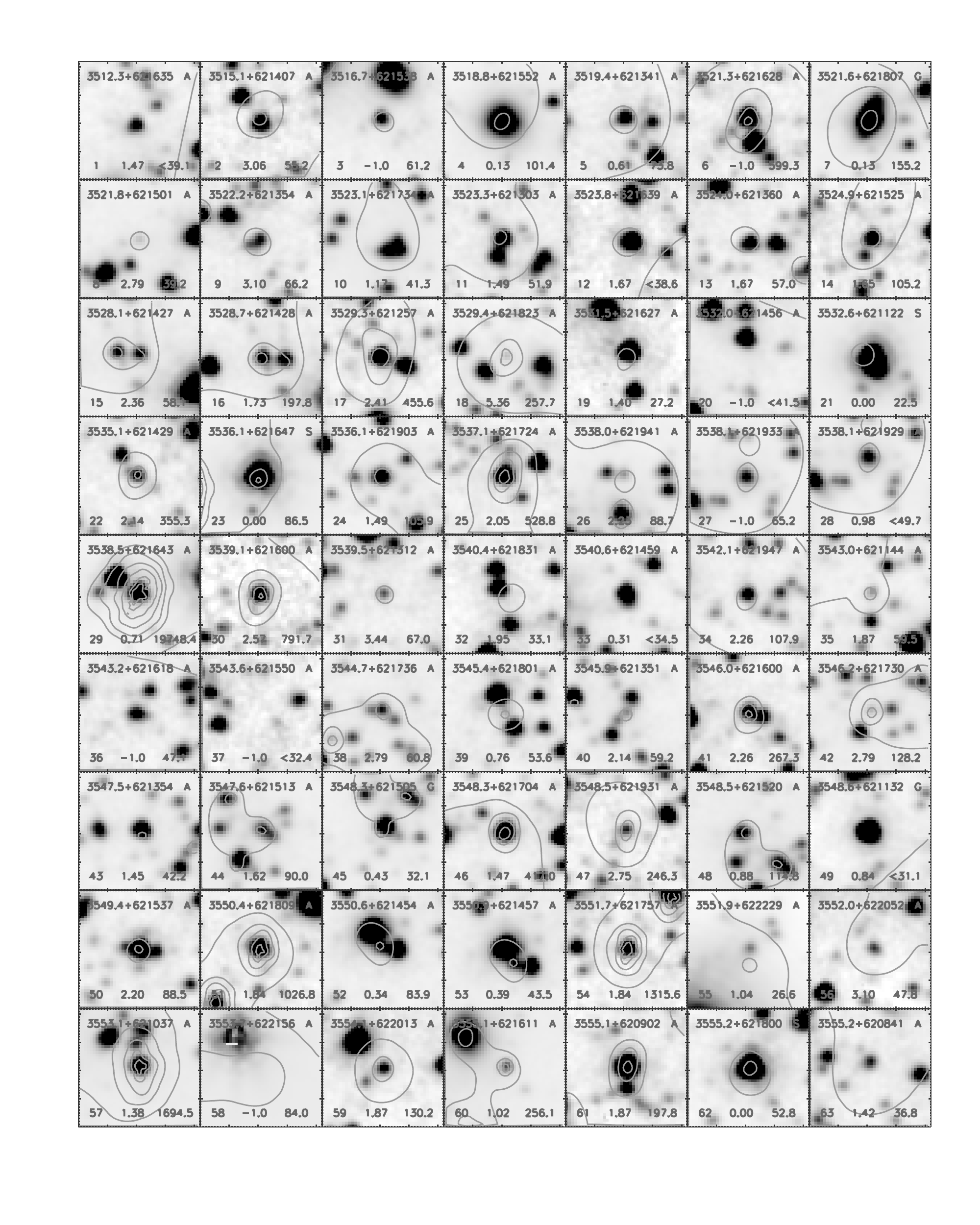}}
\figcaption{Same as Figure~\ref{fig:cdfn-capakr-stamps-page1}, but for the SEDS IRAC 3.6~$\mu$m band (Ashby \etal 2013).
\\(An extended version of this figure is available in the online journal.)
\label{fig:cdfn-irac-stamps-page1}}
\end{figure*}

Figures~\ref{fig:cdfn-capakr-stamps-page1}--\ref{fig:cdfn-irac-stamps-page1}
display $25\arcsec\times 25\arcsec$ postage-stamp images 
from the \hhdfn\ Suprime-Cam $R$ band (Capak \etal 2004),
the \goodsn\ WIRCam $K_s$ band (Wang \etal 2010), and
the SEDS IRAC 3.6~$\mu$m band (Ashby \etal 2013),
overlaid with adaptively smoothed full-band \xray\ contours for
the main-catalog sources, respectively.

\subsubsection{Properties of the 196 New Main-Catalog Sources}\label{sec:cdfn-new}

\begin{figure*} 
\centerline{\includegraphics[width=17.5cm]{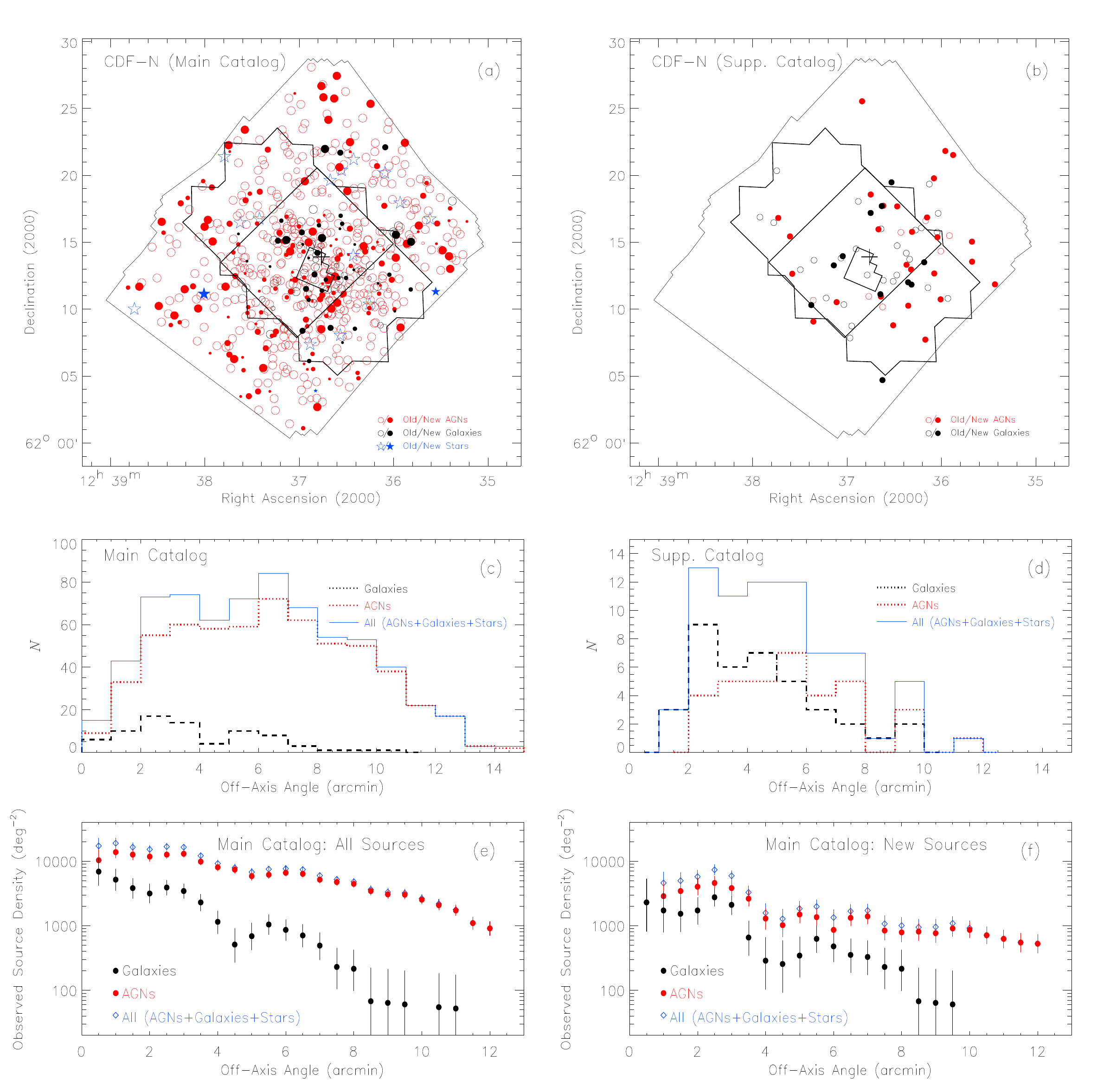}}
\figcaption{(Top) Spatial distributions for (a) the 2~Ms \cdfn\ main-catalog sources
and (b) the supplementary-catalog sources.
Sources classified as AGNs, galaxies, and stars
are plotted as red, black, and blue symbols, respectively.
Open symbols indicate old sources 
that were previously detected in (a) the A03 main catalog
or (b) the A03 main or supplementary catalog, while
filled symbols indicate new sources 
that were not previously detected in the A03 main and/or supplementary catalog.
The regions and the plus sign have the same meanings as those in Fig.~\ref{fig:cdfn-fb-img}.
In panel (a),
larger symbol sizes indicate lower AE binomial no-source probabilities,
ranging from $\log P >-3$, $-4<\log P\le -3$, $-5<\log P\le -4$, to
$\log P\le -5$; while in panel (b), all sources have $\log P >-3$ and are plotted as symbols of the same size.
(Middle) Distributions of off-axis angles for different source types for (c) the main-catalog sources
and (d) the supplementary-catalog sources. 
(Bottom) Observed source densities broken down into different source types as a function of off-axis angle ($\theta$)
for (e) all the 2~Ms \cdfn\ main-catalog sources and (f) the {\it new} main-catalog sources,
which are calculated in bins of $\Delta\theta=1\arcmin$ and whose
$1\sigma$ errors are computed utilizing Poisson statistics.
\label{fig:cdfn-pos}}
\end{figure*}

Figure~\ref{fig:cdfn-pos}(a) displays the spatial distributions
of the 196 new main-catalog sources (i.e., 154 new AGNs,
39 new galaxies, and 3 new stars that are all 
indicated as filled symbols) and 
the 487 old main-catalog sources (indicated as open symbols),
whose colors are coded based on source types
(red for AGNs, black for galaxies, and blue for stars)
and whose symbol sizes represent different $P$ values
(larger sizes denote lower $P$ values and thus 
higher source-detection significances).
The vast majority of both the new and old galaxies 
are located within the \goodsn\ area
that has the deepest exposures (see Fig.~\ref{fig:cdfn-fb-exp}),
as a result of their growing numbers
at the faintest fluxes
(e.g., Bauer \etal 2004; Lehmer \etal 2012).
Both the new and old AGNs 
spread out more evenly within the entire \cdfn\ field.
The above spatial distribution features
are also evident in Figure~\ref{fig:cdfn-pos}(c)
that shows the histograms of
off-axis angles for different source types 
for the main-catalog sources.

Figures~\ref{fig:cdfn-pos}(e) and (f) show the observed
source density as
a function of off-axis angle for all the main-catalog sources
and the new main-catalog sources, respectively.
These two plots reveal, for either all or new sources, that
(1) the source densities decline toward large off-axis angles
due to the decreasing sensitivities (see Section~\ref{sec:cdfn-smap});
(2) overall, observed AGN densities are larger than observed galaxy densities; and
(3) the galaxy source density approaches the AGN source density
toward smaller off-axis angles where lower flux levels are achieved,
due to the observed galaxy number counts having a steeper slope than
the observed AGN number counts (e.g., Bauer \etal 2004; Lehmer \etal 2012).
In the central \cdfn\ area of $\theta\le3\arcmin$,
the observed source densities for all sources, all AGNs, 
and all galaxies reach
$16700_{-1500}^{+1600}$~deg$^{-2}$, 
$12400_{-1300}^{+1400}$~deg$^{-2}$, and
$4200_{-700}^{+900}$~deg$^{-2}$, respectively; and
the observed source densities for all new sources, new AGNs,
and new galaxies reach
$6000_{-900}^{+1000}$~deg$^{-2}$,    
$3700_{-700}^{+800}$~deg$^{-2}$, and
$2300_{-500}^{+700}$~deg$^{-2}$, respectively.

\begin{figure*}
\centerline{\includegraphics[width=17.5cm]{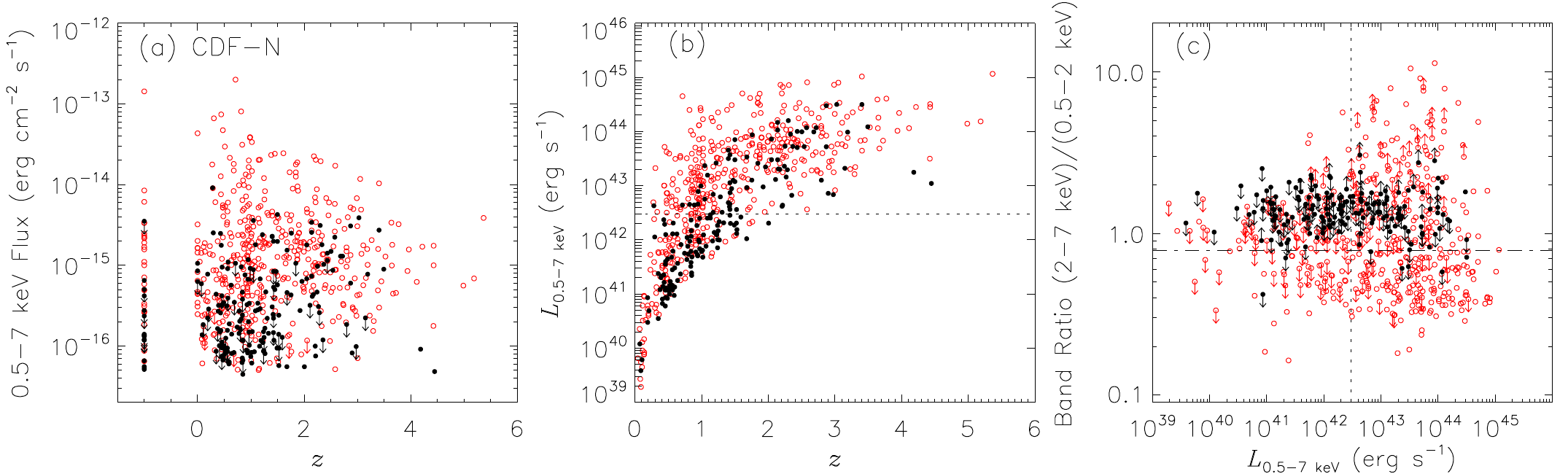}}
\figcaption{(a) Observed-frame full-band flux vs. adopted redshift,
(b) absorption-corrected, rest-frame \hbox{0.5--7 keV} luminosity vs. adopted redshift,
and (c) band ratio vs. absorption-corrected, rest-frame \hbox{0.5--7 keV} luminosity
for the 2~Ms \cdfn\ main-catalog sources.
Red open circles indicate old sources
while black filled circles indicate new sources.
Arrows denote limits.
In panel (b), sources having no redshift estimates are not plotted;
in panel (c), sources having no redshift estimates or sources having only
full-band detections are not plotted.
The dotted lines in panels (b) and (c) and the dashed-dot line in panel (c)
correspond to the threshold values of two AGN-identification criteria,
$L_{\rm 0.5-7\ keV}\ge 3\times 10^{42}$ \hbox{erg s$^{-1}$} and $\Gamma \le 1.0$.
\label{fig:cdfn-f-lx-z-br}}
\end{figure*}

\begin{figure}
\centerline{\includegraphics[width=8.5cm]{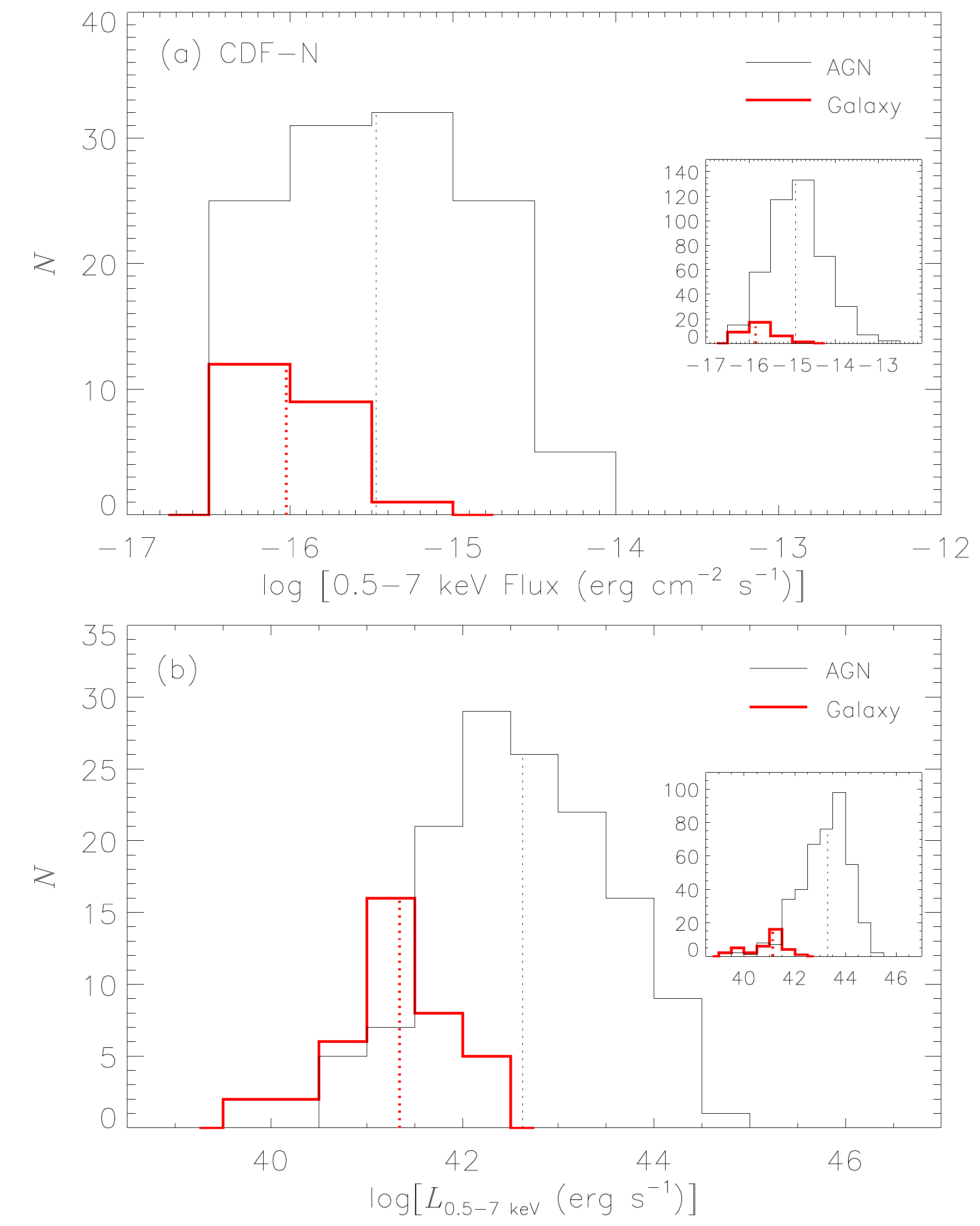}}
\figcaption{Histograms of (a) observed-frame full-band flux and (b) absorption-corrected,
rest-frame \hbox{0.5--7 keV} luminosity for the new 2~Ms \cdfn\ main-catalog sources.
The insets display results for the old main-catalog sources.
The vertical dotted lines indicate the median values.
In panel (a), sources without full-band detections are not included;
in panel (b), sources without redshift estimates are not included.
\label{fig:cdfn-f-lx-hist}}
\end{figure}

Figure~\ref{fig:cdfn-f-lx-z-br} displays 
(a) observed-frame full-band flux vs. adopted redshift,
(b) absorption-corrected, rest-frame \hbox{0.5--7 keV} luminosity
vs. adopted redshift,
and (c) band ratio vs. absorption-corrected, 
rest-frame \hbox{0.5--7 keV} luminosity,
for the new sources (indicated as filled circles) and old sources 
(indicated as open circles), respectively.
We find that
(1) the new sources typically have smaller
\xray\ fluxes and luminosities than the old sources
(also see Figure~\ref{fig:cdfn-f-lx-hist}); and
(2) the median value of 1.40 of band ratios or upper limits on band ratios
of the 128 new sources is larger than the corresponding median value of 0.77 
of the 406 old sources 
(also see Figure~\ref{fig:cdfn-plotstack}).
Following the example provided in Section~10.8.2 of Feigelson \& Babu (2012),
we further quantify the difference in band ratios between 
the above 128 new sources and 406 old sources that involve censored data,
utilizing survival-analysis 2-sample tests 
(the logrank test and the Peto \& Peto modification of the Gehan-Wilcoxon test) that are
implemented in the function {\sc survdiff} in the public domain R statistical software system
(R Core Team 2015).
Both of the 2-sample tests give $p=0.0$ results, indicating that
there is a significant difference in band ratios between the above new and old sources. 
Together, the above observations indicate that our improved 
cataloging methodology allows us to probe fainter 
obscured sources than A03.

Figure~\ref{fig:cdfn-f-lx-hist} presents
histograms of observed-frame full-band flux 
and absorption-corrected, rest-frame \hbox{0.5--7 keV} luminosity
for the new AGNs and galaxies (main panels) as well as the old 
AGNs and galaxies (insets).
It is apparent that 
AGNs and galaxies have disparate distributions of
flux and luminosity,
and overall galaxies become the numerically
dominant population at \hbox{0.5--7 keV} luminosities less than 
$\approx 10^{41.5}$ erg~s$^{-1}$,
no matter whether the new or old sources are considered.

\begin{figure*}
\centerline{\includegraphics[width=14cm]{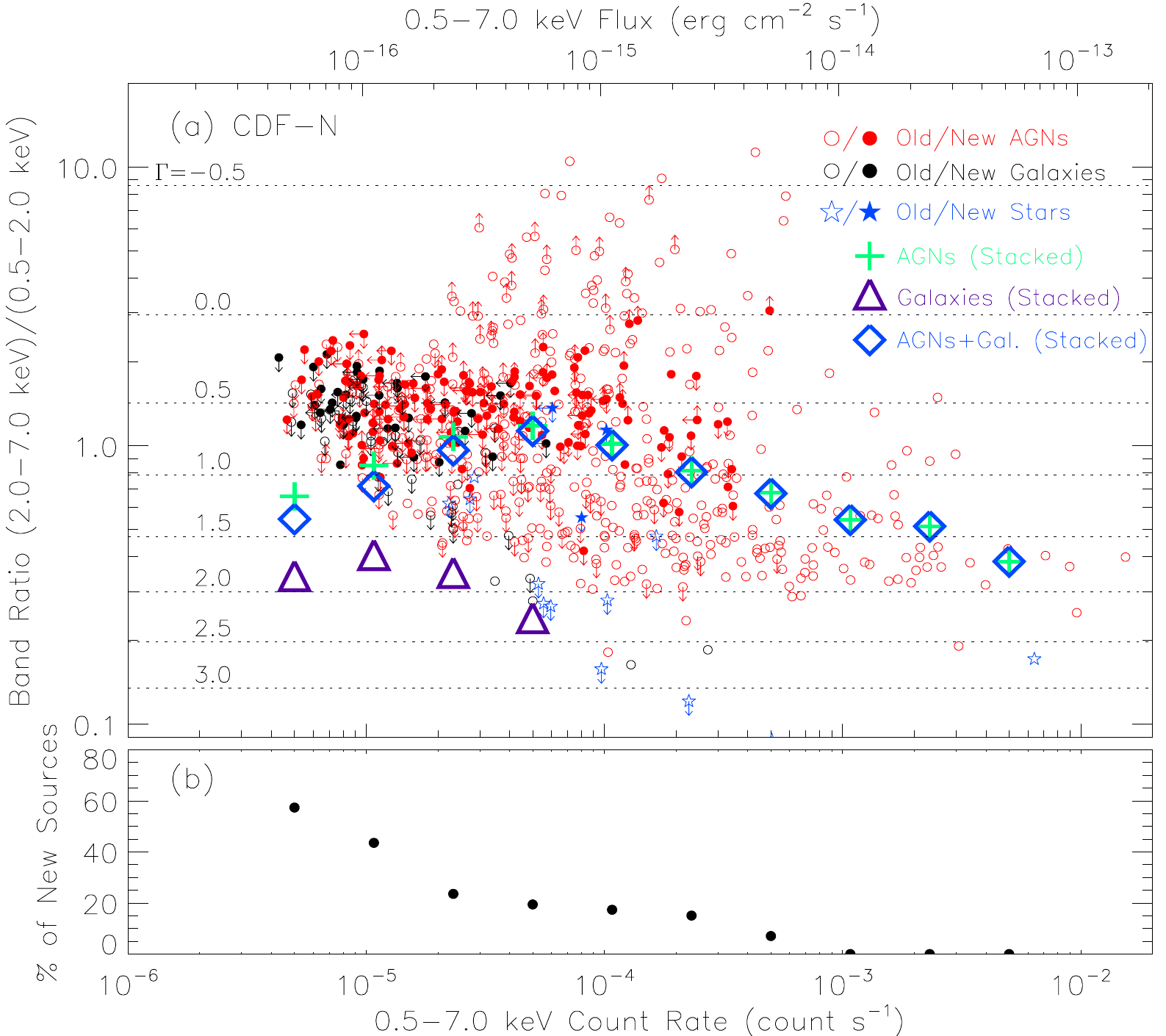}}
\figcaption{(a) Band ratio vs. full-band count rate for the 2~Ms \cdfn\ main-catalog sources. 
For reference,
the top $x$-axis displays representative full-band fluxes that are derived
using full-band count rates given an assumed $\Gamma=1.4$ power law.
The meanings of symbols of different types and colors are indicated by the legend.
Arrows indicate limits.
Sources with only full-band detections are not plotted;
there are only 19 (19/683=2.8\%) such sources, the exclusion of which would not 
affect our results significantly.
Large crosses, triangles, and diamonds denote average/stacked band ratios as a function of full-band count rate that are
derived in bins of $\Delta {\rm log(Count\hspace{0.1cm} Rate)}=0.6$,
for AGNs, galaxies, and both AGNs and galaxies, respectively.
Horizontal dotted lines indicate the band ratios that
correspond to given effective photon indexes.
(b) Fraction of new sources as a function of full-band count rate for the 2~Ms \cdfn\ main-catalog sources,
computed in bins of $\Delta {\rm log(Count\hspace{0.1cm} Rate)}=0.6$.
\label{fig:cdfn-bratio-new}}
\end{figure*}

\begin{figure}
\centerline{\includegraphics[width=8.5cm]{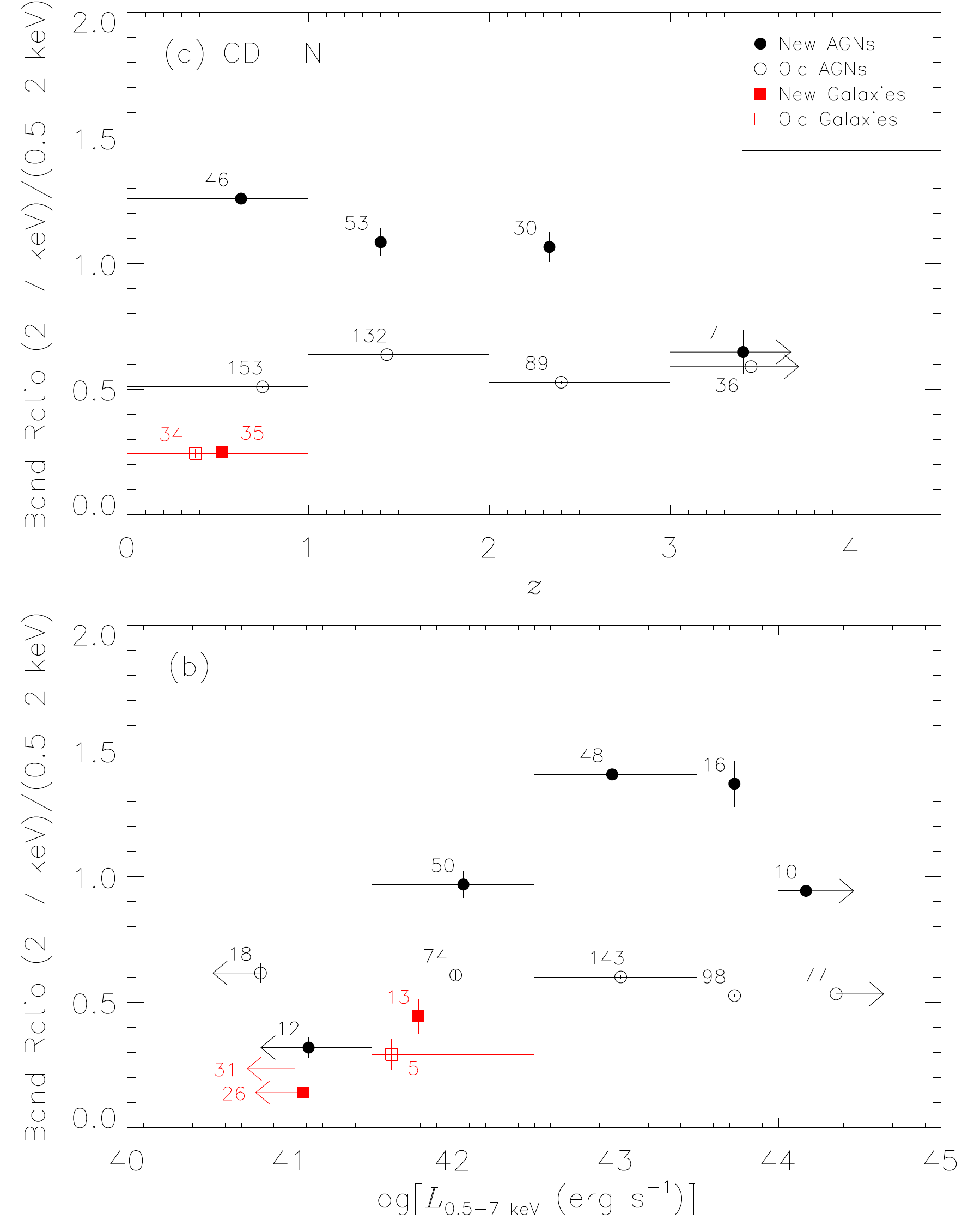}}
\figcaption{Average/stacked band ratios in bins of
(a) redshift ($0<z<1$, $1\le z<2$, $2\le z<3$, and $z\ge 3$) and
(b) absorption-corrected, rest-frame \mbox{0.5--7~keV} luminosity [$\log(L_{\rm X})<41.5$,
$41.5\le \log(L_{\rm X})<42.5$,
$42.5\le \log(L_{\rm X})<43.5$,
$43.5\le \log(L_{\rm X})<44.0$, and
$\log(L_{\rm X})\ge 44.0$] for the new and old 2~Ms \cdfn\ main-catalog sources.
The meanings of symbols are indicated by the legend.
In each bin, the median redshift or \xray\ luminosity is used for plotting;
the number of stacked sources is annotated.
\label{fig:cdfn-plotstack}}
\end{figure}

Figure~\ref{fig:cdfn-bratio-new}(a) displays the band ratio 
as a function of full-band count
rate for the new sources (indicated as filled symbols) 
and the old sources (indicated as open symbols),
with the large crosses, triangles, and diamonds
representing the average (i.e., stacked) band ratios\footnote{We note that, obviously,  
the stacked averages only indicate the mean properties and cannot represent the full 
distribution of the stacked sample.}
for all AGNs, all galaxies, and all sources 
(counting both AGNs and galaxies), respectively.
The overall average band ratio is, as expected, dominated by AGNs,
which has a rising-leveling-off-declining shape
toward low full-band count rates that is
in general agreement with that seen in, e.g., 
Figure~14 of A03 and Figure~18 of X11 (see Section~4.7 of X11
for the relevant discussion on such a shape).
Figure~\ref{fig:cdfn-bratio-new}(b) presents 
the fraction of new sources as a function of
full-band count rate for the sources in the main catalog.
From full-band count rates of $\approx 10^{-3}$ count~s$^{-1}$
to $\approx 5\times 10^{-6}$ count~s$^{-1}$,
the fraction of new sources rises monotonically  
from 0\% to $\approx 57\%$.

Figure~\ref{fig:cdfn-plotstack} presents the average band ratio
in bins of adopted redshift and \xray\ luminosity for the new AGNs, 
old AGNs, new galaxies, and old galaxies, respectively.
A couple of observations can be made, e.g.: 
(1) the new AGNs have larger band ratios than the old AGNs 
no matter which bin of redshift or \xray\ luminosity is considered,
with the only exception of the lowest luminosity bin, 
reflecting the rise of obscured AGNs toward faint fluxes
(e.g., Bauer \etal 2004; Lehmer \etal 2012);
(2) for the lowest redshift bin and the two lowest luminosity bins where
both the AGN and galaxy results are available for comparison,
the AGNs have larger band ratios than the galaxies, no matter being old 
or new ones; and
(3) in the lowest luminosity bin of $\log(L_{\rm X})<41.5$,
the new galaxies have a smaller average band ratio 
than the old galaxies, while in
a higher luminosity bin of $41.5\le \log(L_{\rm X})<42.5$,
the trend is reversed (but note the relatively small numbers
of sources considered in this higher luminosity bin).

\begin{figure*}
\centerline{\includegraphics[width=17cm]{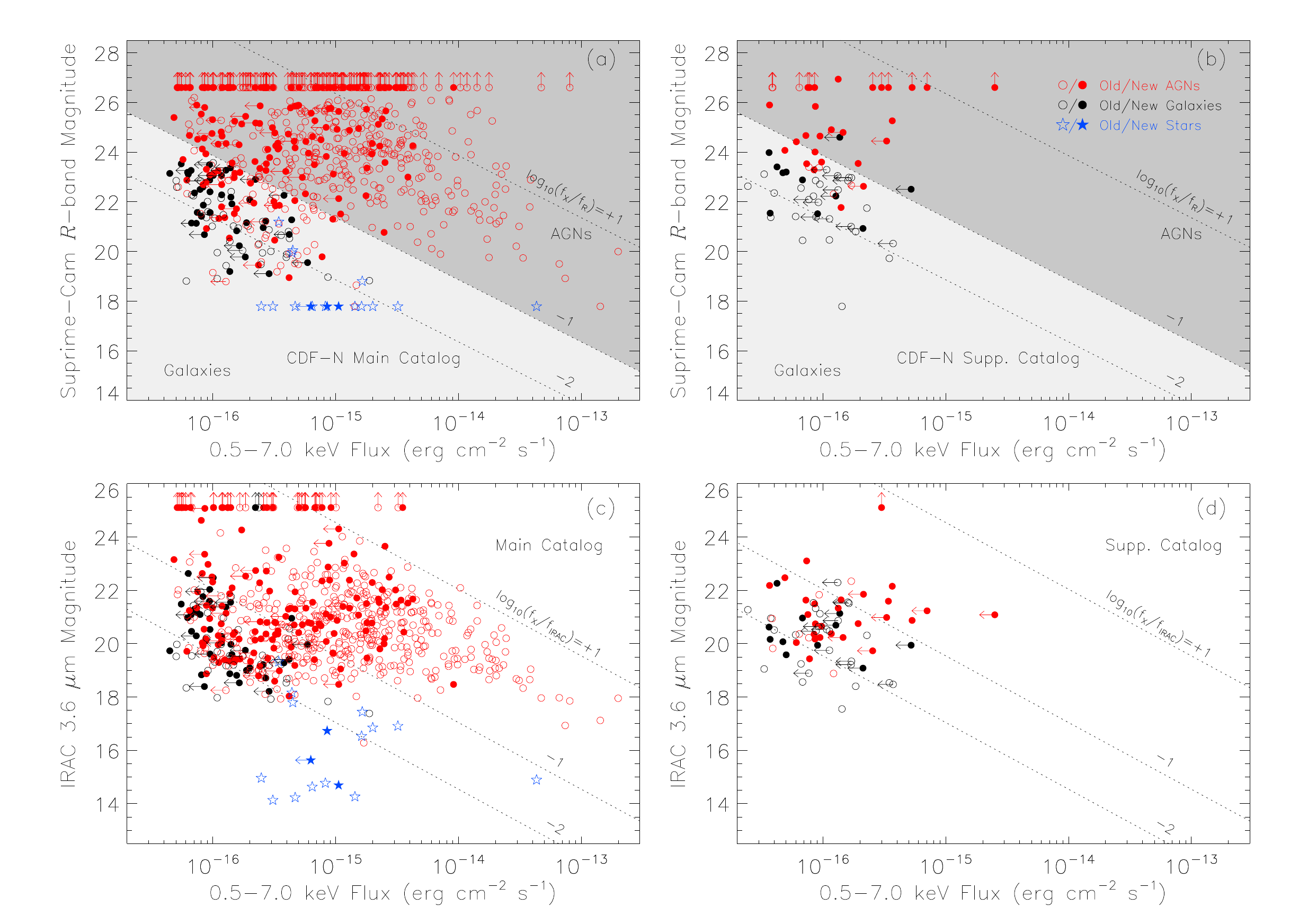}}
\figcaption{(Top) Suprime-Cam $R$-band magnitude vs. full-band flux for
(a) the 2~Ms \cdfn\ main-catalog sources and (b) the supplementary-catalog sources.
(Bottom) IRAC 3.6~$\mu$m magnitude vs. full-band flux for
(c) the 2~Ms \cdfn\ main-catalog sources and (d) the supplementary-catalog sources.
The meanings of symbols of different types and colors are indicated by the legend.
Arrows denote limits.
In panels (a) and (b),
diagonal dotted lines indicate constant full-band-to-$R$ flux ratios, and shaded areas represent 
approximate flux ratios for AGNs (dark gray) and galaxies (light gray).
In panels (c) and (d),
diagonal dotted lines indicate constant full-band-to-IRAC-3.6~$\mu$m flux ratios. 
Note that the stars lined up at $R=17.8$~mag have their $R$-band magnitudes
set to this value for plotting purposes as they are assigned negative
magnitudes in the $R$-band catalog due to saturation 
($R=17.8$ is one magnitude brighter 
than the brightest non-saturated stars in the $R$-band catalog).
\label{fig:cdfn-fox}}
\end{figure*}

Figure~\ref{fig:cdfn-fox}(a) presents the Suprime-Cam $R$-band magnitude versus
the full-band flux for the new sources (indicated as filled symbols) 
and old sources (indicated as open symbols), as well as
the approximate flux ratios for AGNs and galaxies
(see the description of Column~71 for AGN identification),
where the sources are color-coded with red for AGNs, black for galaxies,
and blue for stars, respectively. 
As a comparison, Figure~\ref{fig:cdfn-fox}(c) presents the IRAC 3.6~$\mu$m 
magnitude versus the full-band flux
for the new sources and old sources, since a larger fraction of 
the main-catalog sources have IRAC 3.6~$\mu$m-band counterparts 
than Suprime-Cam $R$-band counterparts
(i.e., 87.0\% vs. 68.5\%;
see the description of \hbox{Columns~23--40}).
We note that the flux ratio of $\log (f_{\rm X}/f_{\rm 3.6~\mu m})$
can also be used to separate AGNs from galaxies when 
the classification threshold is carefully calibrated (e.g.,
Wang \etal 2013).
Overall, a total of 591 (86.5\%) of the sources in the main catalog
are likely AGNs,
the majority of which lie in the region expected for
relatively luminous AGNs that have $\log (f_{\rm X}/f_{\rm R})>-1$
(i.e., dark gray areas in Fig.~\ref{fig:cdfn-fox}a);
among these 591 AGNs, 154 (26.1\%) are new.
A total of 75 (11.0\%) of the sources in the main catalog are likely galaxies,
and by selection all of them lie in the region expected for
normal galaxies, starburst galaxies, and low-luminosity AGNs
that have $\log (f_{\rm X}/f_{\rm R})\le -1$ (i.e., light gray areas 
in Fig.~\ref{fig:cdfn-fox}a);
among these 75 sources, 39 (52.0\%) are new.
Only 17 (2.5\%) of the sources in the main catalog are likely stars,
with all but one having low \hbox{X-ray}-to-optical flux ratios;
among these 17 stars, 3 are new.
Among the new sources, normal and starburst galaxies
total a fraction of 19.9\%,
as opposed to 7.4\% if the old sources are considered, which
is expected due to galaxies having a steeper number-count slope
than AGNs (e.g., Bauer \etal 2004; Lehmer \etal 2012).

\begin{figure}
\centerline{\includegraphics[width=8.5cm]{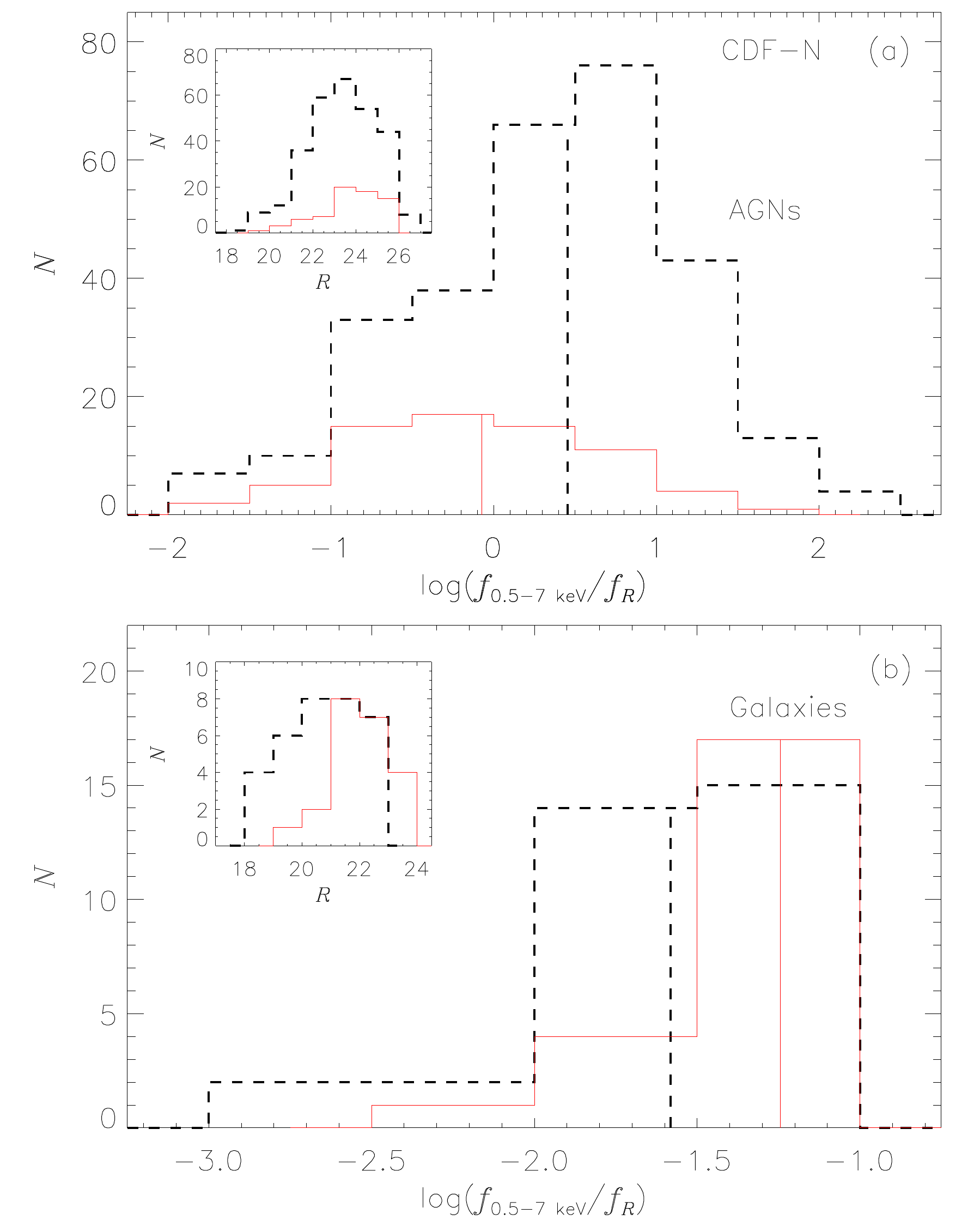}}
\figcaption{Histograms of \xray-to-optical ($R$ band) flux ratio for (a) the new 2~Ms \cdfn\ main-catalog AGNs (solid histogram)
and old AGNs (dashed histogram) and (b) new galaxies (solid histogram) and
old galaxies (dashed histogram), with median flux ratios denoted by vertical lines.
The insets display the histograms of $R$-band magnitude for new sources
(solid histograms) and old sources (dashed histograms).
Only sources with both full-band and $R$-band detections are plotted.
\label{fig:cdfn-x-to-R}}
\end{figure}

Figure~\ref{fig:cdfn-x-to-R} presents the
histograms of \xray-to-optical flux ratio for the new AGNs, old AGNs, 
new galaxies, and old galaxies, respectively.
It is apparent that
(1) the new AGNs have a similar overall distribution of $R$-band magnitude 
to the old AGNs, but generally have smaller \xray-to-optical flux ratios 
than the old AGNs;
and (2) the new galaxies generally have fainter \mbox{$R$-band} magnitudes
and larger \xray-to-optical flux ratios than the old galaxies.

\subsection{Supplementary Near-Infrared Bright \chandra\ Source Catalog}\label{sec:cdfn-supp}

\subsubsection{Supplementary Catalog Production}

Among the 320 (i.e., $1003-683=320$) candidate-list \hbox{X-ray} sources that
do not satisfy the main-catalog source-selection criterion of $P<0.004$,
167 are of moderate significance with $0.004\le P<0.1$.
In order to retrieve genuine \xray\ sources from this sample of 167 sources,
we create a supplementary catalog that consists of
the subset of these sources having bright near-infrared counterparts,
because such \chandra\ sources are likely real, thanks to the
comparatively low spatial density of near-infrared bright sources. 
Similar prior-based source-searching methods have been widely used (e.g., Richards \etal 1998;
A03; L05; Luo \etal 2008; X11), and they allow for detections of real \xray\ sources
with lower significances.  
We match these 167 \chandra\ sources with the $K_s\le 22.9$~mag sources 
in the \goodsn\ WIRCam $K_s$-band catalog utilizing a matching radius of $1\farcs2$.
The choices of $0.004\le P<0.1$, the $K_s$-band catalog and
cutoff magnitude, and the matching radius
are made to maximize the number of included sources while keeping the
expected number of false sources reasonably low.
A total of 72 near-infrared bright \hbox{X-ray} sources are identified this way,
with $\approx 6.0$ false matches expected (i.e., a false-match rate of 8.3\%).
Our supplementary catalog includes
7 A03 main-catalog sources that are not recovered in our main catalog
and 27 A03 supplementary optically bright (i.e., \mbox{$R\lsim 23.0$~mag}) sources,
thus resulting in a total of $72-7-27=38$ new supplementary-catalog sources that are not present in either of the A03 catalogs.
A point worth noting is that 
the vast majority (72 out of 79; 91.1\%) of the A03 supplementary optically
bright sources are included either in our main catalog (45 sources) 
or supplementary near-infrared bright catalog (the aforementioned 27 sources).

Our 72-source supplementary catalog is
presented in Table~\ref{tab:cdfn-supp}, in the same format as Table~\ref{tab:cdfn-main}
(see Section~\ref{sec:cdfn-maincat} for the details of each column).
A source-detection criterion of $P<0.1$ is adopted for photometry-related calculations
for the supplementary-catalog sources;
and the multiwavelength identification-related columns
(i.e., Columns~18--22) are set to the WIRCam $K_s$-band matching results.

\begin{table*}
%\tabletypesize{\scriptsize}
%\tablewidth{0pt}
\caption{2~Ms \cdfn\ Supplementary Near-Infrared Bright {\it Chandra} Source Catalog}
\begin{tabular}{lllcccccccccc}\hline\hline
No. & $\alpha_{2000}$ & $\delta_{2000}$ & $\log P$ & {\sc wavdetect} & Pos Err & Off-axis & FB & FB Upp Err & FB Low Err & SB & SB Upp Err & SB Low Err \\
(1) & (2) & (3) & (4) & (5) & (6) & (7) & (8) & (9) & (10) & (11) & (12) & (13) \\ \hline
1 & 12 35 25.28 & +62 11 53.8 &     $-$1.1 &  $-$5 &  2.2 &   9.60 &    18.2 &  $-$1.0 &  $-$1.0 &     \phantom{0}3.9 &   4.3 &   2.9 \\
2 & 12 35 39.87 & +62 15 05.7 &     $-$2.4 &  $-$5 &  0.8 &   7.75 &    24.5 &  11.7 &  10.2 &    11.5 &   7.3 &   5.9 \\
3 & 12 35 39.87 & +62 13 35.9 &     $-$2.4 &  $-$5 &  0.7 &   7.68 &    36.2 &  15.7 &  14.5 &    14.1 &   9.2 &   8.0 \\
4 & 12 35 51.76 & +62 21 34.5 &     $-$2.2 &  $-$5 &  0.9 &   9.86 &    28.2 &  13.2 &  12.0 &    11.3 &   7.9 &   6.7 \\
5 & 12 35 54.31 & +62 15 33.1 &     $-$2.2 &  $-$5 &  0.6 &   6.19 &    26.0 &  12.4 &  11.2 &    11.4 &   7.8 &   6.5 \\\hline
\end{tabular}
The full table contains 72~columns of information for the 72 \xray\ sources.\\
(This table is available in its entirety in a machine-readable form in the online journal. A portion is shown here for guidance regarding its form and
content.)
\label{tab:cdfn-supp}
\end{table*}

\subsubsection{Properties of Supplementary-Catalog Sources}

Figure~\ref{fig:cdfn-pos}(b) displays the spatial distribution of the 72  
supplementary-catalog sources, with the 38 new
sources denoted as filled symbols;
and Figure~\ref{fig:cdfn-pos}(d) presents the histograms of off-axis angles
for different source types for the supplementary-catalog sources.
Figures~\ref{fig:cdfn-fox}(b) and (d) present 
the Suprime-Cam $R$-band magnitude and
the SEDS IRAC 3.6~$\mu$m magnitude 
versus the full-band flux for the supplementary-catalog sources, 
respectively.
Among the 72 supplementary-catalog sources,
34 (47.2\%) are likely AGNs; 
38 (52.8\%) are likely galaxies, which by selection are all located
in the region expected for normal galaxies, starburst galaxies, and
low-luminosity AGNs; and there are no likely stars.
A total of 69 (95.8\%) of these 72 sources have either
\zs's or \zp's, ranging from 0.083 to 3.583 with a median
redshift of 0.857.

\subsection{Completeness and Reliability Analysis}\label{sec:cdfn-comp}

We have carried out simulations to make an assessment of the completeness and reliability of our main catalog.

\subsubsection{Generation of Simulated Data}\label{sec:cdfn-simdata}

First, we construct a mock catalog covering the entire \cdfn\ field and
extending well below its detection limits (see Section~\ref{sec:cdfn-smap}). 
In this mock catalog,
we follow Miyaji \etal (2007) to include realistic source clustering 
when assigning source coordinates. 
We randomly assign simulated AGN and galaxy fluxes following
the soft-band log~$N$--log~$S$ relations 
in the Gilli \etal (2007) AGN population-synthesis model
and the Ranalli \etal (2005) galaxy ``peak-M'' model, respectively.
We convert soft-band fluxes of simulated AGNs and galaxies
into full-band fluxes assuming $\Gamma=1.4$
and $\Gamma=2.0$ power-law spectra, respectively.
Second, we utilize the MARX simulator to construct source event lists from 
20 simulated ACIS-I observations of the mock catalog, 
each under the same observational configuration (e.g., aim point,
roll angle, exposure time, aspect solution file) as
one of the \cdfn\ observations.
Third, we extract corresponding background event files
from the real \cdfn\ event files
by masking all events relevant to the main- and supplementary-catalog
sources and then filling the masked regions with events that
obey the local background distribution.
We randomly remove
$\approx 0.7\%$\footnote{This $\approx 0.7\%$ is a typical ratio
between the summed full-band counts of undetectable faint sources
and the 2~Ms \cdfn\ total full-band background counts 
(see Table~\ref{tab:cdfn-bkg} in Section~\ref{sec:cdfn-bmap}), 
assuming a range of
reasonable effective photon indexes for the flux-count-rate conversions.}  
of the events in each background event file
in order to avoid counting twice the contribution of
undetectable faint sources that is present in
both the source and background event files.
We then combine the above corresponding source and background 
event files to produce a set of 20 simulated ACIS-I observations that
closely mimic the real \cdfn\ observations.
Finally, following Section~\ref{sec:cdfn-img-cand}, 
we obtain a simulated merged 2~Ms \cdfn\ event file,
construct images for the three standard bands,
run {\sc wavdetect} (sigthresh=$10^{-5}$) to generate a candidate-list catalog,
and make use of AE to extract photometry (including $P$ values) for the candidate-list sources.

\subsubsection{Completeness and Reliability}\label{sec:cdfn-comprel}

\begin{figure*}
\centerline{
\includegraphics[scale=0.64]{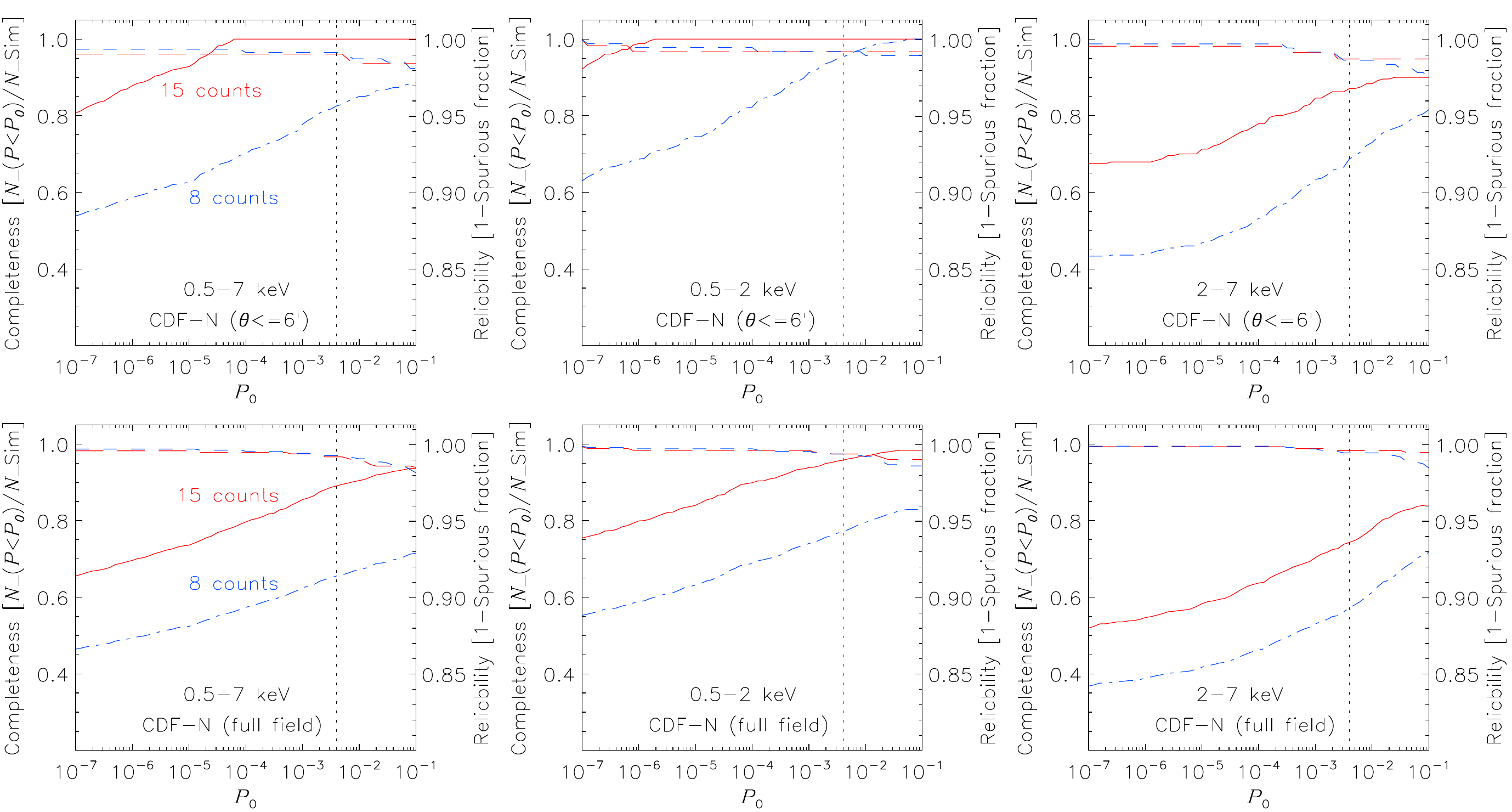}
}
\figcaption{(Top) The $\theta\le 6\arcmin$ case in the 2~Ms \cdfn:
completeness (solid and dashed-dot curves; left $y$-axis) and reliability
(long dashed and short dashed curves; right $y$-axis)
as a function of $P_0$ ($P<P_0$ as the source-detection criterion)
for the simulations in the full, soft, and hard bands,
for sources with $\ge 15$~counts (red solid and long dashed curves)
and $\ge 8$~counts (blue dashed-dot and short dashed curves), respectively.
The vertical dotted lines denote our adopted main-catalog source-detection
threshold of $P_0=0.004$.
(Bottom) Same as top panels, but for the case of the full \cdfn\ field.\label{fig:cdfn-comp-rel}
}
\end{figure*}

We define completeness as
the ratio between the number of detected sources 
that satisfy a specific detection criterion of \hbox{$P<P_0$} 
and the number of input simulated sources,
above a given source-count limit that applies to both
detected sources and input simulated sources.
We define reliability as
1 minus the ratio between the number of spurious sources
that arise from noise fluctuations
and the number of input simulated sources, 
above the same given source-count limit.
Figure~\ref{fig:cdfn-comp-rel} displays 
the completeness and reliability as a function of the AE-computed
binomial no-source probability within the central $\theta\le 6\arcmin$ region 
and the entire \cdfn\ field,
for the simulations in the full, soft, and hard bands,
for sources with at least 15 and 8 counts, respectively.
The case of 8 counts is roughly our on-axis (i.e., $\theta\lsim3\arcmin$) source-detection limit in the full and hard bands.

It seems clear from Fig.~\ref{fig:cdfn-comp-rel} that
(1) in all panels, as expected, 
each completeness curve goes up
and each reliability curve goes down toward large $P$ threshold
values, and the completeness level for the case of 15~counts is
higher than that for the case of 8~counts; and
(2) the completeness level for the case of either 15~counts 
or 8~counts within the central $\theta\le 6\arcmin$ region  
is higher than the corresponding completeness level 
in the entire \cdfn\ field.
At our adopted main-catalog $P$ threshold of 0.004,
the completeness levels within the central $\theta\le 6\arcmin$ region are
100.0\% and 82.5\% (full band), 100.0\% and 95.4\% (soft band),
and 87.1\% and 68.7\% (hard band) for sources with $\ge 15$ and $\ge 8$~counts, respectively.
The completeness levels for the entire \cdfn\ field are
89.1\% and 65.6\% (full band), 95.8\% and 77.2\% (soft band),
and 74.4\% and 57.3\% (hard band) for sources with $\ge 15$ and $\ge 8$~counts, respectively.
At our adopted main-catalog $P$ threshold of 0.004,
the reliability level ranges from 98.7\% to 99.6\% for all panels;
we estimate that, in the main catalog (i.e., the entire \cdfn\ field), 
there are
about 5, 3, and 2 false detections with $\ge 15$~counts in the full,
soft, and hard bands, and
about 5, 4, and 3 false detections with $\ge 8$~counts in the full,
soft, and hard bands, respectively.

\begin{figure}
\centerline{
\includegraphics[scale=0.55]{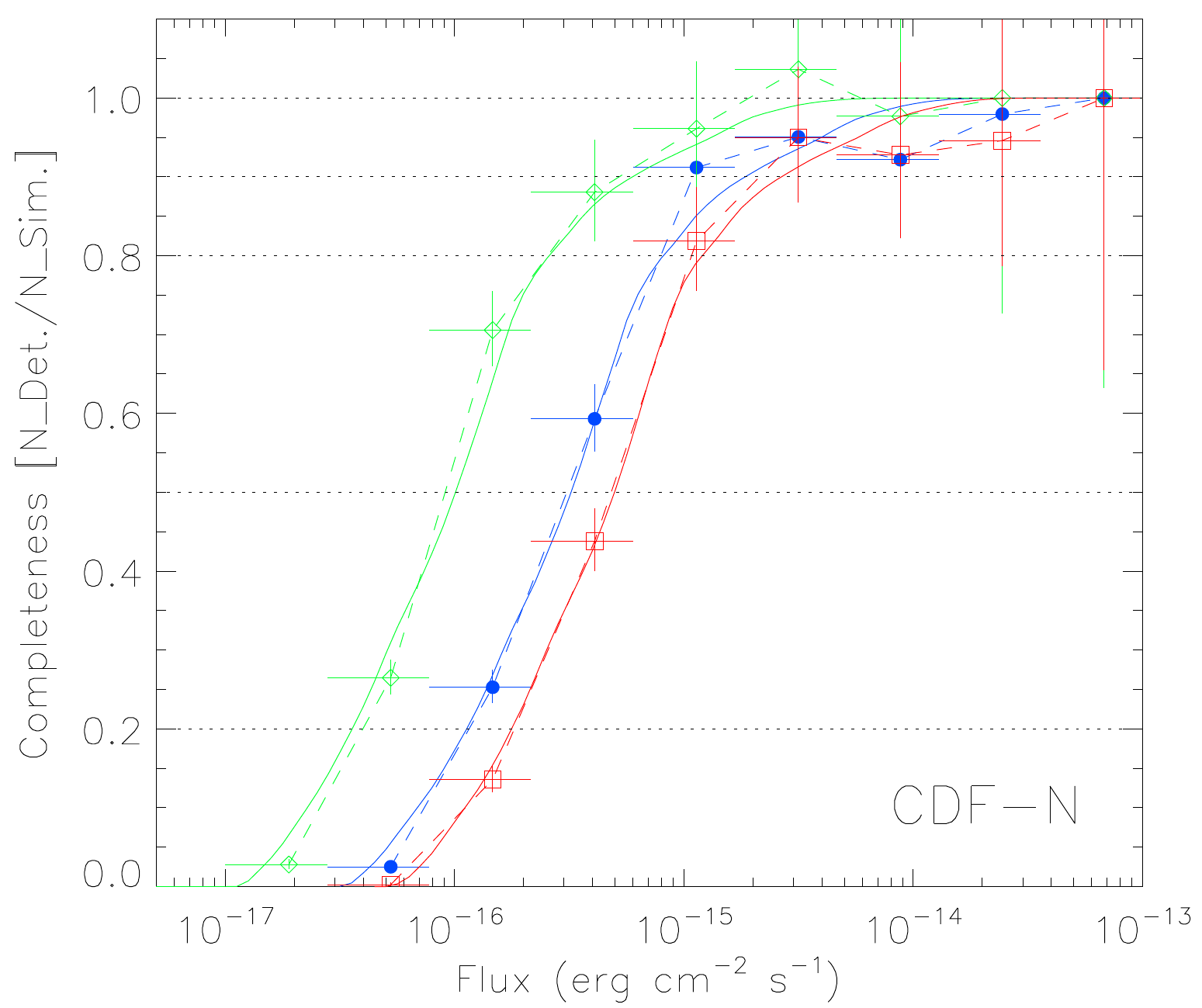}
}
\figcaption{Completeness as a function of flux 
given the 2~Ms \cdfn\ main-catalog $P<0.004$ criterion
for the simulations in the full (blue filled circles), soft (green open diamonds), and hard (red open squares) bands, 
overlaid with the corresponding sky coverage curves (solid curves) 
that are normalized to the maximum sky coverage (see 
Fig.~\ref{fig:cdfn-senhist}).
The dashed lines make connections between the corresponding adjacent cross points.
The horizontal dotted lines denote five completeness levels.
\label{fig:cdfn-completeness-flux}}
\end{figure}

Figure~\ref{fig:cdfn-completeness-flux} presents
the completeness as a function of flux
given the main-catalog $P<0.004$ criterion
for the \mbox{full-,} \mbox{soft-,} and hard-band simulations.
The three curves of completeness versus flux that are derived from the simulations 
(dashed lines) approximately track
the normalized sky coverage curves that are
derived from the real \cdfn\ observations (solid curves).
Table~\ref{tab:cdfn-completeness} presents 
the flux limits corresponding to four specific completeness
levels in the full, soft, and hard bands, which are denoted as horizontal dotted lines in Fig.~\ref{fig:cdfn-completeness-flux}.

\begin{table}[ht]
\caption{2~Ms \cdfn\ Flux Limit and Completeness}
\centering
\begin{tabular}{lccc}\hline\hline
Completeness & $f_{\rm 0.5-7\ keV}$ & $f_{\rm 0.5-2\ keV}$ & $f_{\rm 2-7\ keV}$ \\
(\%) & (\flux) & (\flux) & (\flux) \\ \hline
90 & $1.9\times 10^{-15}$ & $6.0\times 10^{-16}$ & $2.7\times 10^{-15}$ \\
80 & $8.1\times 10^{-16}$ & $2.6\times 10^{-16}$ & $1.2\times 10^{-15}$ \\
50 & $3.2\times 10^{-16}$ & $1.0\times 10^{-16}$ & $5.0\times 10^{-16}$ \\
20 & $1.1\times 10^{-16}$ & $3.6\times 10^{-17}$ & $1.8\times 10^{-16}$ \\\hline
\end{tabular}
\label{tab:cdfn-completeness}
\end{table}

\subsection{Background and Sensitivity Analysis}\label{sec:cdfn-bkg}

\subsubsection{Background Map Creation}\label{sec:cdfn-bmap}

\begin{figure}
\centerline{\includegraphics[width=8.5cm]{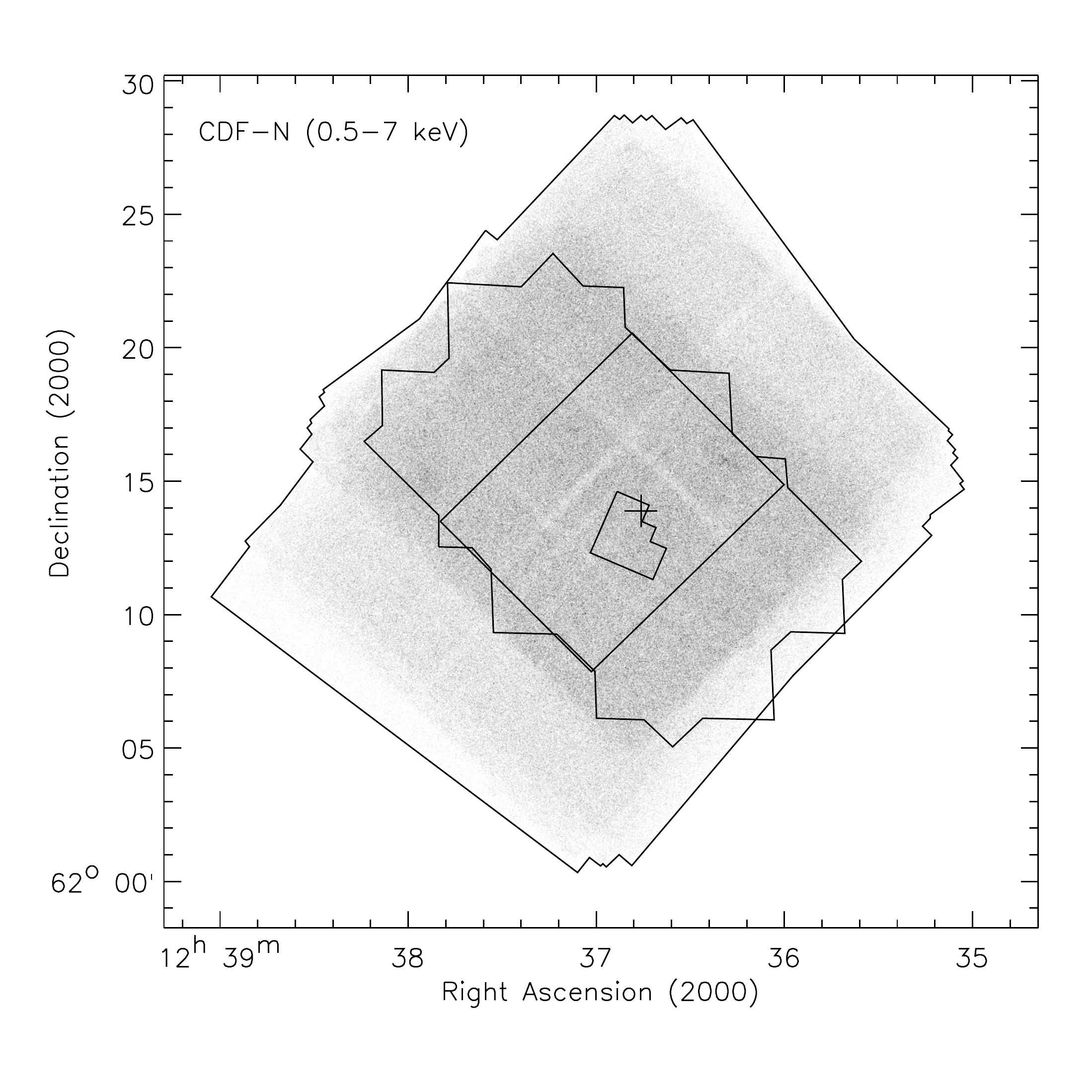}}
\figcaption{Full-band (0.5--7.0 keV) background map of the 2~Ms \cdfn\ rendered using linear gray scales. 
The higher background around the \goodsn\ area is due to the larger effective exposure.
The regions and the plus sign have the same meanings as those in Fig.~\ref{fig:cdfn-fb-img}. 
\label{fig:cdfn-fb-bkg}}
\end{figure}

To create background maps for the three standard-band images,
we first mask the 683 main-catalog sources and the 72 supplementary-catalog
sources, utilizing circular apertures that have
radii of 1.5 (2.0) times the $\approx 99\%$ PSF EEF radii
for sources with full-band counts below (above) 10,000 
(note that there are 4 main-catalog sources with $>$10,000 full-band counts).
We subsequently fill in the masked areas of the sources
with background counts, which obey the local probability distribution of counts 
lying within an annulus that has an inner radius being the aforementioned
masking radius and has an outer radius of 2.5 (3.0) times the \hbox{$\approx 99\%$}
PSF EEF radius for sources with full-band counts below
(above) 10,000.
Table~\ref{tab:cdfn-bkg} summarizes the background properties
including the mean background, total background,
and count ratio between background counts and detected source counts
for the three standard bands.
91.7\%, 97.1\%, and 94.2\% of the pixels
have zero background counts in the background maps
for the full, soft, and hard bands, respectively.
The values in Table~\ref{tab:cdfn-bkg} 
are systematically slightly lower than those reported in Table~8 of A03, 
mainly due to the facts  
that we adopt a smaller upper energy bound of 7~keV than 
the value of 8~keV
adopted in A03 and that
we adopt a more stringent approach for data filtering
(see Section~\ref{sec:cdfn-obs}).
Figure~\ref{fig:cdfn-fb-bkg} displays the full-band background map.

\begin{figure}
\centerline{\includegraphics[width=8.5cm]{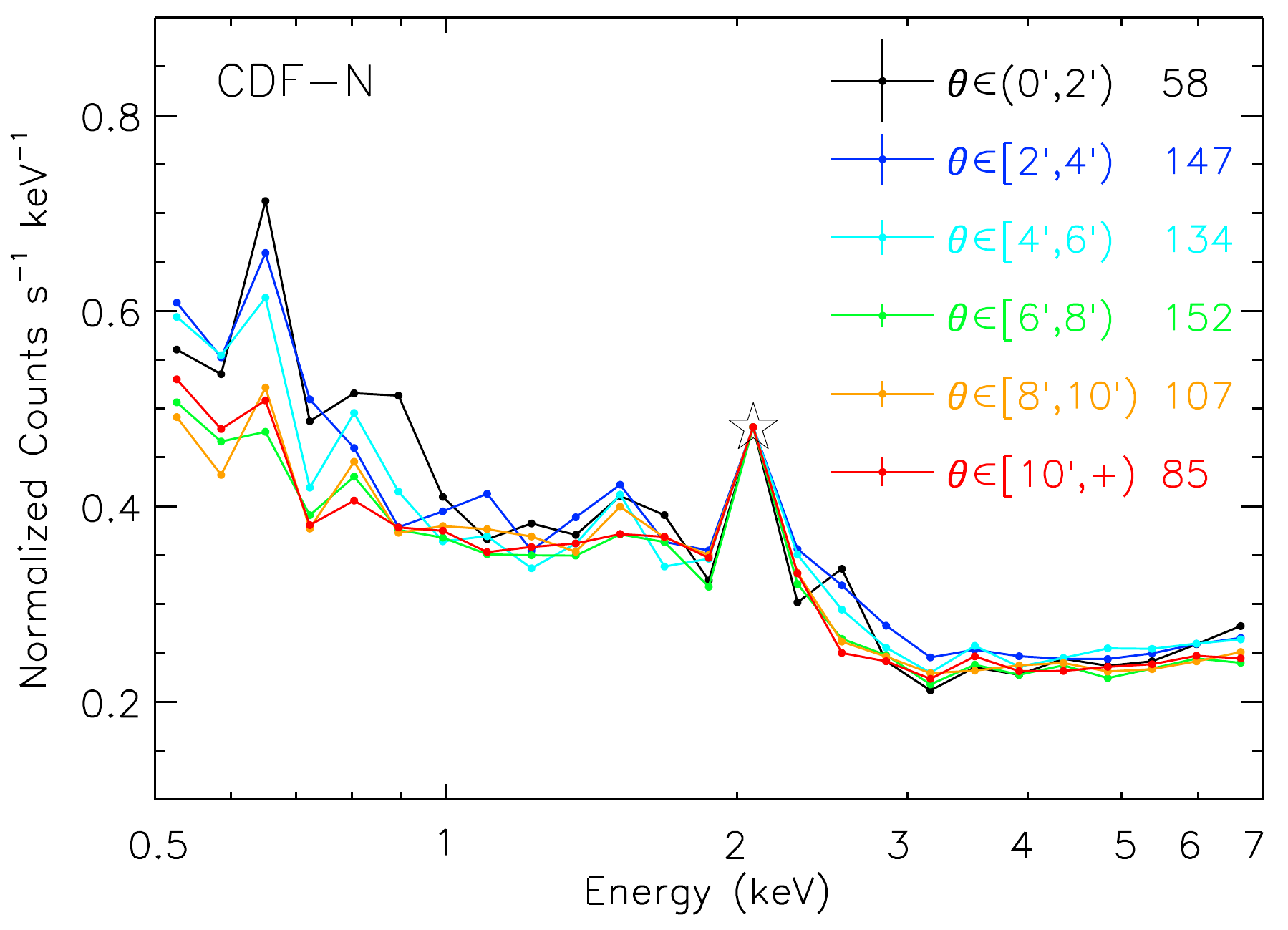}}
\figcaption{Mean background spectra for the 2~Ms \cdfn\ main-catalog sources calculated
in various bins of off-axis angle. The spectra are normalized to have the same value at an
energy slightly above 2~keV, which is indicated by a large 5-pointed star.
For clarity, errors on individual spectral data points are not plotted; 
the typical spectral error value and number of sources in each bin of off-axis angle 
are annotated in the top-right corner. 
\label{fig:cdfn-bkg-spec}}
\end{figure}

Figure~\ref{fig:cdfn-bkg-spec} presents the mean \chandra\ background spectra that are calculated 
for the 683 main-catalog sources in various bins of off-axis angle,
using the individual background spectra extracted in Section~\ref{sec:cdfn-cand}.
We find that
(1) the shapes of the mean \chandra\ background spectra remain largely the same across the  
entire \cdfn\ field given the uncertainties, in particular, 
as far as the $\gsim 1$~keV parts of the spectra are concerned (with $\lsim 10\%$ variations 
between the shapes);
and (2) for the $\lsim 1$~keV parts of the mean background spectra, shape variations seem apparent
(up to $\approx 20\%$), with some hint of the spectra for sources with $\theta<6\arcmin$ lying
slightly above the spectra for sources with $\theta\ge 6\arcmin$.

Our background maps and background spectra have contributions of various origins
that include the unresolved cosmic \xray\ background, particle background,
and instrumental background (e.g., Markevitch \etal\ 2003).
In this work, we are only interested in the total background, thereby not distinguishing between
these different background components.
We refer readers to other works (e.g., Hickox \& Markevitch 2006) 
that carefully characterize, distinguish, and measure
these individual \chandra\ background components that are key to their
specific scientific goals.

\begin{table*}
%\tabletypesize{\small}
%\tablewidth{0pt}
\caption{2~Ms \cdfn: Background Parameters}
\centering
\begin{tabular}{lcccc}\hline\hline
Band (keV) & Mean Background & Mean Background & Total Background$^{\rm c}$ & Count Ratio$^{\rm d}$ \\
 & (count pixel$^{-1}$)$^{\rm a}$ & (count Ms$^{-1}$ pixel$^{-1}$)$^{\rm b}$ & (10$^5$ counts) & (Background/Source) \\ \hline 
Full (0.5--7.0) & 0.171 & 0.167 & 11.4 & \phantom{0}5.3 \\
Soft (0.5--2.0) & 0.057 & 0.055 & \phantom{0}3.8 & \phantom{0}2.7 \\
Hard (2--7) & 0.115 & 0.108 & \phantom{0}7.6 & 10.2 \\\hline
\end{tabular}
\\$^{\rm a}$ The mean numbers of background counts per pixel.
\\$^{\rm b}$ The mean numbers of background counts per pixel
divided by the mean effective exposures.
\\$^{\rm c}$ The total numbers of background counts in the background maps.
\\$^{\rm d}$ Ratio between the total number of background
counts and the total number of detected source counts in the main catalog.
\label{tab:cdfn-bkg}
\end{table*}

\subsubsection{Sensitivity Map Creation}\label{sec:cdfn-smap}

Given the above background maps and
the main-catalog source-detection criterion of $P<0.004$,
we can measure $B_{\rm src}$ and $B_{\rm ext}$
to obtain the minimum number of counts ($S$) required
for a detection according to the binomial no-source probability equation
(i.e., Equation~\ref{equ:bi} in Section~\ref{sec:cdfn-main-select}),
and then create sensitivity maps in the three standard bands 
for the main catalog to assess the sensitivity as a function of position 
across the entire field.
We first determine $B_{\rm src}$ in the background maps 
utilizing circular apertures of $\approx 90\%$ PSF EEF radii.
We then derive $B_{\rm ext}$ as follows to mimic the AE behavior
of extracting background counts for the main-catalog sources:
for a given pixel in the background map, we calculate
its off-axis angle ($\theta_p$) and
set the $B_{\rm ext}$ value to the maximum $B_{\rm ext}$ value 
(corresponding to the highest sensitivity) 
of the main-catalog sources
that lie within an annulus having the inner/outer radius of
$\theta_p-0\farcm 25$/$\theta_p+0\farcm 25$.
Subsequently, we numerically solve Equation~(\ref{equ:bi})
to obtain the minimum counts $S$ in the source-extraction region
that lead to a value of $P<0.004$. 
Finally, we create sensitivity maps for the main catalog utilizing the
exposure maps, under the assumption of a $\Gamma=1.4$ power-law model with Galactic
absorption.
We find that there are 12, 15, and 9 main-catalog sources
in the three standard bands that lie typically $\lsim10\%$
below the corresponding derived
sensitivity limits, respectively, which is likely
due to background fluctuations and/or their real $\Gamma$ values
differing significantly from the assumed $\Gamma=1.4$.

\begin{figure}
\centerline{\includegraphics[width=8.5cm]{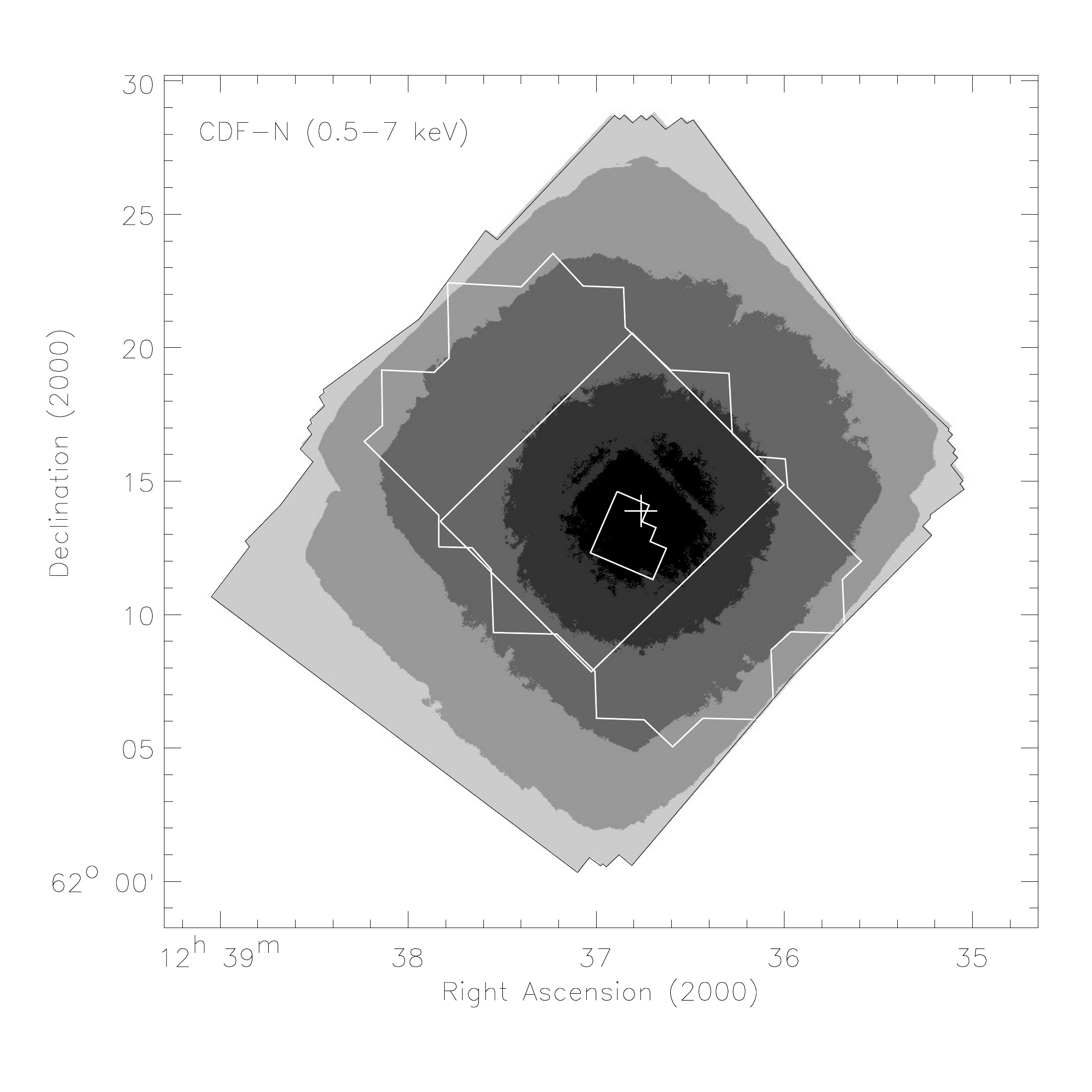}}
\figcaption{Full-band (0.5--7.0 keV) sensitivity map for the 2~Ms \cdfn\ main catalog.
The gray-scale levels, ranging from black to light gray, denote areas with flux limits
of $<5.0\times 10^{-17}$, \hbox{$5.0\times 10^{-17}$} to $10^{-16}$, $10^{-16}$ to
\hbox{$3.3\times10^{-16}$}, \hbox{$3.3\times 10^{-16}$} to $10^{-15}$, and $>10^{-15}$ \flux, respectively.
The regions and the plus sign have the same meanings as those in Fig.~\ref{fig:cdfn-fb-img}.
\label{fig:cdfn-fb-senimg}}
\end{figure}

\begin{figure}
\centerline{\includegraphics[width=8.5cm]{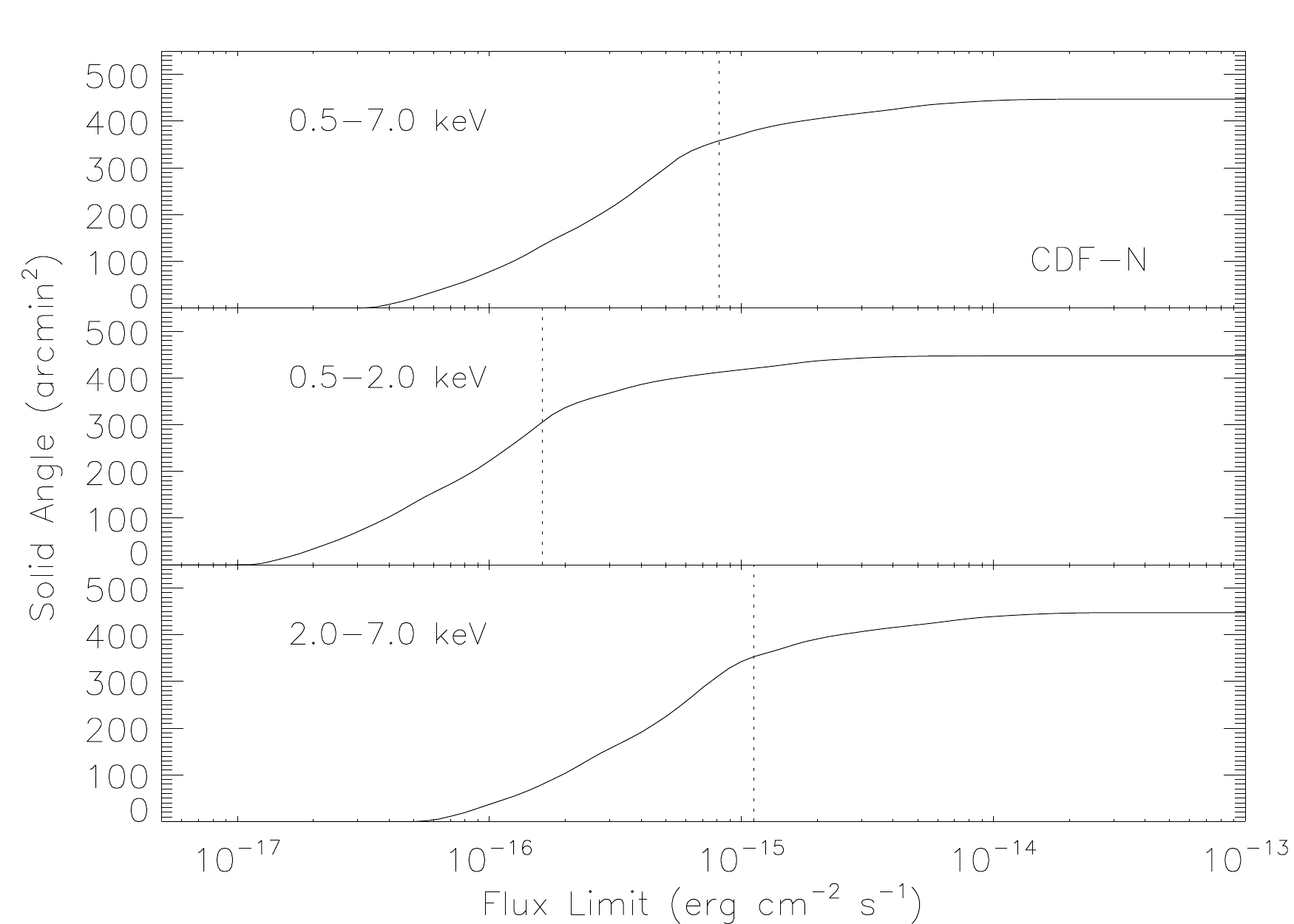}}
\figcaption{Survey solid angle as a function of flux limit in the full, soft, and hard bands
for the 2~Ms \cdfn\ main catalog.
The vertical dotted lines indicate the median fluxes of the main-catalog sources detected in the three bands.
\label{fig:cdfn-senhist}}
\end{figure}

Figure~\ref{fig:cdfn-fb-senimg} displays the full-band 
sensitivity map for the main catalog, and 
Figure~\ref{fig:cdfn-senhist} presents plots of survey solid angle
versus flux limit in the three standard bands given $P<0.004$.
It is clear that higher sensitivities are reached
at smaller off-axis angles and thus within smaller
survey solid angles.
We find the mean sensitivity limits achieved in
the central $\approx$1~arcmin$^2$ area at the average aim point to be 
$\approx 3.5\times 10^{-17}$, $1.2\times 10^{-17}$, and $5.9\times 10^{-17}$ \flux\
for the full, soft, and hard bands, respectively,
which represent a factor of $\approx 2$ improvement over those of A03,
due to the facts that we adopt a sensitive two-stage source-detection 
procedure and that A03 adopted a different methodology 
for sensitivity calculations.

\section{Production of the improved 250~ks \ecdfs\ point-source catalogs}\label{sec:ecdfs}

The overall production procedure is similar to that
used for the 2~Ms \cdfn\ cataloging (see Fig.~\ref{fig:cataloging_flow}) 
detailed in Section~\ref{sec:cdfn}
and to that described in X11.
To avoid unnecessary repetition,
we provide here the salient details when appropriate and
refer readers to Section~\ref{sec:cdfn} for essential details
and X11 for full details.
In addition, we make our 250~ks \ecdfs\ data products publicly
available.$^{\ref{ft:datalink}}$

\subsection{Observations and Data Reduction}\label{sec:ecdfs-obs}

The \ecdfs\ is composed of four distinct and contiguous 
$\approx$250~ks \chandra\ pointings
(hereafter Fields 1, 2, 3, and 4) 
that flank the \cdfs\ proper, 
consisting of a total of 9 separate observations
taken between 2004 February 29 and 2004 November 20 (see Table~1 of L05
for the journal of these 9 \ecdfs\ observations).
The 9 \ecdfs\ observations made use of ACIS,
with the four \mbox{ACIS-I} CCDs being in operation throughout the 9
\ecdfs\ observations and the \mbox{ACIS-S} CCD S2 being operated 
for observations 5019--5022 and 6164.
We do not use the data taken with the \mbox{ACIS-S} CCD S2 
due to its large off-axis angle and consequently its low sensitivity.
For all 9 \ecdfs\ observations,
the focal-plane temperature was $-120\degr$C
and Very Faint mode was adopted.

The background light curves for all the 9 \ecdfs\ observations
were examined utilizing ChIPS.
During observation 5015,
there are two significant flares in the background, with
each lasting $\approx 1.0$~ks and 
being $\gsim 1.5$ times higher than nominal. 
The background increased to $\gsim 1.5$ times the nominal rate
and remained above this level toward the end of observation 5017,
affecting an exposure of $\approx 10.0$~ks.
All the other observations are free from strong flaring
and are stable within $\approx 20$\% of typical quiescent \chandra\ values,
except for a number of short moderate ``spikes'' (up to $\approx 1.5$ times
higher than nominal).
To remove these significant flares and moderate spikes, 
we utilize {\sc lc\_sigma\_clip}
with \mbox{3.5-sigma} clippings adopted for all the 9 \ecdfs\ observations.
We then follow Section~\ref{sec:cdfn-obs-data} for subsequent
data reduction.

The entire 250~ks \ecdfs\ covers a total region of 1128.6 arcmin$^2$,
slightly smaller than four times the \mbox{ACIS-I} field of view
due to overlapping of observation field edges.
The aim points are 
($\alpha_{\rm J2000.0}=03^{\rm h}33^{\rm m}05.^{\rm s}6$,
$\delta_{\rm J2000.0}=-27\degr41\arcmin08\farcs 8$),
($\alpha_{\rm J2000.0}=03^{\rm h}31^{\rm m}51.^{\rm s}4$,
$\delta_{\rm J2000.0}=-27\degr41\arcmin38\farcs 8$),
($\alpha_{\rm J2000.0}=03^{\rm h}31^{\rm m}49.^{\rm s}9$,
$\delta_{\rm J2000.0}=-27\degr57\arcmin14\farcs 6$), and
($\alpha_{\rm J2000.0}=03^{\rm h}33^{\rm m}02.^{\rm s}9$,
$\delta_{\rm J2000.0}=-27\degr57\arcmin16\farcs 1$)
for Fields 1--4, respectively.

\subsection{Images, Exposure Maps, and Candidate-List Catalog}\label{sec:ecdfs-img-cand}

We follow Section~\ref{sec:cdfn-img} to
construct the raw images and effective-exposure maps
for the three standard bands as well as
the exposure-weighted smoothed images
in the \mbox{0.5--2}, \mbox{2--4}, and \mbox{4--7~keV} bands that
are subsequently combined into a false-color composite.
When registering the individual observations to a common astrometric frame,
we match \mbox{X-ray} centroid positions to 
the $K_s\le 21.0$~mag sources in
the TENIS WIRCam $K_s$-band catalog (Hsieh \etal 2012)
rather than the VLA 1.4~GHz \ecdfs\ radio sources (Miller \etal 2013)
that were adopted in X11,
because we find the astrometric frame of the $K_s$-band catalog 
in better agreement with that of other multiwavelength
catalogs that are used for our \xray\ source identifications
in Section~\ref{sec:ecdfs-id}.

\begin{figure}
\centerline{\includegraphics[width=8.5cm]{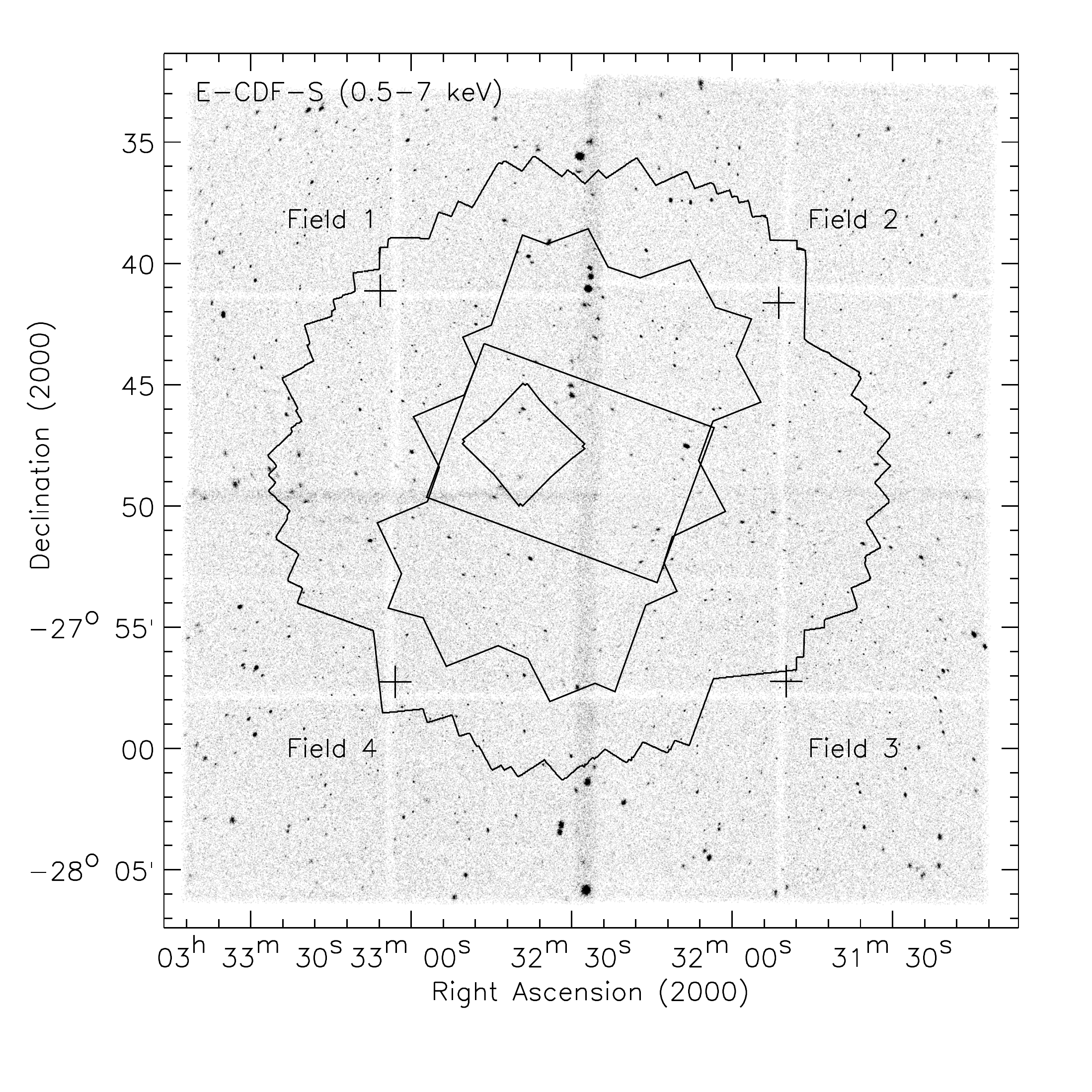}}
\figcaption{Full-band (0.5--7.0~keV) raw image of the 250~ks \ecdfs\ rendered using linear gray scales.
The \ecdfs\ consists of four \chandra\ observational fields (Fields 1--4)
that flank the \cdfs\ proper (1~Ms \cdfs, Giacconi \etal 2002; 2~Ms \cdfs, Luo \etal 2008; 4~Ms \cdfs, X11)
indicated by the outermost segmented boundary;
note the increase in background where these fields overlap.
The aim points of the four fields
are indicated as plus signs within the fields.
The large polygon, the rectangle, and the small polygon indicate the regions for the
\goodss\ (Giavalisco \etal 2004),
the CANDELS \goodss\ deep (Grogin \etal 2011; Koekemoer \etal 2011),
and the HUDF (Beckwith \etal 2006), respectively.
The light grooves running through the image are caused by the \mbox{ACIS-I} CCD gaps,
thereby having lower effective exposures than the nearby non-gap areas
(clearly revealed in Fig.~\ref{fig:ecdfs-fb-exp}).
\label{fig:ecdfs-fb-img}}
\end{figure}

\begin{figure}
\centerline{\includegraphics[width=8.5cm]{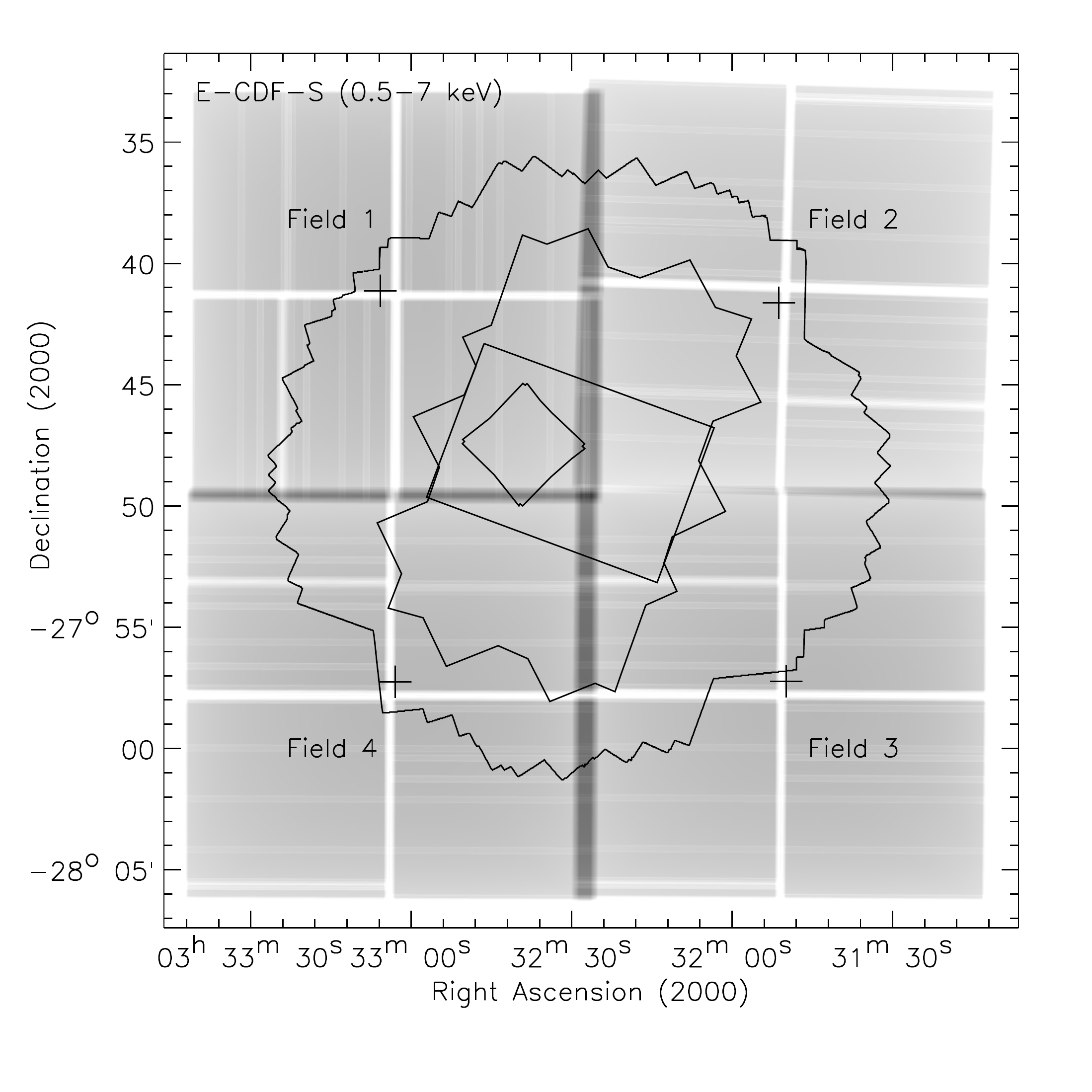}}
\figcaption{Full-band (0.5--7.0~keV) effective-exposure map of the 250~ks \ecdfs\ rendered using linear gray scales.
The darkest areas indicate the highest effective exposure times;
the high effective exposures between fields are due to overlap of observations.
The \hbox{ACIS-I} CCD gaps can be clearly identified as the white grooves.
The regions and the plus signs have the same meanings as those in Fig.~\ref{fig:ecdfs-fb-img}.
\label{fig:ecdfs-fb-exp}}
\end{figure}

\begin{figure}
\centerline{\includegraphics[width=8.5cm]{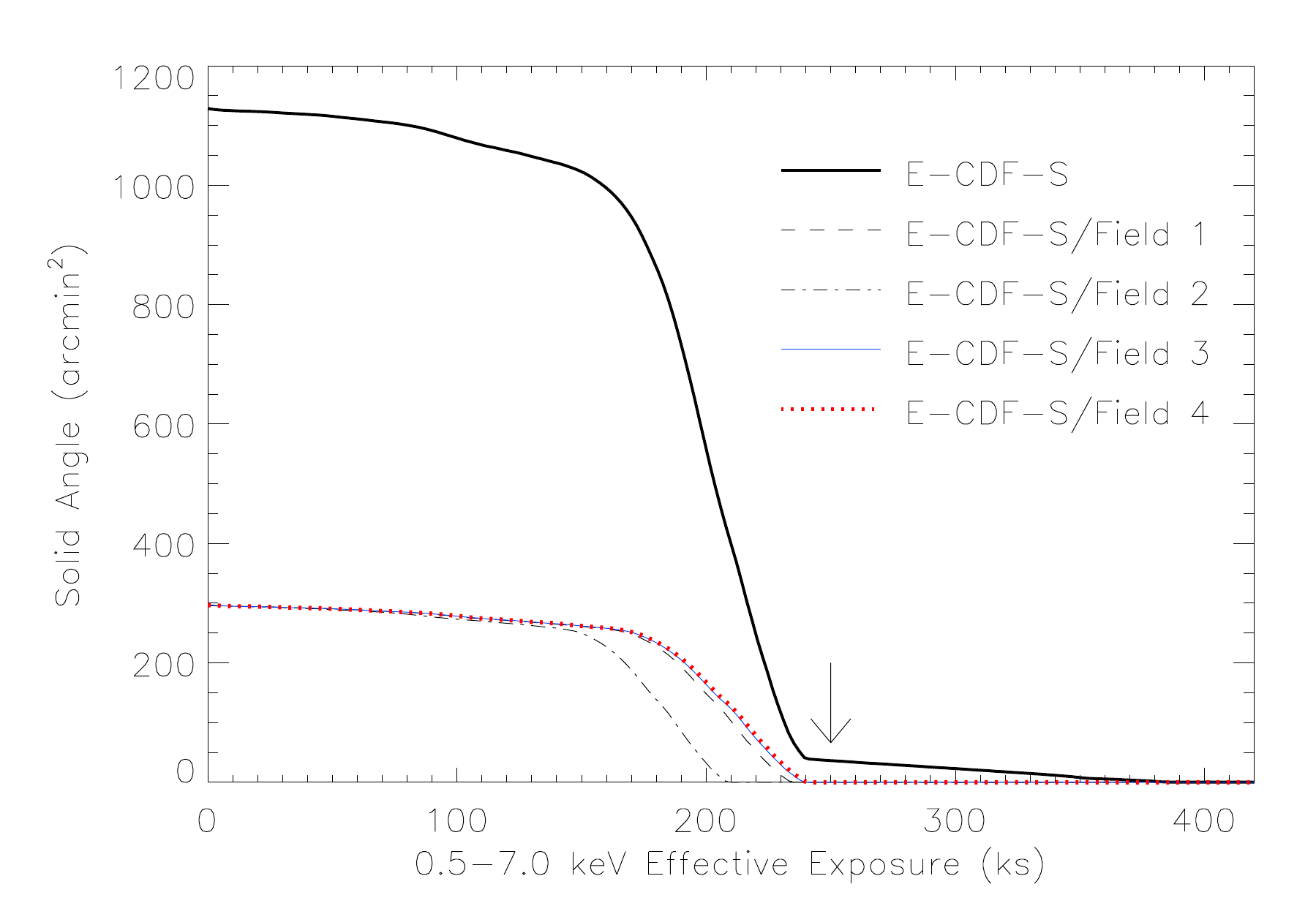}}
\figcaption{Survey solid angle as a function of minimum full-band \mbox{(0.5--7.0~keV)} effective exposure for the entire 250~ks \ecdfs\ (thick solid curve) and the four observational fields (dashed, dashed-dot, solid, and dotted curves).
Note that the dashed-dot curve for Field~2 appears different
from the curves for the other fields, due to the fact that
the nominal summed exposure of Field~2 is \mbox{$\approx13$--19~ks}
shorter than that of the other fields. 
The entire \ecdfs\ covers a total area of 1128.6 arcmin$^2$, roughly four times \hbox{ACIS-I} field of view (16\farcm9$\times$16\farcm9).
The ``tail'' with exposures $>250$~ks (i.e., the portion of the thick solid curve on the right of the downward arrow
signifying the 250~ks exposure) corresponds to regions where observational fields overlap (see Fig.~\ref{fig:ecdfs-fb-exp}).
\label{fig:ecdfs-solid-exp}}
\end{figure}

\begin{figure}
\centerline{\includegraphics[width=8.5cm]{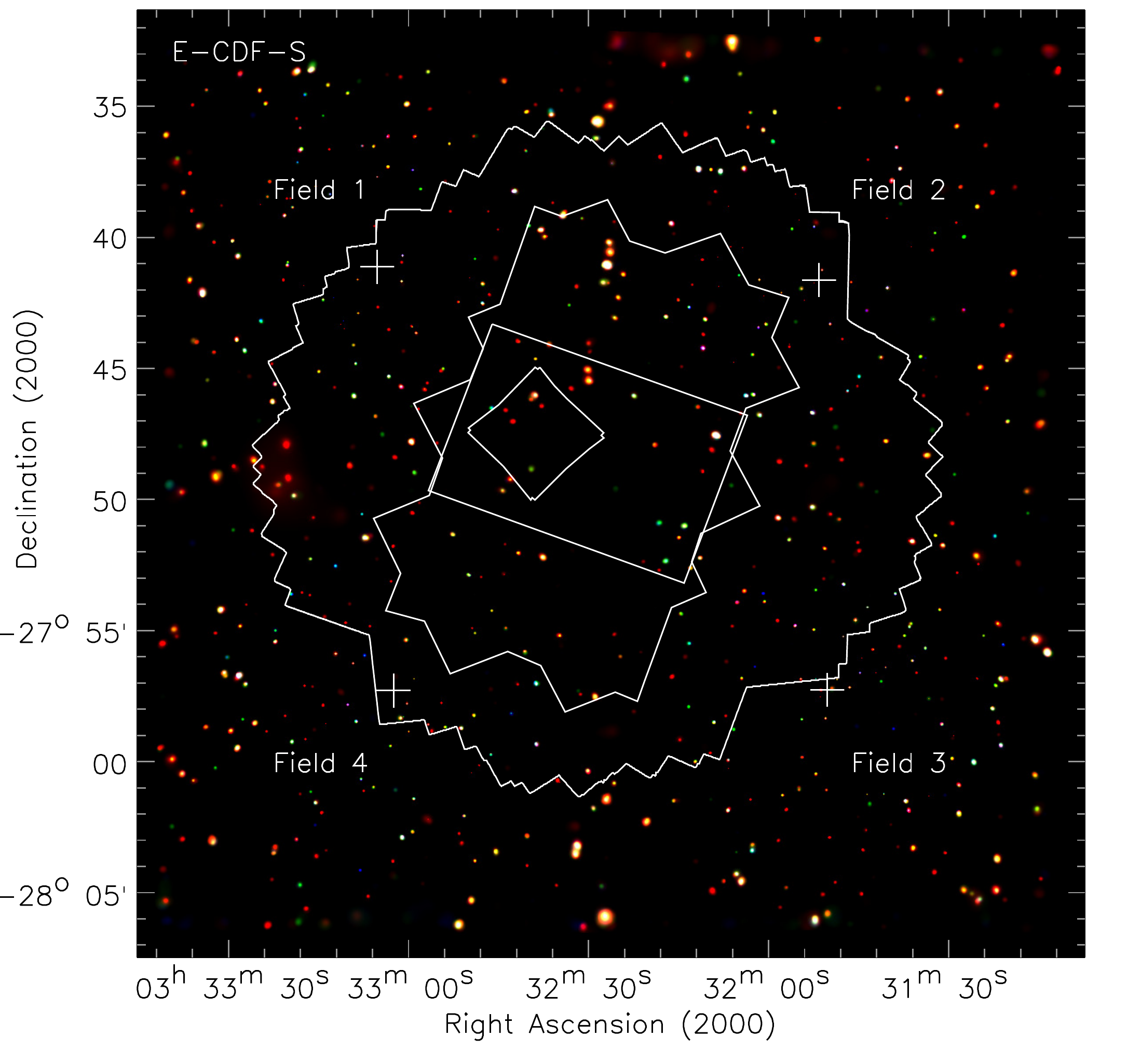}}
\figcaption{False-color image of the 250~ks \ecdfs\ that
is a color composite of the exposure-weighted and adaptively smoothed images
in the 0.5--2.0 keV ({\it red}), 2--4 keV ({\it green}), and 4--7 keV ({\it blue}) bands.
Near the aim points of the four observational fields, the seemingly 
smaller sizes and lower brightnesses of sources
are due to the smaller PSF size on-axis.
The regions and the plus signs have the same meanings as those in Fig.~\ref{fig:ecdfs-fb-img}.
\label{fig:ecdfs-smooth-false}}
\end{figure}

Figures~\ref{fig:ecdfs-fb-img} and
\ref{fig:ecdfs-fb-exp} present the full-band raw image
and effective-exposure map, respectively.
Figure~\ref{fig:ecdfs-solid-exp} displays the survey solid angle as a
function of the minimum full-band effective exposure,
and Figure~\ref{fig:ecdfs-smooth-false}
presents a false-color composite of the 250~ks \ecdfs.

Following the blind-source search in Section~\ref{sec:cdfn-cand},
we use {\sc wavdetect}
(with the key parameters set to sigthresh$=10^{-5}$,
energy=1.497~keV, ECF=0.393, and minimum PSF map size, respectively)
to detect sources in the combined raw images 
in the three standard bands,
utilize the AE-computed centroid and matched-filter positions
to improve the {\sc wavdetect} source positions,
and make use of AE to perform reliable \xray\ photometry extractions.
Our candidate-list catalog consists of 1434 \ecdfs\ source candidates,
with each being detected in at least one of the three standard bands.

\subsection{Main \chandra\ Source Catalog}\label{sec:ecdfs-main}

\subsubsection{Selection of Main-Catalog Sources}\label{sec:ecdfs-main-select}

\begin{figure*}
\centerline{\includegraphics[width=13cm]{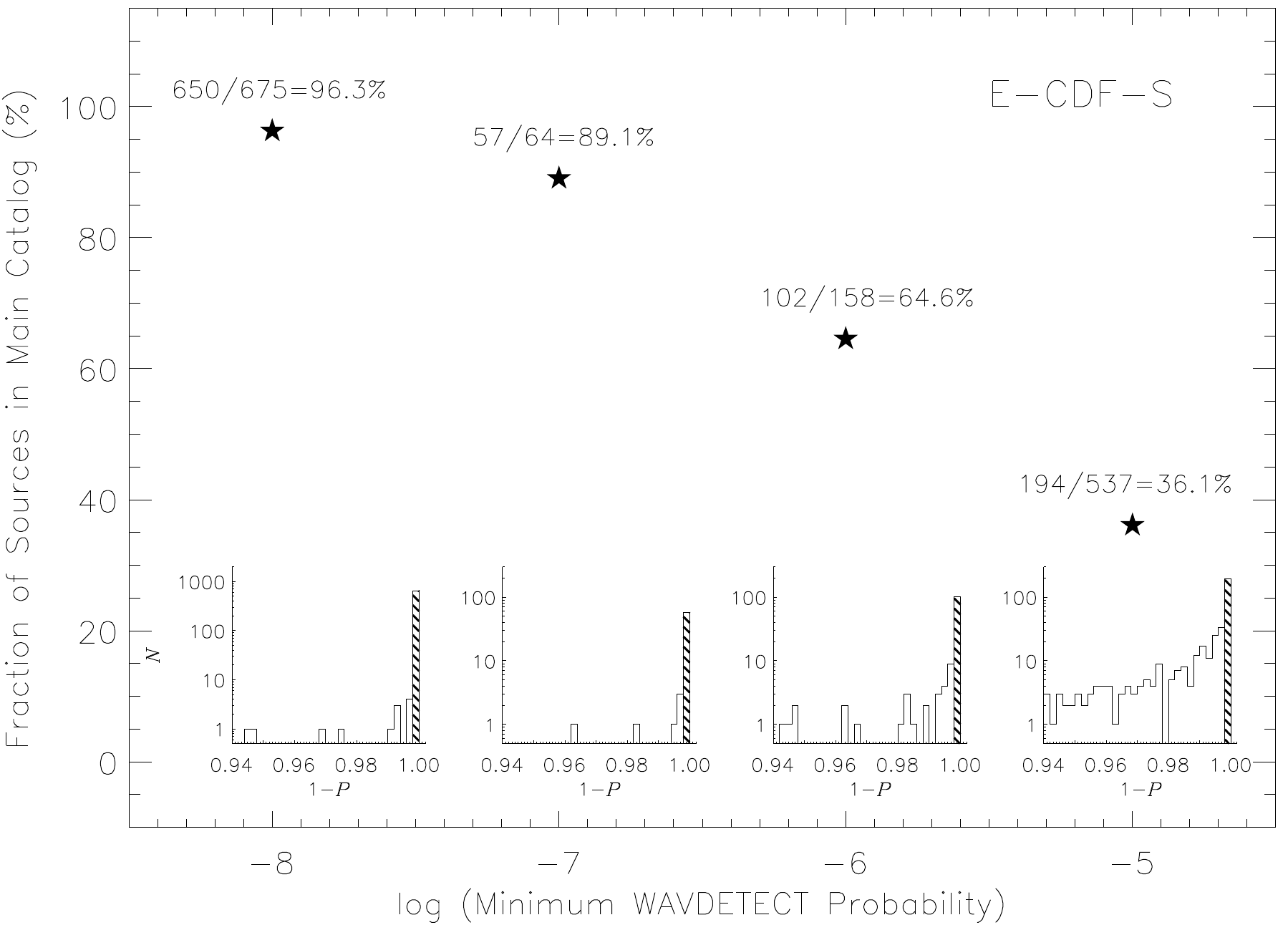}}
\figcaption{Fraction of the candidate-list catalog sources having an AE
binomial no-source probability $P<0.002$ that are included in
the 250~ks \ecdfs\ main catalog, as a function of minimum {\sc wavdetect} probability$^{\ref{ft:siglev}}$ 
(denoted as five-pointed stars).
The number of sources having $P<0.002$ versus the number of
candidate-list catalog
sources detected at each minimum {\sc wavdetect} probability are
displayed (note that there are
650+57+102+194=1003 main-catalog sources and 675+64+158+537=1434 candidate-list catalog sources).
The fraction of candidate-list catalog sources included in the
main catalog falls from 96.3\% to 36.1\%
between minimum {\sc wavdetect} probabilities of $10^{-8}$
and $10^{-5}$.
The insets present the $1-P$ distributions for the
candidate-list catalog sources at each minimum {\sc wavdetect}
probability, and the shaded areas highlight those included in
the main catalog (i.e., satisfying $1-P>0.998$).
\label{fig:ecdfs-prob-siglev}}
\end{figure*}

To discard spurious candidate-list catalog sources,
we include a candidate source into the main catalog only if
it has $P < 0.002$ in at least one of the three standard bands.
The choice of the $P < 0.002$ criterion results from a balance
between keeping the fraction of spurious sources small and
recovering the maximum possible number of real sources,
primarily based on joint maximization of the total number of sources and
minimization of the fraction of sources without significant
multiwavelength counterparts (see Section~\ref{sec:ecdfs-id}).
Our main catalog consists of a total of 1003 sources given this $P < 0.002$
criterion.
Figure~\ref{fig:ecdfs-prob-siglev} presents the fraction of 
candidate-list sources that satisfy the $P<0.002$ main-catalog 
source-selection criterion
and the $1-P$ distribution of candidate-list sources 
as a function of the minimum {\sc wavdetect} probability.$^{\ref{ft:siglev}}$

\subsubsection{X-ray Source Positional Uncertainty}\label{sec:ecdfs-dpos}

\begin{figure}
\centerline{\includegraphics[scale=0.6]{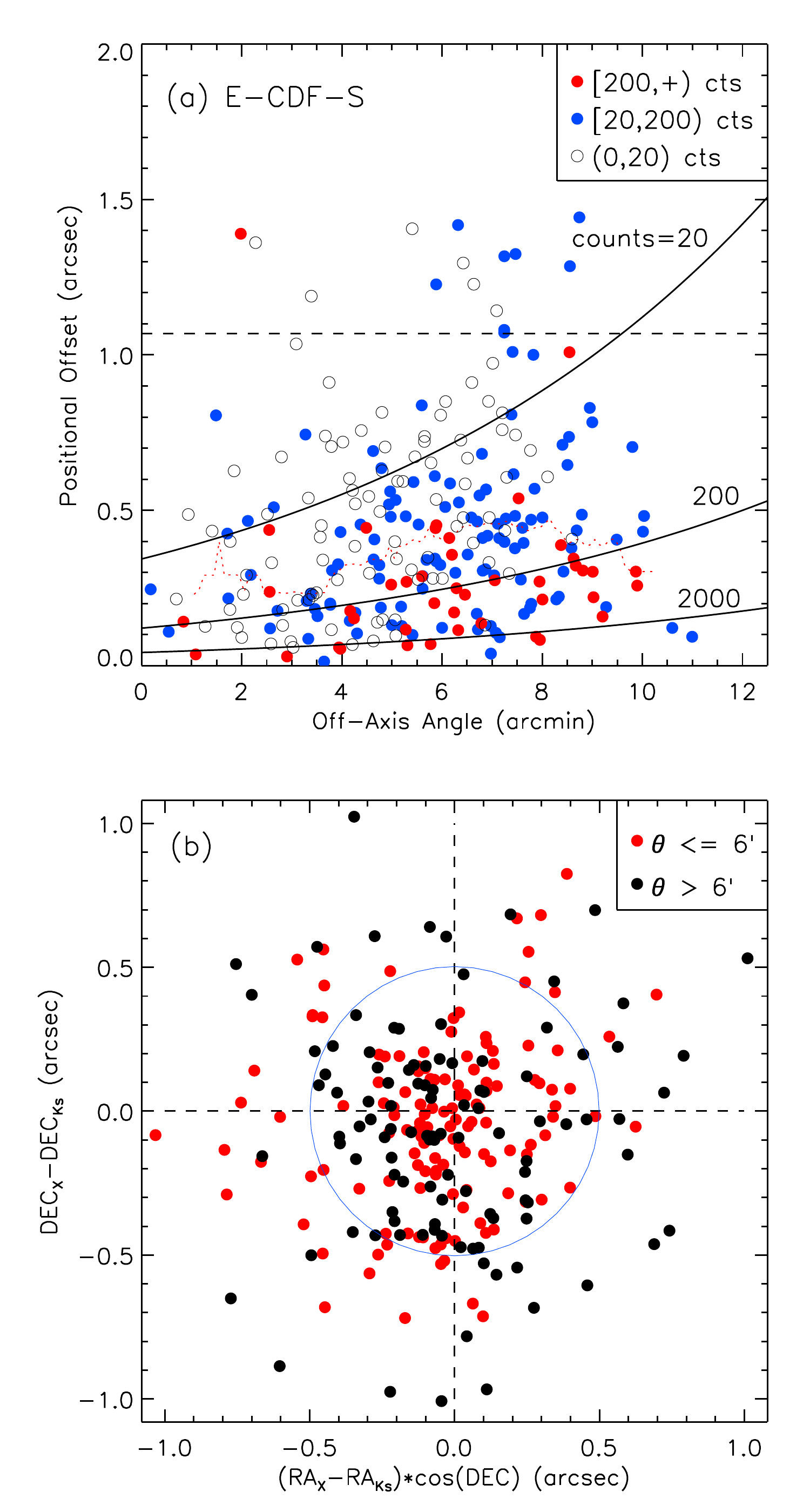}}
\figcaption{(a) Positional offset vs. off-axis angle for
the 257 250~ks \ecdfs\ main-catalog sources that have $K_s\le 20.0$~mag counterparts in the
TENIS WIRCam $K_s$-band catalog (Hsieh \etal 2012) utilizing a matching radius of $1\farcs5$ (see Section~\ref{sec:ecdfs-dpos}
for the description of an apparent outlier,
i.e., the red filled circle located at the top-left corner,
that deviates significantly from the relation defined as 
Equation~\ref{equ:dpos}).
Red filled, blue filled, and black open circles indicate
\mbox{X-ray} sources with $\ge200$, $\ge20$, and $<20$
counts in the energy band that is used to determine the source position, respectively.
The red dotted curve denotes the running median of positional offset in
bins of $2\arcmin$.
The horizontal dashed line represents the median offset ($1\farcs07$)
of the false matches expected.
The three solid curves correspond to the $\approx68\%$ confidence-level \xray\ positional
uncertainties (derived according to Equation~\ref{equ:dpos}) for sources with 20, 200 and 2000 counts.
(b) Positional residuals between the \xray\ and $K_s$-band positions 
for the remaining 256 \xray-$K_s$-band matches.
Red and black filled circles represent sources with an off-axis angle
of $\le 6\arcmin$ and $>6\arcmin$, respectively.
A blue circle with a $0\farcs5$ radius is drawn at the center as visual guide.
\label{fig:ecdfs-dposfit}}
\end{figure}

\begin{figure}
\centerline{\includegraphics[width=8.5cm]{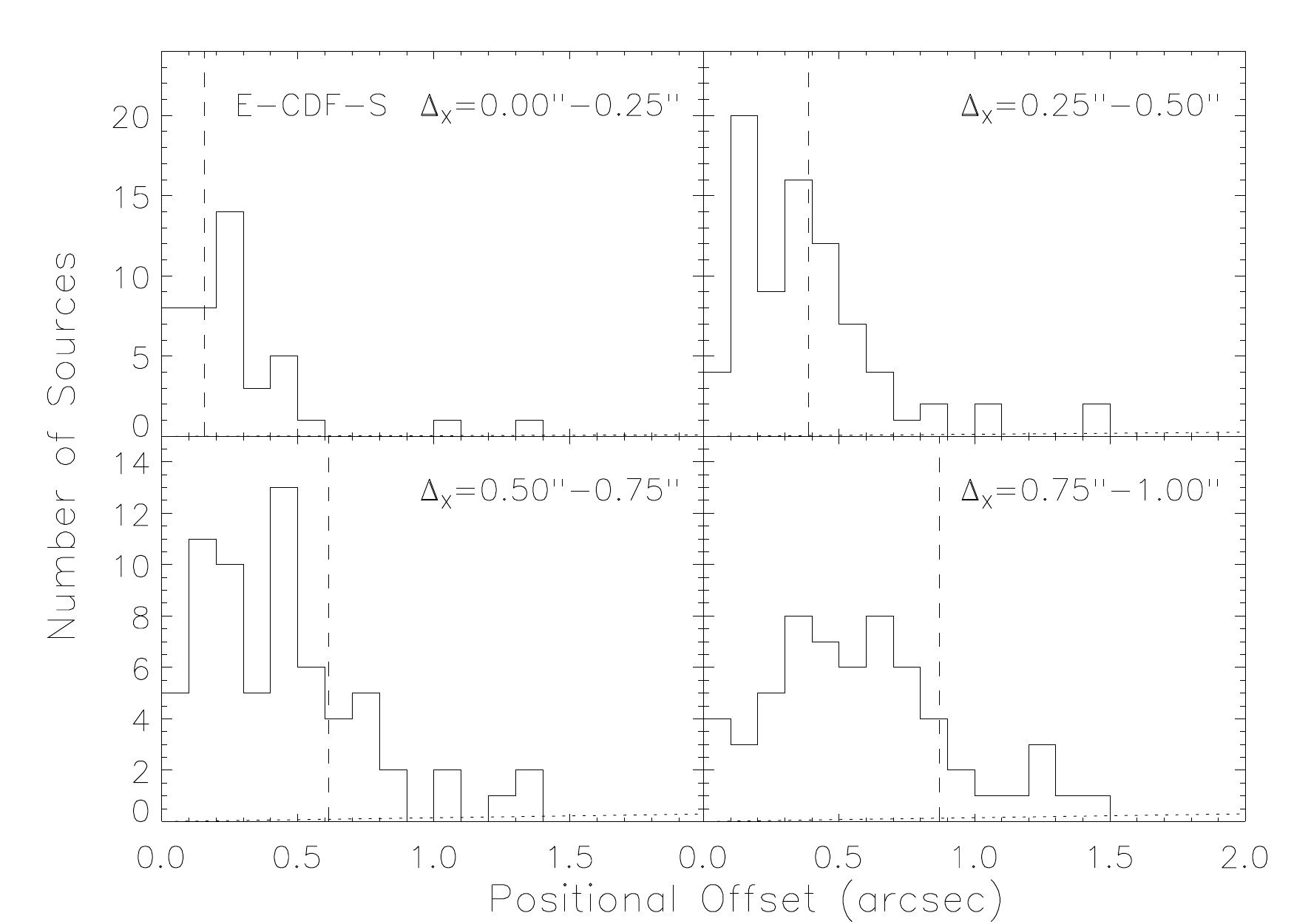}}
\figcaption{Histograms of \xray-$K_s$-band positional offsets for the 256 250~ks \ecdfs\ main-catalog
sources that are matched to the TENIS WIRCam $K_s\le 20.0$~mag sources (Hsieh \etal 2012)
utilizing a matching radius of $1\farcs5$ (note that one source
is excluded; see Fig.~\ref{fig:ecdfs-dposfit}(a) and Section~\ref{sec:ecdfs-dpos}).
Based on their \xray\ positional
uncertainties estimated with Equation~(\ref{equ:dpos}),
these matched sources are divided into four bins of $0\arcsec$--$0\farcs25$, $0\farcs25$--$0\farcs50$,
$0\farcs50$--$0\farcs75$, and $0\farcs75$--$1\arcsec$.
In each panel (bin), the vertical dashed line denotes the median \xray\ positional uncertainty;
the dotted line (almost indistinguishable from the bottom $x$-axis) displays
the expected numbers of false matches
assuming a uniform spatial distribution of $K_s$-band sources
as a function of \xray-$K_s$-band positional offset.
\label{fig:ecdfs-poshist}}
\end{figure}

We find 257 matches between the 1003 main-catalog sources and
the $K_s\le 20.0$~mag sources in the TENIS WIRCam $K_s$-band catalog using 
a matching radius of $1\farcs5$. 
We estimate on average $\approx 6.5$ (2.3\%) false matches and
a median offset of $1\farcs07$ for these false matches.
Figure~\ref{fig:ecdfs-dposfit}(a) presents the positional offset between
these 257 \hbox{X-ray}-$K_s$-band matches
(the median offset is $0\farcs38$) as a function of off-axis angle.
The source indicated as a red filled circle at the top-left corner
is a mismatch (its real counterpart is a nearby fainter source
that is resolved in the {\it HST} image but not resolved in
the $K_s$-band image) and 
is therefore not included in the following analysis of 
\xray\ positional uncertainty.
Figure~\ref{fig:ecdfs-dposfit}(b) presents 
the positional residuals between the \xray\
and \mbox{$K_s$-band} positions for the remaining 256 sources,  
which appear roughly symmetric.
We find that
the empirical formula of the 68\% confidence-level \xray\
positional uncertainty with off-axis angle and source-count dependencies
that is derived for the 2~Ms \cdfn\ main-catalog
sources (i.e., Equation~\ref{equ:dpos} in Section~\ref{sec:cdfn-dpos}) 
is fully applicable to the 256 250~ks \ecdfs\ main-catalog sources. 
Figure~\ref{fig:ecdfs-poshist} shows the distributions of
\xray-$K_s$-band positional
offsets in four bins of X-ray positional uncertainty.
For the analysis here, as in Section~\ref{sec:cdfn-dpos},
we allow for positional uncertainties arising from
the $K_s$-band sources that are typically $\lsim$0\farcs1.

\subsubsection{Multiwavelength Identifications}\label{sec:ecdfs-id}

We implement the likelihood-ratio matching procedure
to identify the primary ONIR counterparts
for the main-catalog \xray\ sources.
We adopt, in order of priority, 
nine ONIR catalogs for identification purposes.
\begin{enumerate}

\item The VLA 1.4 GHz \ecdfs\ radio catalog (denoted as ``VLA'';
Miller \etal 2013), with a 5$\sigma$ limiting flux density of 
$\approx 20~\mu$Jy.

\item The \mbox{GOODS-S} {\it HST} version 2.0
$z$-band catalog (denoted as ``\mbox{GOODS-S}''; Giavalisco \etal 2004), 
with a $5\sigma$ limiting magnitude of $28.2$.

\item The GEMS {\it HST} $z$-band catalog (denoted as ``GEMS''; 
Caldwell \etal 2008), with a $5\sigma$ limiting magnitude of $27.3$.

\item The CANDELS \goodss\ WFC3 $H$-band catalog (denoted as ``CANDELS''; 
Grogin \etal 2011; Koekemoer \etal 2011),
with a $5\sigma$ limiting magnitude of $28.0$. 

\item The \goodss\ MUSIC catalog (denoted as ``MUSIC''; Grazian \etal 2006;
we adopt the $K$-selected sources in the V2 catalog 
presented in Santini \etal 2009)
based on the Retzlaff \etal (2010) VLT/ISAAC data,
with a limiting $K$-band magnitude of $23.8$ at 90\% completeness.

\item The TENIS WIRCam $K_s$-band catalog (denoted as ``TENIS'';
Hsieh \etal 2012), with a $5\sigma$ limiting magnitude of $25.0$
in the inner 400 arcmin$^2$ region.

\item The ESO 2.2-m WFI $R$-band catalog (denoted as ``WFI'';
Giavalisco \etal 2004),
with a $5\sigma$ limiting magnitude of $27.3$.

\item The MUSYC $K$-band catalog (denoted as ``MUSYC''; Taylor \etal 2009),
with a $5\sigma$ limiting magnitude of $22.4$.

\item The SIMPLE IRAC 3.6 $\mu$m catalog (denoted as ``SIMPLE''; 
Damen \etal 2011), with a
$5\sigma$ limiting magnitude of 23.8.

\end{enumerate}

We shift the above ONIR source positions
appropriately to be consistent with the TENIS
WIRCam $K_s$-band astrometry (see Section~\ref{sec:ecdfs-img-cand}).
We identify primary ONIR counterparts for 958 (95.5\%) of 
the 1003 main-catalog sources.
Utilizing the Monte Carlo approach mentioned in Section~\ref{sec:cdfn-id}, 
we estimate the false-match rates for the above nine catalogs 
in the listed order to be
0.2\%, 4.4\%, 4.6\%, 3.4\%, 2.9\%, 2.4\%, 5.6\%, 2.1\%, and 2.6\%,
respectively, with a weighted mean false-match rate of 3.3\%.

We visually examine the \hbox{X-ray} images of
the 45 main-catalog sources without highly significant multiwavelength
counterparts, and find that
the majority of them have apparent or even strong \hbox{X-ray} signatures.
Of these 45 sources,
three are located near a very bright star
(their counterparts might thus be hidden by light of the
bright stars), with one having 40.7
full-band counts and having no
associations with any background flares or cosmic-ray afterglows; 
one has a large off-axis angle of $8\farcm4$ and 12.5 full-band counts; 
two have full-band counts of 14.7 and 16.6, respectively;
and all the other 39 sources have $<10$ full-band counts
(some of them are thus likely false detections).
Only 4 out of these 45 sources are also present in the L05 main catalog.

\subsubsection{Main-Catalog Details}\label{sec:ecdfs-maincat}

Our main catalog consists of a total of 97~columns,
the vast majority of which are similar to the columns presented in the
2~Ms \cdfn\ main catalog (see Table~\ref{tab:cdfn-cols} in 
Section~\ref{sec:cdfn-maincat}),
with some additional distinct columns including
\zs\ quality flag, corresponding L05 and X11 source information, and
observation field.
We present the main catalog itself in Table~\ref{tab:ecdfs-main}.
Below we give the details of these 97~columns.

\begin{table*}
%\tabletypesize{\scriptsize}
%\tablewidth{0pt}
\caption{250~ks \ecdfs\ Main {\it Chandra} Source Catalog}
\begin{tabular}{lllcccccccccc}\hline\hline
No. & $\alpha_{2000}$ & $\delta_{2000}$ & $\log P$ & {\sc wavdetect} & Pos Err & Off-axis & FB & FB Upp Err & FB Low Err & SB & SB Upp Err & SB Low Err \\
(1) & (2) & (3) & (4) & (5) & (6) & (7) & (8) & (9) & (10) & (11) & (12) & (13) \\ \hline
1 & 03 31 11.32 & $-$27 33 36.9 &    $-$10.1 &  $-$8 &  1.0 &  11.97 &    \phantom{0}46.8 &  10.5 &   \phantom{0}9.3 &    \phantom{0}32.2 &   \phantom{0}8.1 &   \phantom{0}6.9 \\
2 & 03 31 12.63 & $-$27 57 18.3 &     \phantom{0}$-$4.0 &  $-$5 &  1.0 &   \phantom{0}8.24 &    \phantom{0}15.7 &   \phantom{0}6.9 &   \phantom{0}5.6 &     \phantom{0}\phantom{0}9.9 &  $-$1.0 &  $-$1.0 \\
3 & 03 31 12.99 & $-$27 55 48.8 &    $-$99.0 &  $-$8 &  0.3 &   \phantom{0}8.28 &   333.4 &  21.2 &  20.0 &   157.1 &  14.6 &  13.4 \\
4 & 03 31 13.06 & $-$27 32 51.9 &    $-$11.6 &  $-$5 &  0.9 &  12.22 &    \phantom{0}58.0 &  12.0 &  10.8 &    \phantom{0}36.4 &   \phantom{0}8.6 &   \phantom{0}7.3 \\
5 & 03 31 13.64 & $-$27 49 49.0 &     \phantom{0}$-$3.5 &  $-$5 &  1.0 &  10.93 &    \phantom{0}29.6 &  11.0 &   \phantom{0}9.8 &    \phantom{0}17.6 &  $-$1.0 &  $-$1.0 \\ \hline
\end{tabular}
The full table contains 97~columns of information for the 1003 \xray\ sources.\\
(This table is available in its entirety in a machine-readable form in the online journal. A portion is shown here for guidance regarding its form and
content.)
\label{tab:ecdfs-main}
\end{table*}

\begin{enumerate}

\item Column 1 gives the source sequence number (i.e., XID) in this work. 
Sources are sorted in order of increasing right ascension.

\item Columns 2 and 3 give the J2000.0 right ascension and declination 
of the \mbox{X-ray} source, respectively.

\item Columns 4 and 5 give the minimum value of $\log P$
among the three standard bands,
and the logarithm of the minimum {\sc wavdetect} false-positive 
probability detection threshold, respectively.
For sources with $P=0$, we set $\log P=-99.0$.
We find a median value of $\log P=-10.4$ for the main-catalog sources, being
much smaller than the main-catalog selection threshold value of 
\mbox{$\log P<-2.7$} (i.e., $P<0.002$; see Section~\ref{sec:ecdfs-main-select}).
We find that 650, 57, 102, and 194 sources have minimum {\sc wavdetect} 
probabilities$^{\ref{ft:siglev}}$ of $10^{-8}$, $10^{-7}$, $10^{-6}$, and $10^{-5}$, 
respectively (see Fig.~\ref{fig:ecdfs-prob-siglev}).

\item Column 6 gives  the \mbox{X-ray}
positional uncertainty in units of arcseconds at the $\approx68\%$ confidence level,
which is computed utilizing Equation~(\ref{equ:dpos}).
For the main-catalog sources, the positional uncertainty ranges from
$0\farcs10$ to $1\farcs30$, with a median value of $0\farcs63$.

\item Column 7 gives the off-axis angle of each \hbox{X-ray} source in units
of arcminutes that is the angular separation between the \hbox{X-ray} source
and the aim point of the corresponding field 
(given in Section~\ref{sec:ecdfs-obs}).
For the main-catalog sources, the off-axis angle ranges from
$0\farcm18$ to $12\farcm22$, with a median value of $5\farcm47$.

\item Columns 8--16 give the aperture-corrected net 
(i.e., background-subtracted) source counts and
the associated $1\sigma$ upper and lower statistical errors 
for the three standard bands, respectively.
We treat a source as being ``detected'' for photometry purposes
in a given band only if it satisfies $P<0.002$ in that band.
We calculate upper limits for sources not detected in a given band, 
and set the associated errors to $-1.00$.

\item Column 17 gives a flag indicating whether a source shows
a radial profile consistent with that of the local PSF.
Of the 1003 main-catalog sources, 
we find that all but 60 have radial profiles consistent with that of
their corresponding PSFs above a 95\% confidence level,
and set the value of this column to 1;
the 60 sources have the value of this column set to 0.
These 60 sources are located across the entire \ecdfs\ field and
show no pattern of spatial clustering.
Moreover, we visually inspect these 60 sources and do not find any significant signature of extension.

\item Columns 18 and 19 give the right ascension and declination
of the primary ONIR counterpart (shifted accordingly to be consistent with the TENIS WIRCam $K_s$-band astrometric frame; see Section~\ref{sec:ecdfs-id}).
Sources without ONIR counterparts have these two columns
set to \hbox{``00 00 00.00''} and \hbox{``$-$00 00 00.0''}.

\item Column 20 gives the offset between the \mbox{X-ray} source and
the primary ONIR counterpart in units of arcseconds. 
Sources without ONIR counterparts have this column set to $-1.00$.

\item Column 21 gives the AB magnitude$^{\ref{ft:radio_mag}}$ of the primary ONIR counterpart
in the counterpart-detection band.
Sources without ONIR counterparts have this column set to $-1.00$.

\item Column 22 gives the name of the ONIR catalog (i.e., VLA,
\goodss, GEMS, CANDELS, MUSIC, TENIS, WFI, MUSYC, or SIMPLE; 
see Section~\ref{sec:ecdfs-id}) where the primary 
counterpart is found.
Sources without ONIR counterparts have a value set to ``...''.

\item Columns 23--49 give the counterpart right ascension, declination, and AB magnitude$^{\ref{ft:radio_mag}}$ from
the above nine ONIR catalogs
that are used for identifications (the coordinates have been shifted accordingly to be consistent with the TENIS WIRCam $K_s$-band astrometric frame; see Section~\ref{sec:ecdfs-id}).
We match the position of the primary ONIR counterpart  
with the nine ONIR catalogs
using a matching radius of $0\farcs5$. 
We set values of right ascension
and declination to \hbox{``00 00 00.00''} and \hbox{``$-$00 00 00.0''}
and set AB magnitudes to $-1.00$ for sources without matches.
We find 16.9\%, 13.3\%, 57.3\%, 9.0\%, 9.3\%, 79.5\%, 72.3\%, 59.4\%, 
and 75.1\% of the main-catalog sources have
VLA, \goodss, GEMS, CANDELS, MUSIC, TENIS, WFI, MUSYC, and SIMPLE
counterparts, respectively.

\item Columns 50--52 give the $z_{\rm spec}$,
\zs\ quality flag, and \zs\ reference.  
\zs's are collected from
(1) Szokoly \etal (2004),
(2) Zheng \etal (2004),            
(3) Ravikumar \etal (2007),        
(4) Treister \etal (2009),         
(5) Balestra \etal (2010),         
(6) Silverman \etal (2010),        
(7) Bonzini \etal (2012),          
(8) Cooper \etal (2012),           
(9) Coppin \etal (2012),           
(10) Georgantopoulos \etal (2013), 
(11) Le F{\`e}vre \etal (2013),        
(12) Taylor \etal (2009),          
(13) Kriek \etal (2008),           
(14) Hsu \etal (2014),                        
(15) Skelton \etal (2014),                   
(16) Luo \etal (2010), and                       
(17) Cardamone \etal (2010).                 
We match the positions of primary ONIR counterparts
with the above \zs\ catalogs utilizing a $0\farcs5$ matching radius.
For the 958 main-catalog sources with ONIR counterparts, we find that
476 (49.7\%) have \zs\ measurements 
(384/476=80.7\% have $R\le 24$~mag and 92/476=19.3\% have $R>24$~mag).
394 (82.8\%) of these 476 \zs's are secure,
being flagged as ``Secure'' in Column~51;
82 (17.2\%) of these 476 \zs's are insecure,
being flagged as ``Insecure'' in Column~51.
Sources without \zs\ have these three columns set to
$-1.000$, ``None'', and $-1$, respectively.
The \zs\ histogram is shown in Fig.~\ref{fig:ecdfs-zphot-comp}(a).

\begin{figure*}
\centerline{\includegraphics[width=17.5cm]{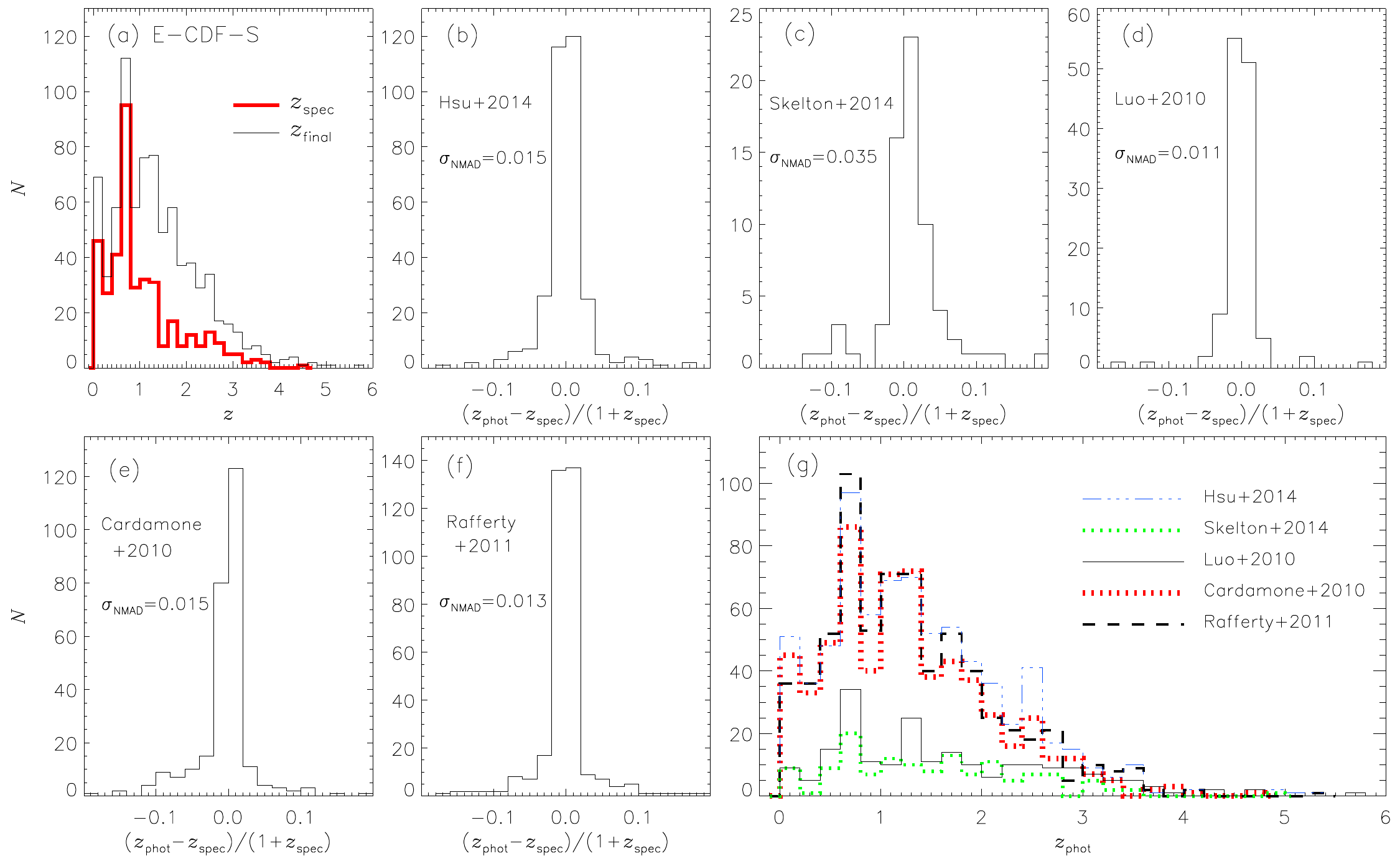}}
\figcaption{Redshift information for the 250~ks \ecdfs\ main-catalog sources.
(a) Histograms of \zs\ (476 sources; 476/1003=47.5\%) and $z_{\rm final}$ (810 sources; 810/1003=80.8\%). 
(b) Histogram of (\zp-\zs)/(1+\zs) from Hsu \etal (2014; 424 sources) with \nmad\ annotated.
(c) Histogram of (\zp-\zs)/(1+\zs) from Skelton \etal (2014; 99 sources) with \nmad\ annotated.
(d) Histogram of (\zp-\zs)/(1+\zs) from Luo \etal (2010; 161 sources) with \nmad\ annotated.
(e) Histogram of (\zp-\zs)/(1+\zs) from Cardamone \etal (2010; 387 sources) with \nmad\ annotated.
(f) Histogram of (\zp-\zs)/(1+\zs) from Rafferty \etal (2011; 431 sources) with \nmad\ annotated.
(g) Histograms of $z_{\rm phot}$ from Hsu \etal (2014; 748 sources), Skelton \etal (2014; 136 sources), Luo \etal (2010; 221 sources), Cardamone \etal (2010; 624 sources), and Rafferty \etal (2011; 692 sources).
\label{fig:ecdfs-zphot-comp}}
\end{figure*}

\item Columns 53--75 give the \zp\  
information compiled from the literature.
Columns 53--56 give the \zp, the associated $1\sigma$
lower and upper bounds, and the alternative \zp\ 
(set to $-1.000$ if not available) 
from the CANDELS/\goodss, \cdfs, and \ecdfs\ \zp\ 
catalog of Hsu \etal (2014).
Columns 57--61 give the \zp, the associated $1\sigma$
lower and upper bounds, $Q_z$, and 
the likely photometric classification
(``Galaxy'' or ``Star'') from the CANDELS/3D-\hst\ \zp\ catalog
of Skelton \etal (2014). 
Columns 62--65 give the \zp, the associated $1\sigma$
lower and upper bounds, and the alternative \zp\ (set to
$-1.000$ if not available) from Luo \etal (2010).
Columns 66--71 give the \zp, the associated $1\sigma$
lower and upper bounds, $Q_z$, the stellarity index
(ranging from 0 to 1; a value of $-1.00$ indicating information 
not available), and the flag of whether the source prefers 
the photometric fitting result using the stellar templates
(the values of 1, 0, and $-1$ indicating preferring the 
stellar templates, not preferring the stellar templates, and information
not available, respectively)
from Cardamone \etal (2010).
Columns 72--75 give the \zp, the associated $1\sigma$
lower and upper bounds, and
the likely photometric classification
(``Hybrid'', ``Galaxy'', or ``Star'', with the former
indicating preferring a mixture of AGN and galaxy templates)
from Rafferty \etal (2011).
We match the positions of primary ONIR counterparts 
with the above \zp\ catalogs utilizing a $0\farcs5$ matching radius.
Of the 958 main-catalog sources with ONIR counterparts,
748 (78.1\%), 136 (14.2\%), 221 (23.1\%), 624 (65.1\%), and 692 (72.2\%)
have \zp\ estimates from
Hsu \etal (2014), Skelton \etal (2014), Luo \etal (2010),
Cardamone \etal (2010), and Rafferty \etal (2011), respectively.
Sources without \zp's have all these columns set to $-1.000$ or ``...'' correspondingly.
Figures~\ref{fig:ecdfs-zphot-comp}(b--g) present
the histograms of $(z_{\rm phot}-z_{\rm spec})/(1+z_{\rm spec})$ and
\zp\ for these five \zp\ catalogs;
we caution that the quoted \zp\ qualities, as indicated by values of \nmad\
annotated in \mbox{Figures~\ref{fig:ecdfs-zphot-comp}(b--f)},
do not necessarily represent realistic estimates because
those \zp\ qualities are not derived using blind tests 
and in some cases ``training biases'' are involved in \zp\ derivation.

\item Column 76 gives the \zf\ adopted in this work.
We choose \zf\ for a source in the following order of preference:
(1) secure \zs's;
(2) insecure \zs's that are in agreement with
any \zp\ estimate available [i.e., $|(z_{\rm spec}-z_{\rm phot})/(1+z_{\rm spec})|\le 0.15$];
(3) \zp's from Hsu \etal (2014);
(4) \zp's from Skelton \etal (2014);
(5) \zp's from Luo \etal (2010);
(6) \zp's from Cardamone \etal (2010);
(7) \zp's from Rafferty \etal (2011); and
(8) insecure \zs's that are the only redshift information available
(thus being unable to compare with any \zp).
Of the 958 main-catalog sources with ONIR counterparts,
810 (84.6\%) have \zs's or \zp's.

\item Column 77 gives the corresponding source ID number
in the L05 \ecdfs\ catalogs.
We match our \mbox{X-ray} source positions 
to the L05 source positions (shifted accordingly to be 
consistent with the TENIS WIRCam $K_s$-band astrometric frame) 
using a $2\farcs5$ matching radius for sources having
$\theta <6\arcmin$ and a $4\farcs0$ matching radius
for sources having $\theta \ge 6\arcmin$.
Among the 1003 main-catalog sources, we find that
(1) 728 have matches 
in the 762-source L05 main catalog
(the value of this column is that from Column~1 of Table~2 in L05), i.e.,
there are 275 (i.e., $1003-728=275$) new main-catalog sources 
(see Section~\ref{sec:ecdfs-new} for more details of these 275 new sources),
compared to the L05 main catalog;
(2) 21 have matches in the 33-source L05 supplementary
catalog (the value of this column is that from Column~1 of Table~6 in 
L05 added with a prefix of ``SP\_''); and 
(3) 254 have no match 
in either the L05 main or supplementary
catalog, which are detected now thanks to our two-stage source-detection
approach (the value of this column is set to $-1$).
We refer readers to Section~\ref{sec:ecdfs-comp-old}
for the information of the 34 L05 main-catalog sources that
are not included in our main catalog.

\item Columns 78 and 79 give the right ascension and declination of
the corresponding L05 source (shifted accordingly to be         
consistent with the TENIS WIRCam $K_s$-band astrometric frame).
Sources without an L05 match have these two columns
set to \hbox{``00 00 00.00''} and \hbox{``$-$00 00 00.0''}.

\item Column 80 gives the corresponding source ID number
in the X11 4~Ms \cdfs\ catalogs.
For the 1003 main-catalog sources, we find that
(1) 273 have matches in the 740-source X11 main catalog 
(the value of this column is that from Column~1 of Table~3 in X11),
(2) 12 have matches in the 36-source X11 supplementary
catalog (the value of this column is that from Column~1 of Table~6 in  
X11 added with a prefix of ``SP\_''); and   
(3) 718 have no match   
in either the X11 main or supplementary
catalog, mainly due to their spatial locations being not covered by
the 4~Ms \cdfs\ (the value of this column is set to $-1$).

\item Columns 81 and 82 give the right ascension and declination of
the corresponding X11 source (shifted accordingly to be         
consistent with the TENIS WIRCam $K_s$-band astrometric frame).
Sources without an X11 match have these two columns
set to \hbox{``00 00 00.00''} and \hbox{``$-$00 00 00.0''}.

\item Columns 83--85 give the effective exposure times in units of seconds
derived from the exposure maps 
for the three standard bands.

\item Columns 86--88 give the band ratio and the associated 
upper and lower errors, respectively.
Upper limits are computed for sources detected in the soft
band but not the hard band, while lower limits are computed for sources detected
in the hard band but not the soft band;
for these sources, the upper and lower
errors are set to the calculated band ratio.
Band ratios and associated errors are set to $-1.00$ for sources 
with full-band detections only.

\item Columns 89--91 give the effective photon index $\Gamma$ and
the associated upper and lower errors, respectively, 
assuming a power-law model with the Galactic
column density that is given in Section~\ref{sec:intro}.
Upper limits are computed for sources detected in the hard band 
but not the soft band, while lower limits are computed for sources detected in the soft band but not the hard band;
for these sources, the upper and lower errors are set to 
the calculated $\Gamma$.
A value of $\Gamma=1.4$ is assumed for low-count sources
(as defined in Section~\ref{sec:cdfn-maincat}), 
and the associated upper and lower errors are set to 0.00.

\item Columns 92--94 give observed-frame fluxes in units of \flux\ 
for the three standard bands.
Negative flux values denote upper limits.

\item Column 95 gives a basic estimate of the absorption-corrected, 
rest-frame \hbox{0.5--7~keV} luminosity
($L_{\rm 0.5-7\ keV}$ or $L_{\rm X}$) in units of \hbox{erg s$^{-1}$}.
Note that $L_{\rm 0.5-8\ keV}$=1.066$\times L_{\rm 0.5-7\ keV}$
and $L_{\rm 2-10\ keV}$=0.721$\times L_{\rm 0.5-7\ keV}$, given the assumed $\Gamma_{\rm int}=1.8$ (see the description of Column~70 of the 2~Ms \cdfn\ main catalog in Section~\ref{sec:cdfn-maincat} for details).
Sources without \zf\ have this column set to $-1.000$.

\item Column 96 gives a basic estimate of likely source type: 
``AGN'', ``Galaxy'', or ``Star''. 
We use the same classification scheme detailed in
Section~4.4 of X11 (see the description of Column~78 of
the X11 main catalog), which makes use of additional
spectroscopic and photometric data available in the \cdfs/\ecdfs.
There are 909 (90.6\%), 67 (6.7\%), and 27 (2.7\%)
of the 1003 main-catalog sources identified as AGNs, galaxies, and stars, respectively.

\item Column 97 gives brief notes on the sources.
Sources in close doubles or triples are annotated with ``C'' 
(a total of 29 such sources) and 
sources lying at the field edge 
are annotated with ``E'' (only one such source); otherwise,
sources are annotated with ``...''.

\end{enumerate}

\subsubsection{Comparison with the L05 Main-Catalog Sources}\label{sec:ecdfs-comp-old}

Table~\ref{tab:ecdfs-det} summarizes the source detections 
in the three standard bands for the main catalog.
Of the 1003 main-catalog sources,
929, 769, and 655 are detected in the full, soft, and hard bands, 
respectively;
as a comparison,
of the 762 L05 main-catalog sources,
689, 598, and 453 are detected in the full, soft, and hard bands, 
respectively (note that L05 adopt an upper energy bound of 8~keV).
As stated in Section~\ref{sec:ecdfs-maincat} 
(see the description of Column~77),
728 of the main-catalog sources have matches in the L05 main catalog.
For these 728 common sources, we find that the \hbox{X-ray} photometry
derived in this work is in general agreement with that in L05,
e.g., the median ratio between our and the L05
soft-band count rates for the soft-band detected common sources
is 1.04, with an interquartile range of 0.96--1.12.
The significant increase in the number of main-catalog sources,
i.e., an increase of $1003-728=275$ new main-catalog sources, 
is mainly due to the improvements of our cataloging methodology
that are summarized in Table~\ref{tab:impro}, in particular,
due to our two-stage source-detection approach.
Indeed, we are able to detect fainter sources than L05
that are yet reliable,  
with median detected counts (see Table~\ref{tab:ecdfs-det}) 
in the three standard bands being $\approx 70\%$ of those of L05
(see their Table~4).

\begin{table}
%\tabletypesize{\small}
%\tablewidth{0pt}
\caption{250~ks \ecdfs\ Main Catalog: Summary of Source Detections}
\centering
\begin{tabular}{lccccc}\hline\hline
Band & Number of & Maximum & Minimum & Median & Mean \\
(keV) & Sources & Counts & Counts & Counts & Counts \\\hline
Full (0.5--7.0)   & 929 & 4010.6 & 3.3 & 27.1 & 87.3 \\
Soft (0.5--2.0)   & 769 & 2802.6 & 2.2 & 18.9 & 64.5 \\
Hard (2--7)       & 655 & 1210.8 & 3.4 & 20.4 & 46.0 \\ \hline
\end{tabular}
\label{tab:ecdfs-det}
\end{table}

Thirty-four (i.e., $762-728=34$) of the L05 main-catalog sources 
are not recovered in our main catalog,
among which 6 are recovered in our supplementary catalog 
(see Section~\ref{sec:ecdfs-supp}).
Among the 28 L05 main-catalog sources that are not recovered in our main or
supplementary catalogs,
(1) 2 sources were the fainter sources in pairs in L05
but now fail our source-detection criterion of $P<0.002$;
(2) 1 source was in a triplet in L05
but is now removed based on visual inspection of the \xray\ images
compared to the local PSF size, 
thus degrading the previous triplet into a doublet;
(3) 4 sources barely fail our source-detection criterion,
with $0.002<P<0.004$;
(4) 12 sources not only have faint \xray\ signatures,
but also have multiwavelength counterparts, thus being 
likely real \hbox{X-ray} sources, although they do not satisfy 
our $P<0.002$ source-selection criterion; and
(5) the remaining 9 sources have marginal \xray\ signatures
and have no multiwavelength counterparts, 
thus being likely false detections.

Table~\ref{tab:ecdfs-undet} summarizes 
the number of sources detected in one band but not another in the main catalog
(cf. Table~5 of L05).
There are 57, 50, and 24 sources detected only in the full, soft, and hard bands, 
in contrast to 68, 58, and 15 sources in the L05 main catalog, respectively.

\begin{table}
%\tabletypesize{\small}
%\tablewidth{0pt}
\caption{250~ks \ecdfs\ Main Catalog: Sources Detected in One Band but not Another}
\centering
\begin{tabular}{lccc}\hline\hline
Detection Band & Nondetection & Nondetection & Nondetection \\
(keV) & Full Band & Soft Band & Hard Band \\\hline
Full (0.5--7.0)  & \ldots & 210 & 298 \\
Soft (0.5--2.0)  &  ~~~~~50~~~~ & \ldots & 291 \\
Hard (2--7)      &  ~~~~~24~~~~ & 177 & \ldots \\ \hline
\end{tabular}
\label{tab:ecdfs-undet}
\end{table}

\subsubsection{Properties of Main-Catalog Sources}\label{sec:ecdfs-prop}

\begin{figure}
\centerline{\includegraphics[width=8.5cm]{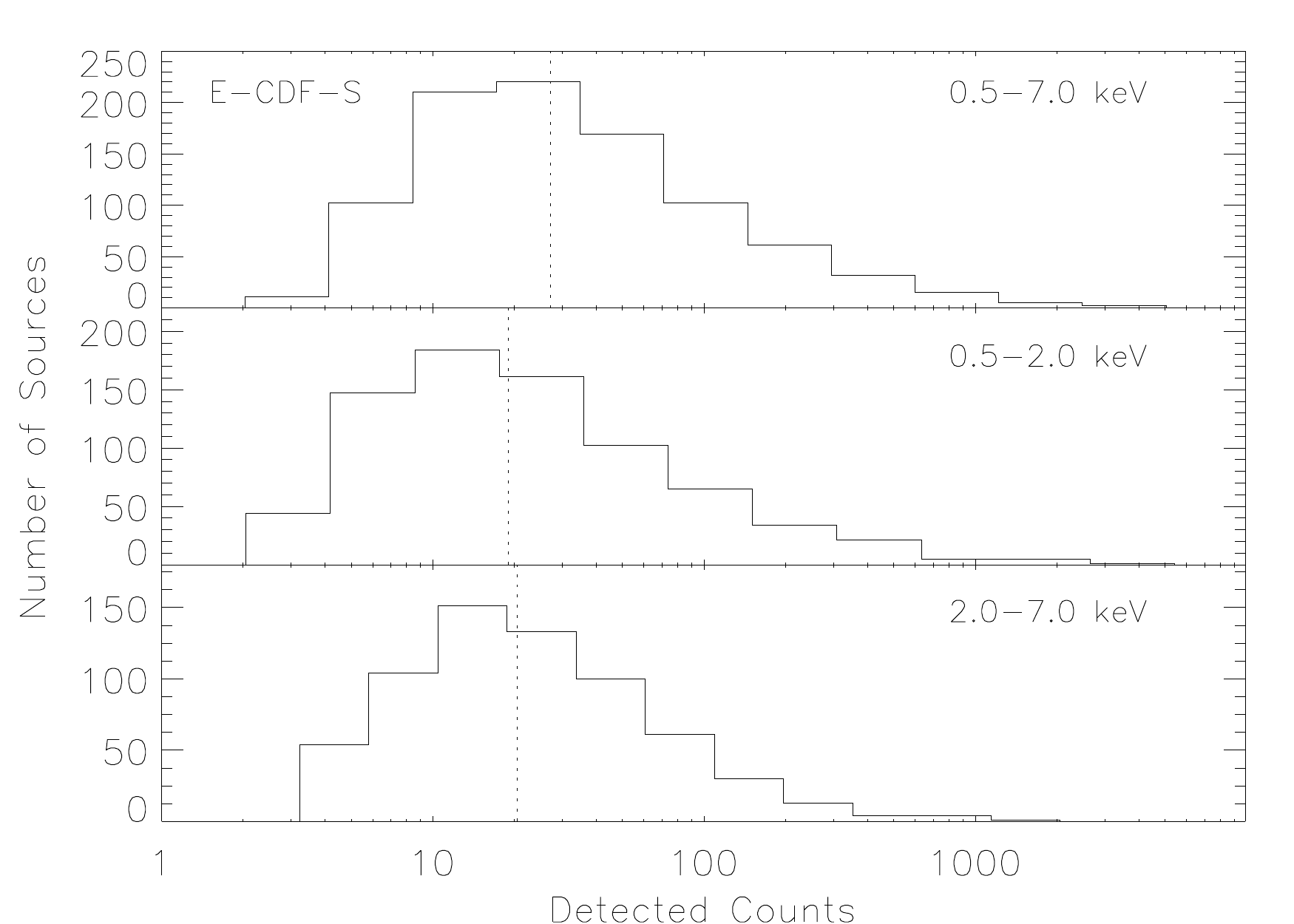}}
\figcaption{Distributions of detected source
counts for the 250~ks \ecdfs\ main-catalog sources
in the full, soft, and hard bands.
Sources with upper limits are not plotted.
The vertical dotted lines indicate the median detected counts
of 27.1, 18.9, and 20.4, for the full, soft, and hard bands, respectively
(detailed in Table~\ref{tab:ecdfs-det}).
\label{fig:ecdfs-cnthist}}
\end{figure}

Figure~\ref{fig:ecdfs-cnthist} presents the histograms of detected
source counts in the three standard bands for the sources in the main catalog.
The median detected counts are 27.1, 18.9, and 20.4
for the full, soft, and hard bands, respectively;
and there are 165, 77, 29, and 9 sources having $>100$, $>200$, $>500$,
and $>1000$ full-band counts, respectively.

\begin{figure}
\centerline{\includegraphics[width=8.5cm]{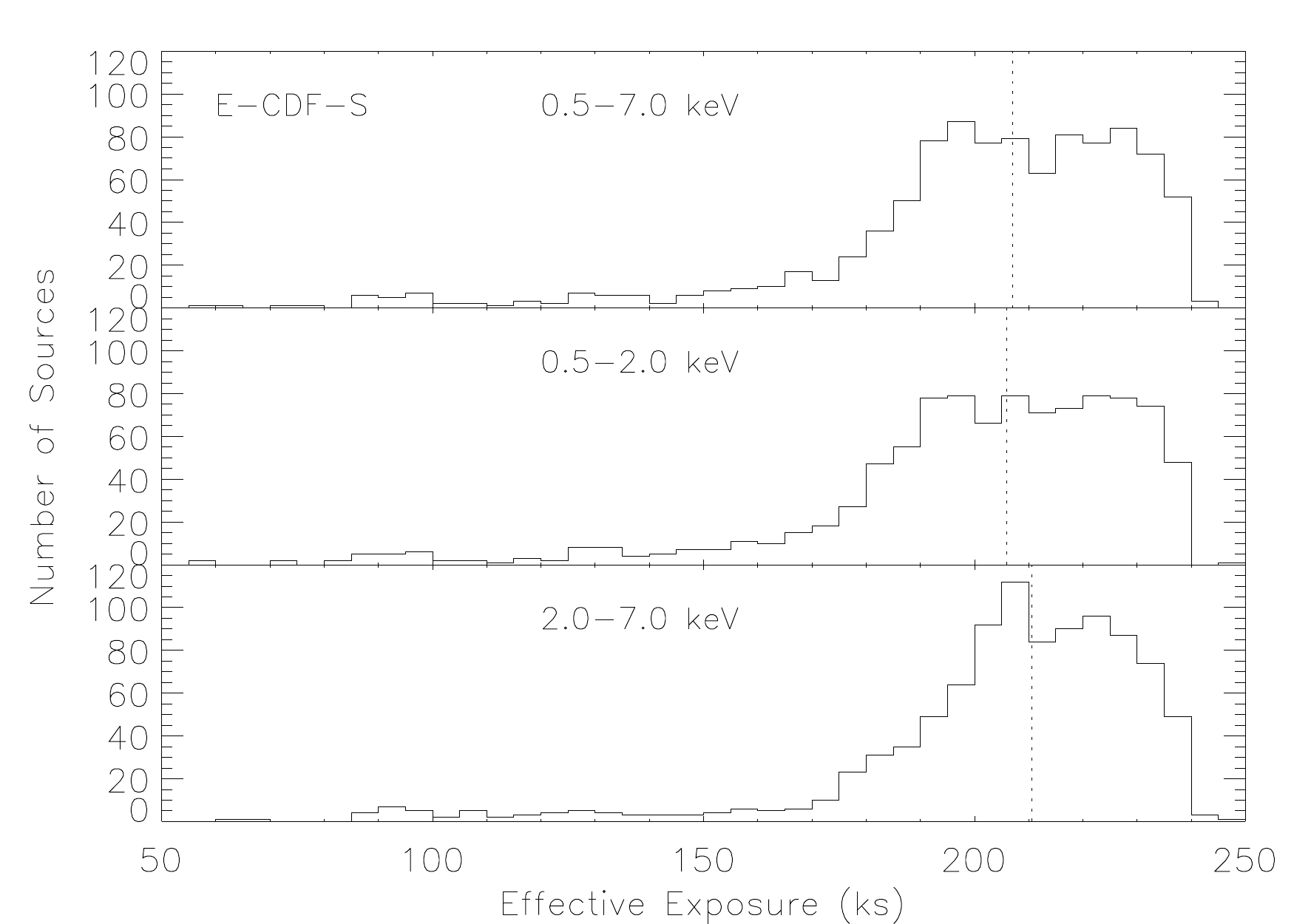}}
\figcaption{Distributions of effective exposure times
for all the 1003 \ecdfs\ main-catalog sources
in the full, soft, and hard bands.
The vertical dotted lines indicate the median effective exposures
of 207.1, 206.0, and 210.6~ks, for the full, soft, and hard bands, respectively.
\label{fig:ecdfs-exphist}}
\end{figure}

Figure~\ref{fig:ecdfs-exphist} presents the histograms of 
effective exposure times in the three standard bands for all the 1003 main-catalog sources.
The median effective exposures are
207.1, 206.0, and 210.6~ks for the full, soft, and hard bands, respectively.

\begin{figure}
\centerline{\includegraphics[width=8.5cm]{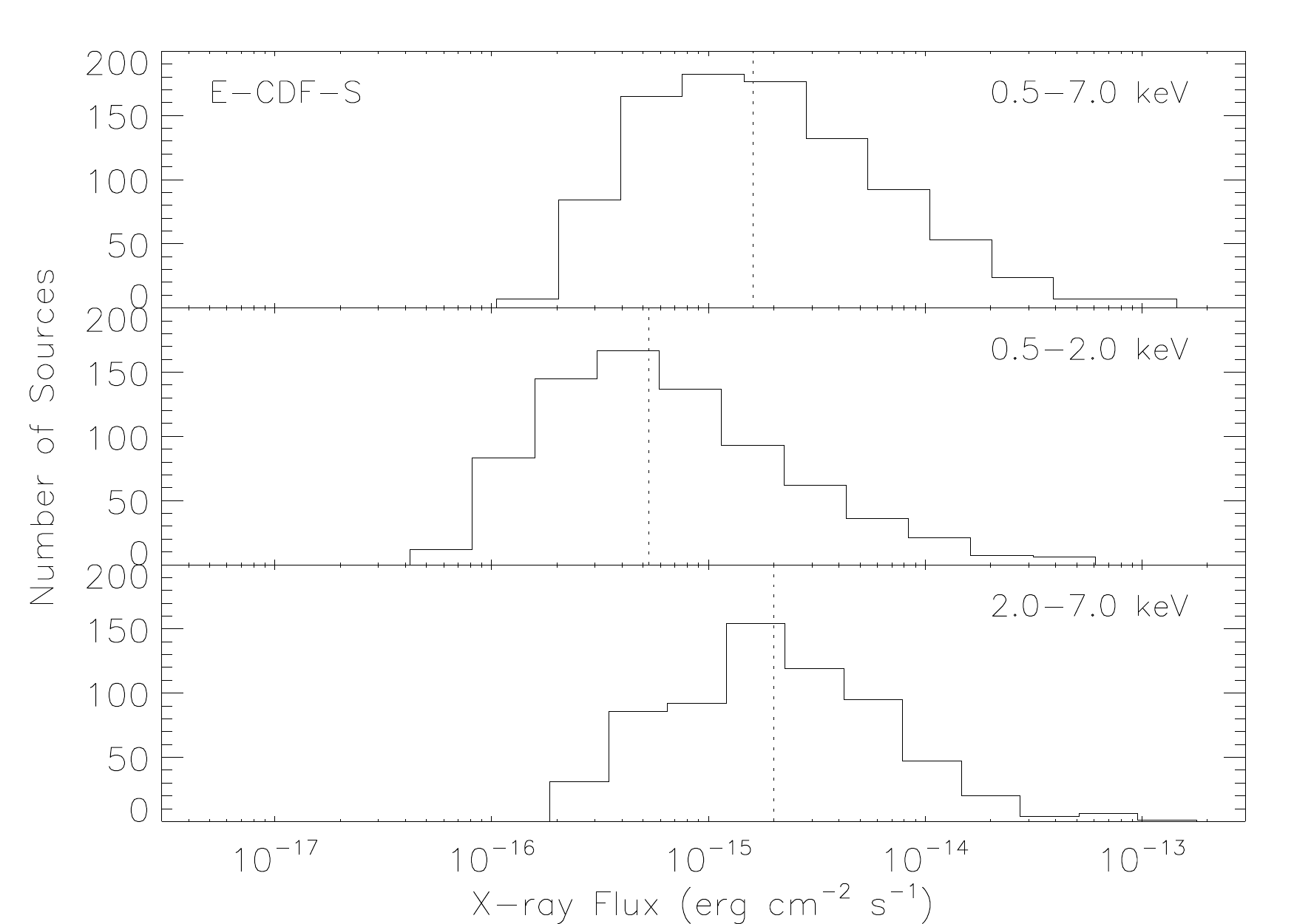}}
\figcaption{Distributions of \hbox{X-ray} fluxes for the 250~ks \ecdfs\ main-catalog
sources in the full, soft, and hard bands.
Sources with upper limits are not plotted.
The vertical dotted lines denote the median fluxes of
$1.6\times10^{-15}$, $5.3\times10^{-16}$ and $2.0\times10^{-15}$ \flux\
for the full, soft, and hard bands, respectively.
\label{fig:ecdfs-fluxhist}}
\end{figure}

Figure~\ref{fig:ecdfs-fluxhist} presents the histograms of observed-frame \hbox{X-ray} fluxes
in the three standard bands for the sources in the main catalog.
The \hbox{X-ray} fluxes distribute over three orders of
magnitude, with a median value of
$1.6\times10^{-15}$, $5.3\times10^{-16}$ and $2.0\times10^{-15}$ \flux\
for the full, soft, and hard bands, respectively.

\begin{figure}
\centerline{\includegraphics[width=8.5cm]{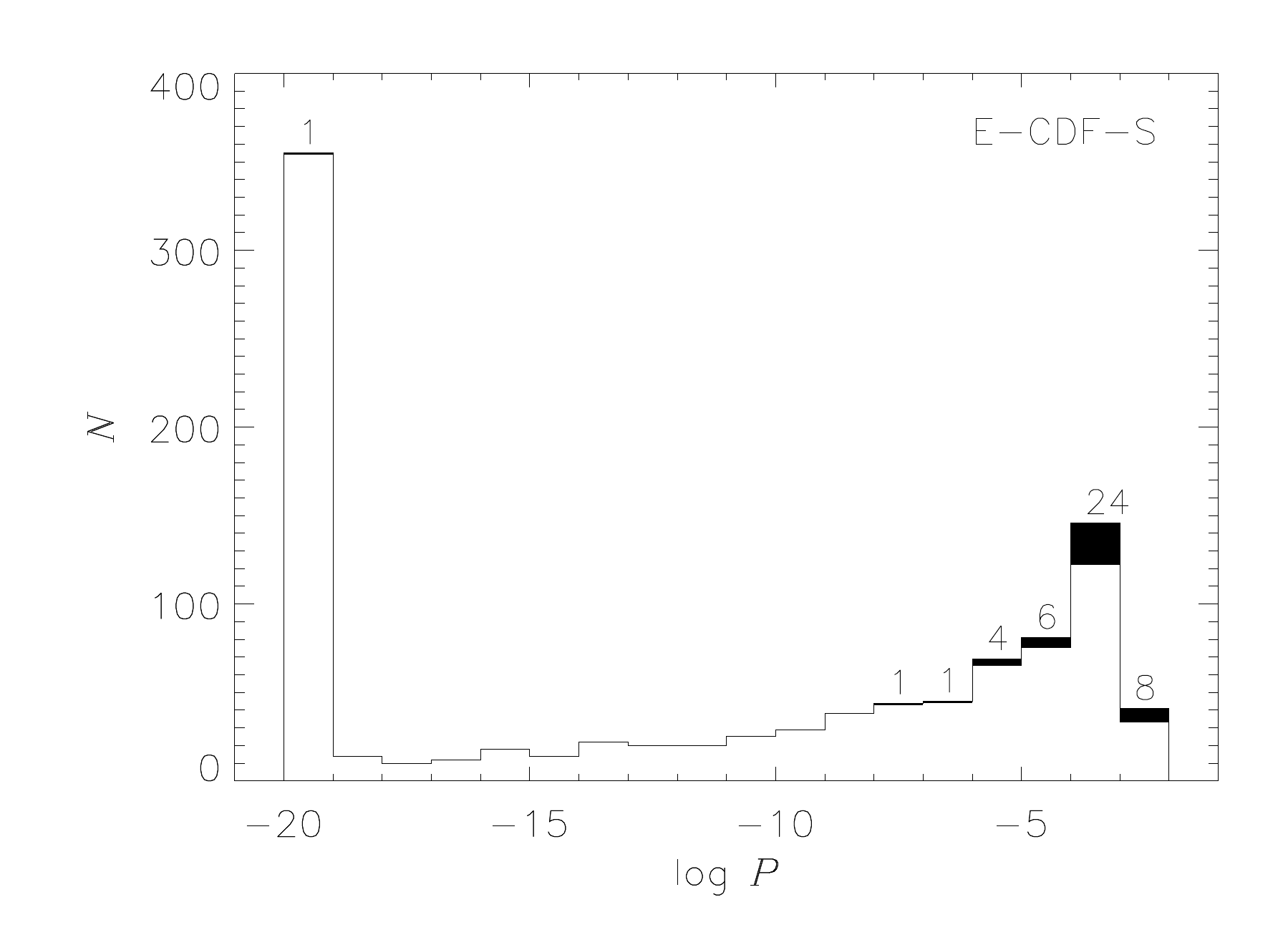}}
\figcaption{Distribution of the AE-computed binomial
no-source probability, $P$, for the 250~ks \ecdfs\
main-catalog sources.
The values of $\log P<-20$ are set to $\log P=-20$ for easy illustration.
The shaded areas denote sources without multiwavelength counterparts,
with their corresponding numbers annotated.
\label{fig:ecdfs-p-idrate}}
\end{figure}

Figure~\ref{fig:ecdfs-p-idrate} presents the histogram of the AE-computed binomial
no-source probability $P$ for the sources in the main catalog,
with a total of 45 sources having no multiwavelength counterparts highlighted by shaded areas.
The majority of the main-catalog sources have low $P$ values that indicate significant detections,
with a median $P$ of $3.68\times 10^{-11}$ and an interquartile range of 
$1.28\times 10^{-32}$ to $2.04\times 10^{-5}$.
We find that 1.0\% of the $\log P\le -5$ sources
have no ONIR counterparts, in contrast to 14.2\% of $\log P> -5$
sources lacking ONIR counterparts.
Given the small false-match rate estimated in Section~\ref{sec:ecdfs-id},
a main-catalog source with a secure ONIR counterpart is almost certain to be real 
(note that sources without ONIR counterparts are more likely
but not necessarily false detections).

\begin{figure*}
\centerline{\includegraphics[width=17.5cm]{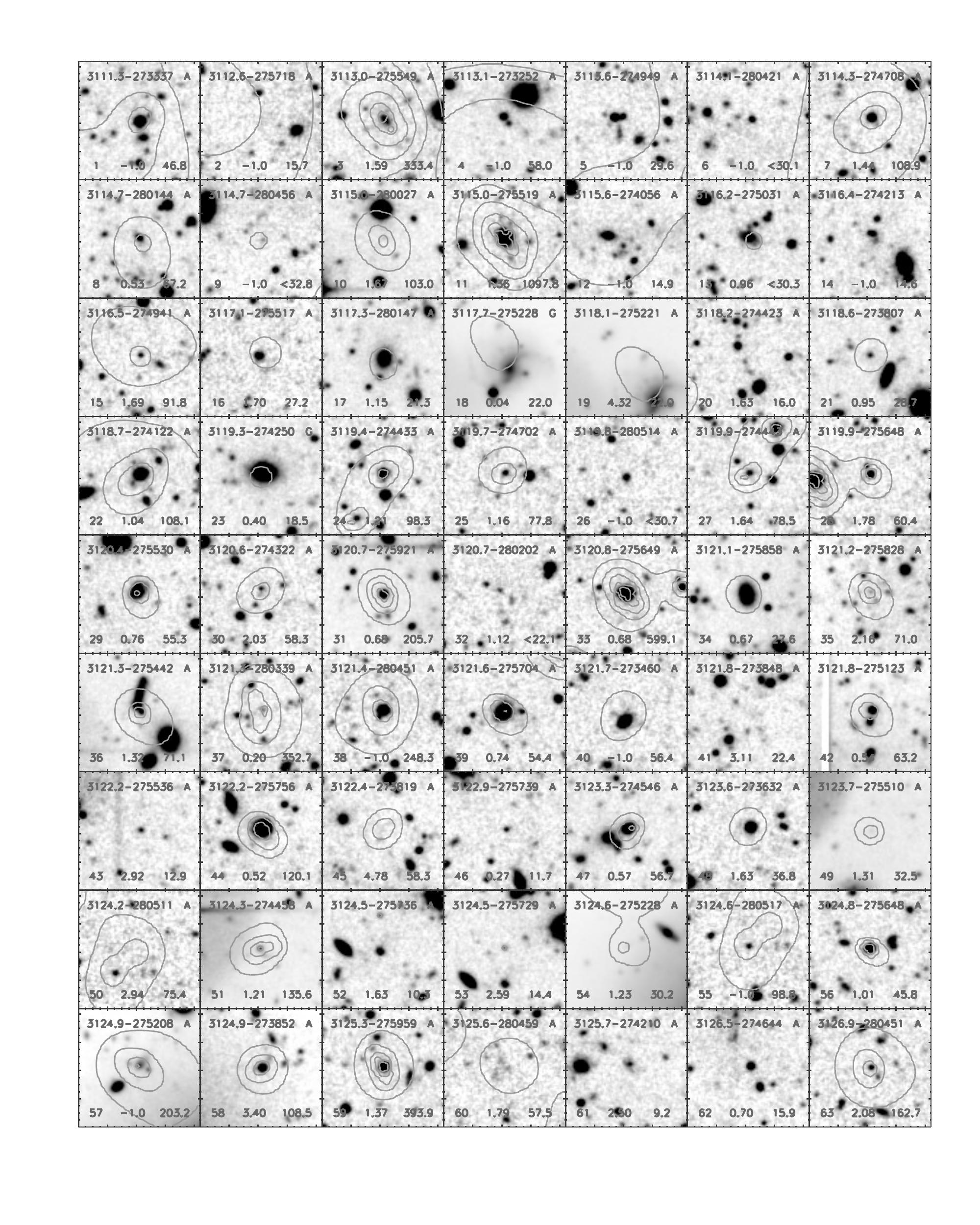}}
\figcaption{$25\arcsec\times 25\arcsec$ postage-stamp images from the WFI $R$ band (Giavalisco \etal 2004)
for the 250~ks \ecdfs\ main-catalog sources that are
centered on the \xray\ positions, overlaid
with full-band adaptively smoothed \hbox{X-ray} contours
that have a logarithmic scale and range from \hbox{$\approx$0.003\%--30\%}
of the maximum pixel value.
In each image, the labels at the top are the
source name (the hours ``03'' of right ascension are omitted for succinctness)
and source type (A=AGN, G=Galaxy, and S=Star);
the bottom numbers
indicate the source \xray\ ID number, adopted redshift,
and full-band counts or upper limit (with a ``$<$'' sign).
%There are cutouts (i.e., nearly plain white portions) in some images that
%are caused by stellar light-induced saturation.  
In some cases there are
no \hbox{X-ray} contours present,
either due to these sources being not detected in the full band or
having low full-band counts leading to their observable emission in the adaptively
smoothed image being suppressed by {\sc csmooth}.
\\(An extended version of this figure is available in the online journal.)
\label{fig:ecdfs-wfir-stamps-page1}}
\end{figure*}

\begin{figure*}
\centerline{\includegraphics[width=17.5cm]{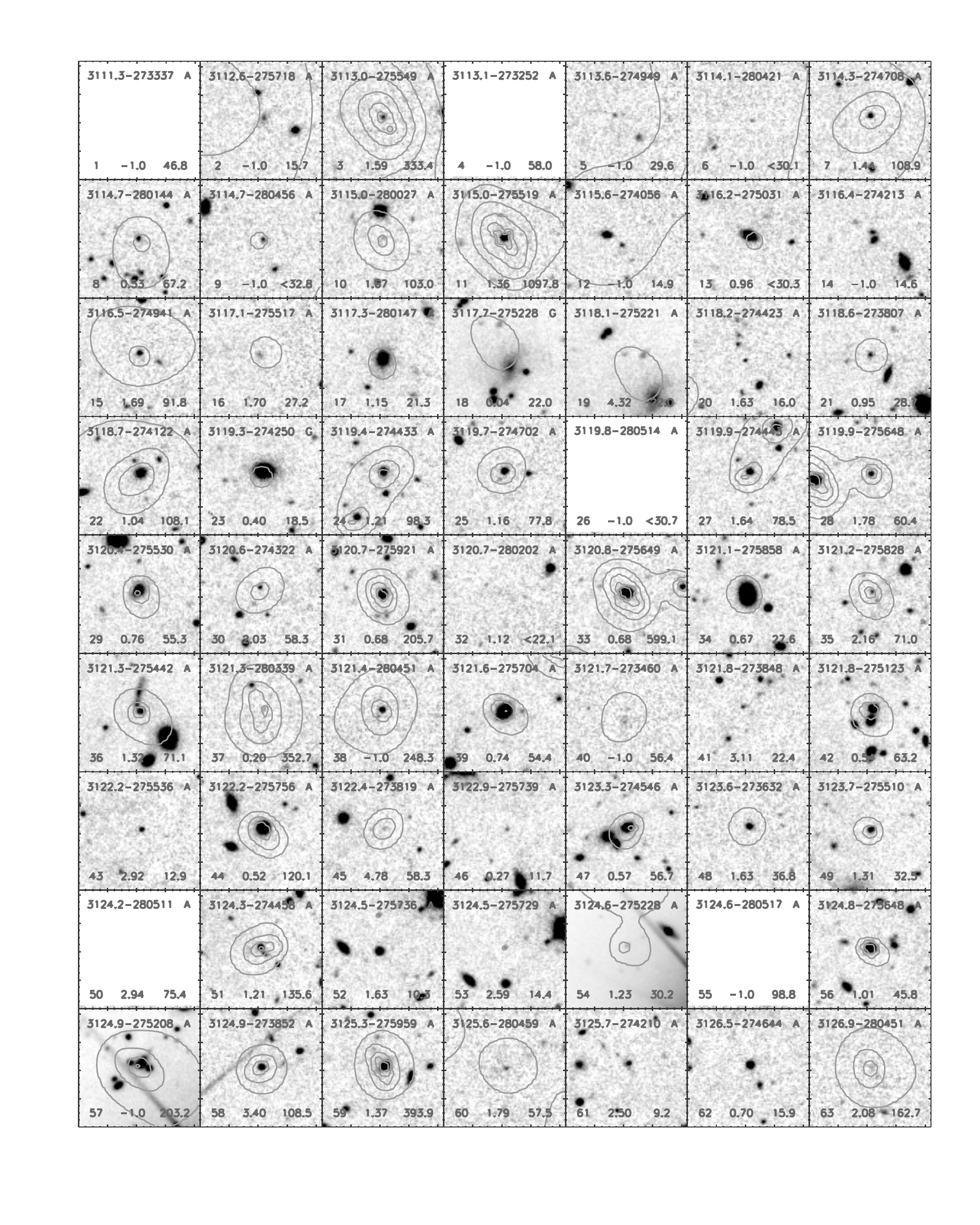}}
\figcaption{Same as Figure~\ref{fig:ecdfs-wfir-stamps-page1}, but for the TENIS WIRCam $K_s$ band (Hsieh \etal 2012).
In some cases there is no $K_s$-band coverage (e.g., XIDs=1, 4, 26, 50, 55).
\\(An extended version of this figure is available in the online journal.)
\label{fig:ecdfs-tenis-stamps-page1}}
\end{figure*}

\begin{figure*}
\centerline{\includegraphics[width=17.5cm]{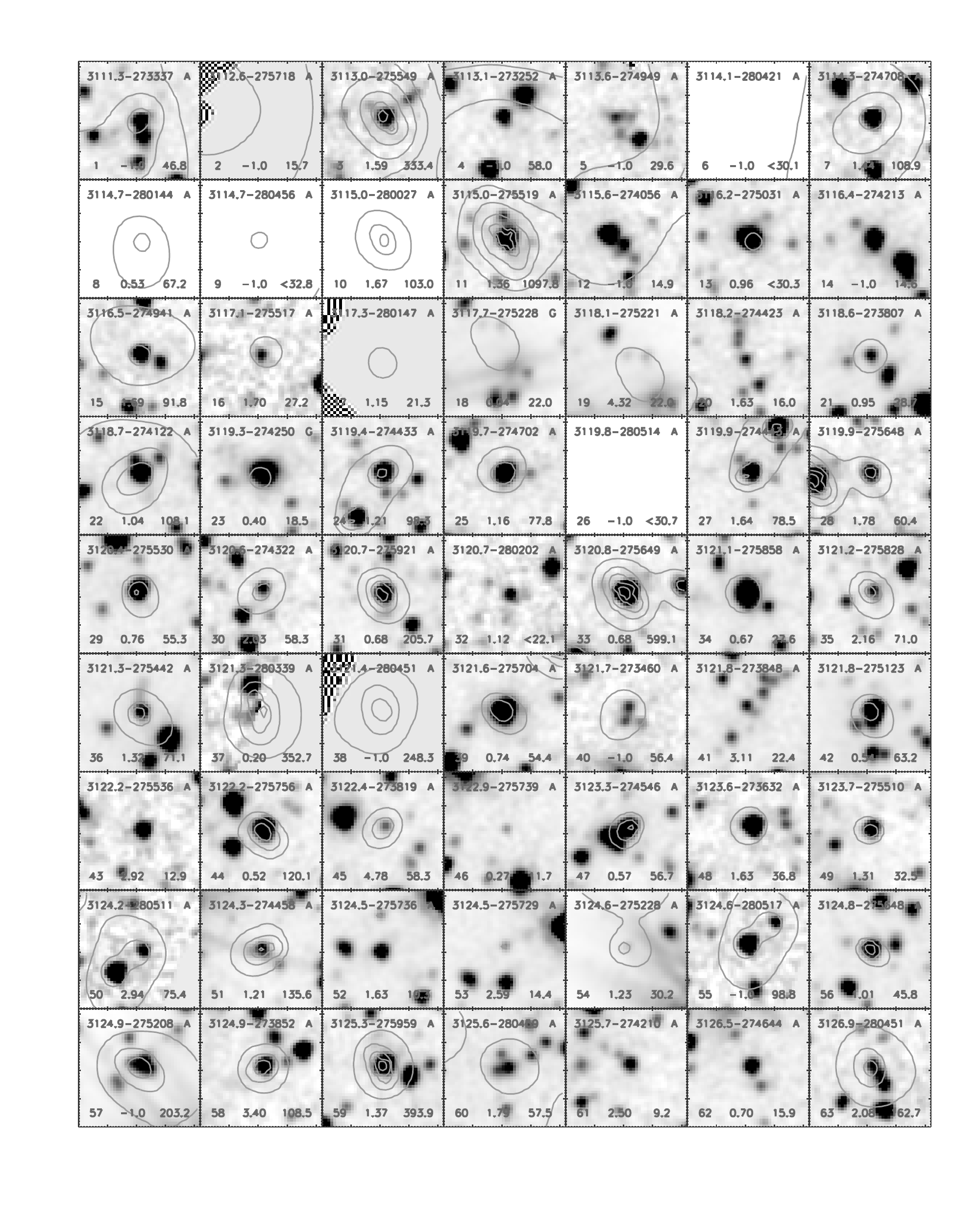}}
\figcaption{Same as Figure~\ref{fig:ecdfs-wfir-stamps-page1}, but for the SIMPLE IRAC 3.6~$\mu$m band (Damen \etal 2011).
In some cases there is partial (e.g., XIDs=2, 17, 38) or no IRAC 3.6~$\mu$m-band coverage (e.g., XIDs=6, 8, 9, 10, 26).
\\(An extended version of this figure is available in the online journal.)
\label{fig:ecdfs-irac-stamps-page1}}
\end{figure*}

Figures~\ref{fig:ecdfs-wfir-stamps-page1}--\ref{fig:ecdfs-irac-stamps-page1}
display $25\arcsec\times 25\arcsec$ postage-stamp images 
from the WFI $R$ band (Giavalisco \etal 2004),
the TENIS WIRCam $K_s$ band (Hsieh \etal 2012), and
the SIMPLE IRAC 3.6~$\mu$m band (Damen \etal 2011),
overlaid with adaptively smoothed full-band \xray\ contours for
the main-catalog sources, respectively.

\subsubsection{Properties of the 275 New Main-Catalog Sources}\label{sec:ecdfs-new}

\begin{figure*}
\centerline{\includegraphics[width=17.5cm]{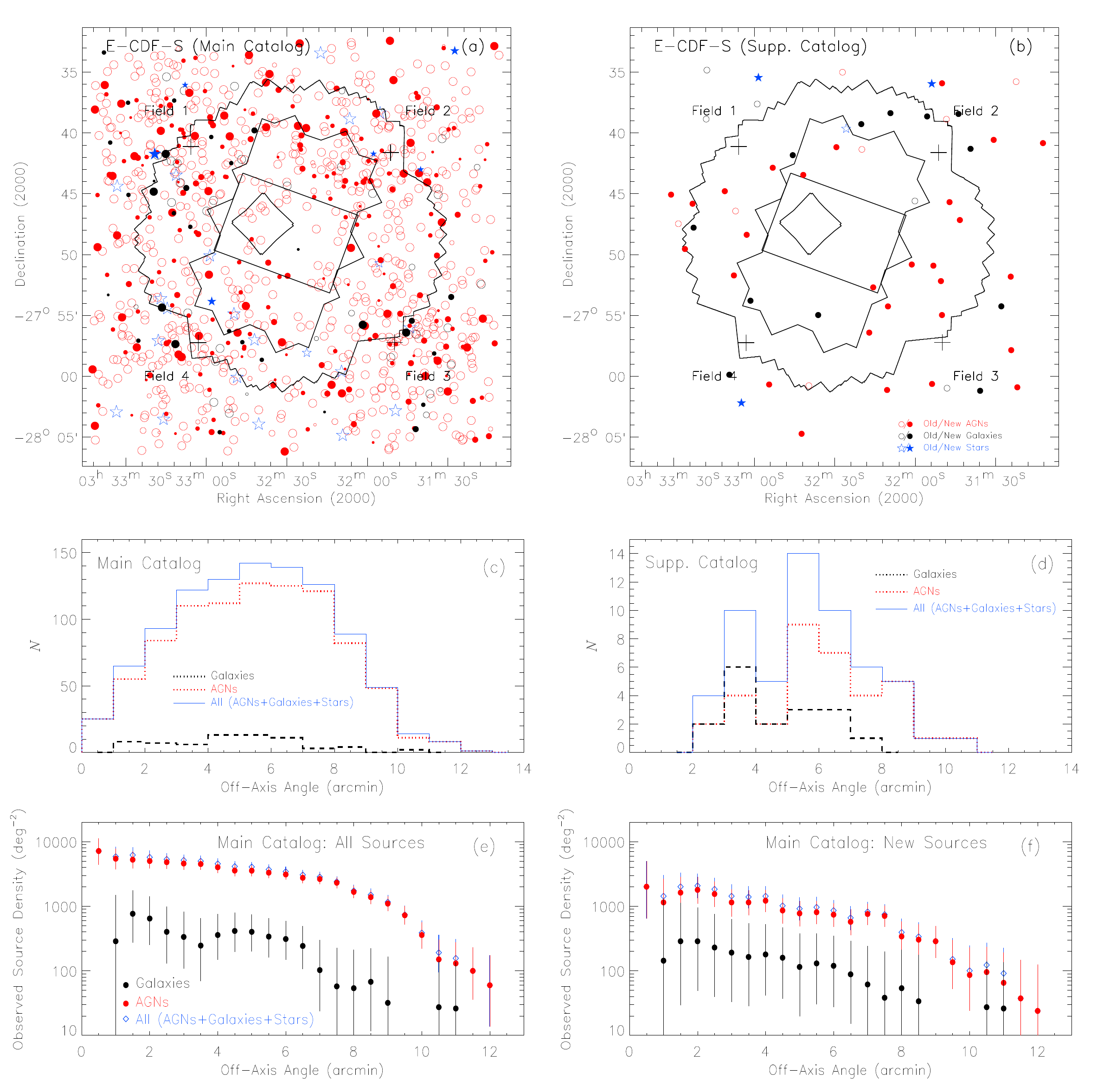}}
\figcaption{(Top) Spatial distributions for (a) the 250~ks \ecdfs\ main-catalog sources
and (b) the supplementary-catalog sources.
Sources classified as AGNs, galaxies, and stars
are plotted as red, black, and blue symbols, respectively.
Open symbols indicate old sources
that were previously detected in (a) the L05 main catalog
or (b) the L05 main or supplementary catalog, while
filled symbols indicate new sources
that were not previously detected in the L05 main and/or supplementary catalog.
The regions and the plus signs have the same meanings as those in Fig.~\ref{fig:ecdfs-fb-img}.
In panel (a),
larger symbol sizes indicate lower AE binomial no-source probabilities,
ranging from $\log P >-3$, $-4<\log P\le -3$, $-5<\log P\le -4$, to
$\log P\le -5$; while in panel (b), all sources have $\log P >-3$ and are plotted as symbols of the same size.
(Middle) Distributions of off-axis angles for different source types for (c) the main-catalog sources
and (d) the supplementary-catalog sources.
(Bottom) Observed source densities broken down into different source types as a function of off-axis angle ($\theta$)
for (e) all the 250~ks \ecdfs\ main-catalog sources and (f) the {\it new} main-catalog sources,
which are calculated in bins of $\Delta\theta=1\arcmin$
and whose $1\sigma$ errors are computed utilizing Poisson statistics.
\label{fig:ecdfs-pos}}
\end{figure*}

Figure~\ref{fig:ecdfs-pos}(a) displays the spatial distributions
of the 275 new main-catalog sources
(i.e., 238 new AGNs,
31 new galaxies, and 6 new stars that are all 
indicated as filled symbols) and 
the 728 old main-catalog sources (indicated as open symbols),
whose colors are coded based on source types
(red for AGNs, black for galaxies, and blue for stars)
and whose symbol sizes represent different $P$ values
(larger sizes denote lower $P$ values and thus 
higher source-detection significances).
Figure~\ref{fig:ecdfs-pos}(c) shows the histograms of
off-axis angles for different source types 
for the main-catalog sources.

Figures~\ref{fig:ecdfs-pos}(e) and (f) show the observed
source density as
a function of off-axis angle for all the main-catalog sources
and the new main-catalog sources, respectively.
These two plots reveal, for either all or new sources, that
(1) the source densities decline toward large off-axis angles
due to the decreasing sensitivities (see Section~\ref{sec:ecdfs-smap}); and
(2) overall, observed AGN densities are larger than observed galaxy densities.
In the central $\theta\le3\arcmin$ areas of the four \ecdfs\ observation
fields,
the averaged observed source densities for all sources, all AGNs, 
and all galaxies reach
$5800_{-900}^{+1000}$~deg$^{-2}$, 
$5200_{-800}^{+1000}$~deg$^{-2}$, and
$500_{-200}^{+400}$~deg$^{-2}$, respectively; and
the averaged observed source densities for all new sources, new AGNs,
and new galaxies reach
$1900_{-500}^{+600}$~deg$^{-2}$,    
$1600_{-500}^{+600}$~deg$^{-2}$, and
$200_{-200}^{+300}$~deg$^{-2}$, respectively.

\begin{figure*}
\centerline{\includegraphics[width=17.5cm]{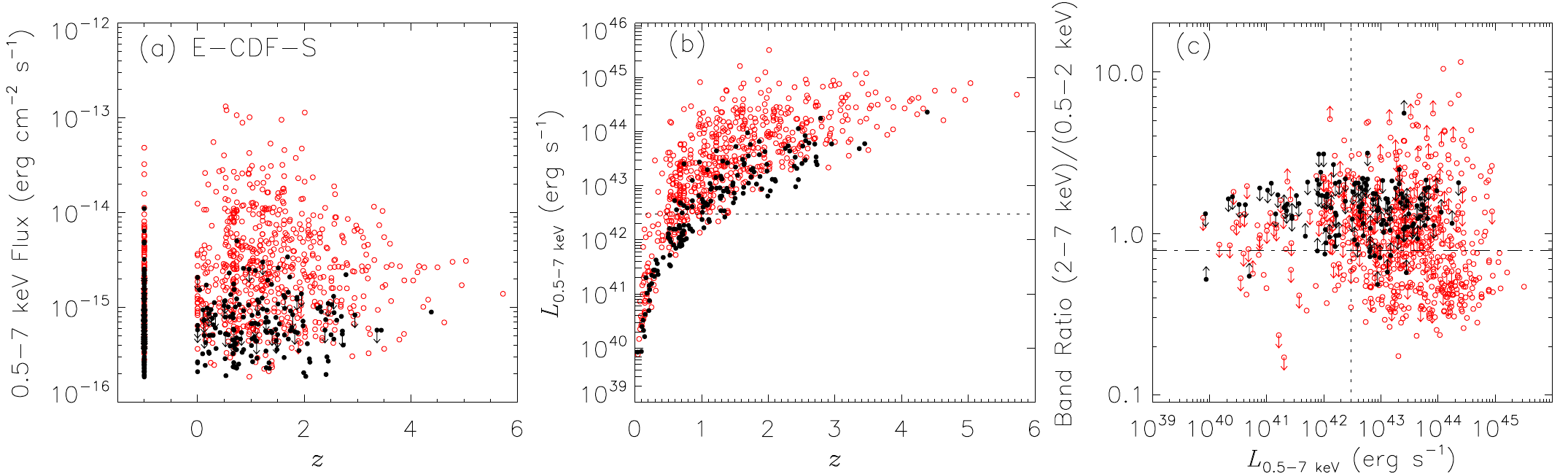}}
\figcaption{(a) Observed-frame full-band flux vs. adopted redshift,
(b) absorption-corrected, rest-frame \hbox{0.5--7 keV} luminosity vs. adopted redshift,
and (c) band ratio vs. absorption-corrected, rest-frame \hbox{0.5--7 keV} luminosity
for the 250~ks \ecdfs\ main-catalog sources.
Red open circles indicate old sources
while black filled circles indicate new sources.
Arrows denote limits.
In panel (b), sources having no redshift estimates are not plotted;
in panel (c), sources having no redshift estimates or sources having only
full-band detections are not plotted.
The dotted lines in panels (b) and (c) and the dashed-dot line in panel (c)
correspond to the threshold values of two AGN-identification criteria,
$L_{\rm 0.5-7\ keV}\ge 3\times 10^{42}$ \hbox{erg s$^{-1}$} and $\Gamma \le 1.0$.
\label{fig:ecdfs-f-lx-z-br}}
\end{figure*}

\begin{figure}
\centerline{\includegraphics[width=8.5cm]{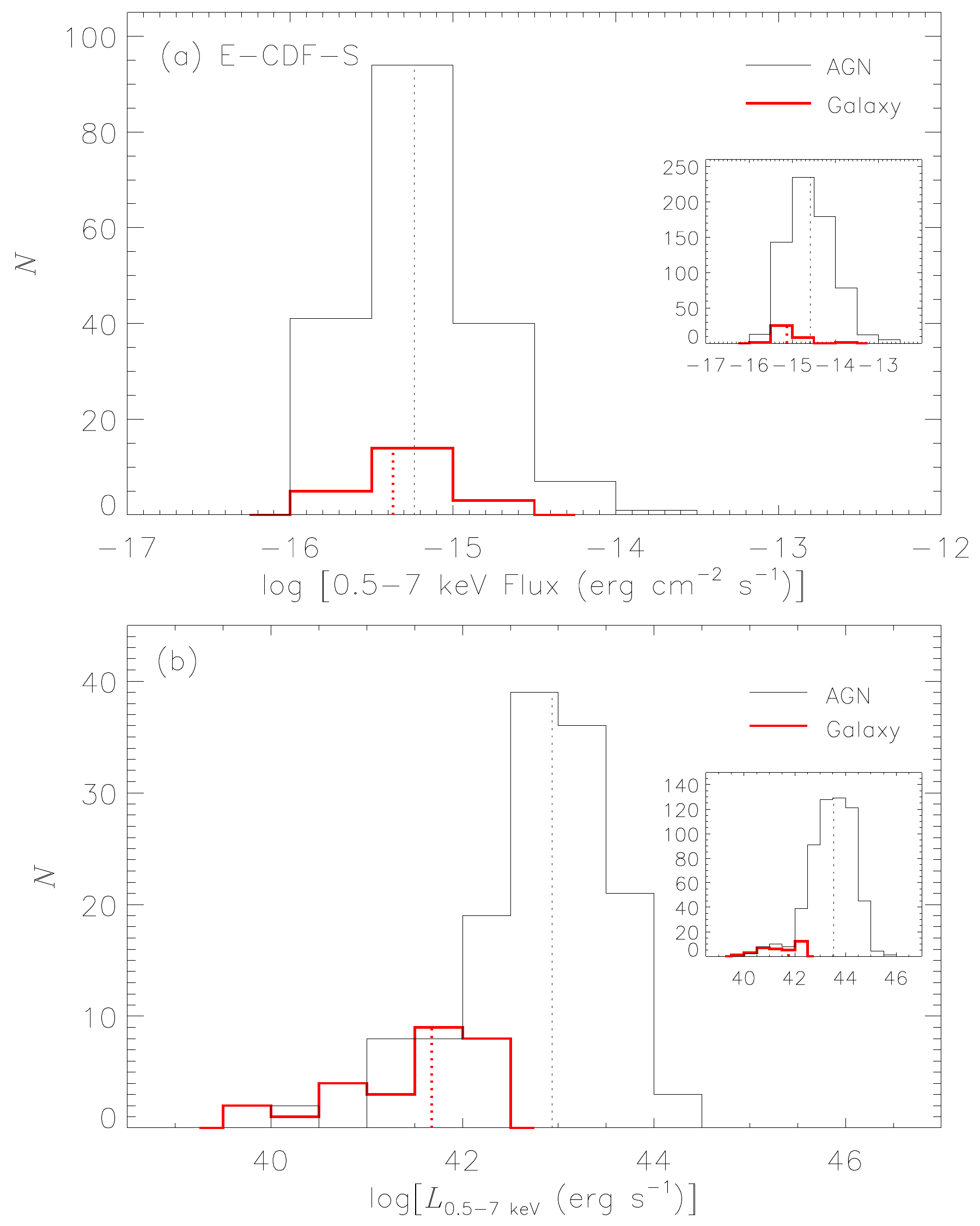}}
\figcaption{Histograms of (a) observed-frame full-band flux and (b) absorption-corrected,
rest-frame \hbox{0.5--7 keV} luminosity for the new 250~ks \ecdfs\ main-catalog sources.
The insets display results for the old main-catalog sources.
The vertical dotted lines indicate the median values.
In panel (a), sources without full-band detections are not included;
in panel (b), sources without redshift estimates are not included.
\label{fig:ecdfs-f-lx-hist}}
\end{figure}

Figure~\ref{fig:ecdfs-f-lx-z-br} displays 
(a) observed-frame full-band flux vs. adopted redshift,
(b) absorption-corrected, rest-frame \hbox{0.5--7 keV} luminosity
vs. adopted redshift,
and (c) band ratio vs. absorption-corrected, 
rest-frame \hbox{0.5--7 keV} luminosity,
for the new sources (indicated as filled circles) and old sources 
(indicated as open circles), respectively.
We find that
(1) the new sources typically have smaller
\xray\ fluxes and luminosities than the old sources
(also see Figure~\ref{fig:ecdfs-f-lx-hist}); and
(2) the median value of 1.71 of band ratios or upper limits on band ratios
of the 83 new sources is larger than the corresponding median value of 0.84 
of the 548 old sources
(also see Figure~\ref{fig:ecdfs-plotstack}).
We further utilize survival-analysis 2-sample tests to 
quantify the difference in band ratios between
the above 83 new sources and 548 old sources that involve censored data,
which give $p=0.0$ results indicating that
there is a significant difference in band ratios between the above new and old sources.
Together, the above observations indicate that our improved 
cataloging methodology allows us to probe fainter 
obscured sources than L05.

Figure~\ref{fig:ecdfs-f-lx-hist} presents
histograms of observed-frame full-band flux 
and absorption-corrected, rest-frame \hbox{0.5--7 keV} luminosity
for the new AGNs and galaxies (main panels) as well as the old 
AGNs and galaxies (insets).
It is apparent that 
AGNs and galaxies have disparate distributions of
flux and luminosity,
no matter whether the new or old sources are considered
(except in the main panel of Figure~\ref{fig:ecdfs-f-lx-hist}a where
the flux distributions for the new AGNs and galaxies are
somewhat similar to each other).

\begin{figure*}
\centerline{\includegraphics[width=14cm]{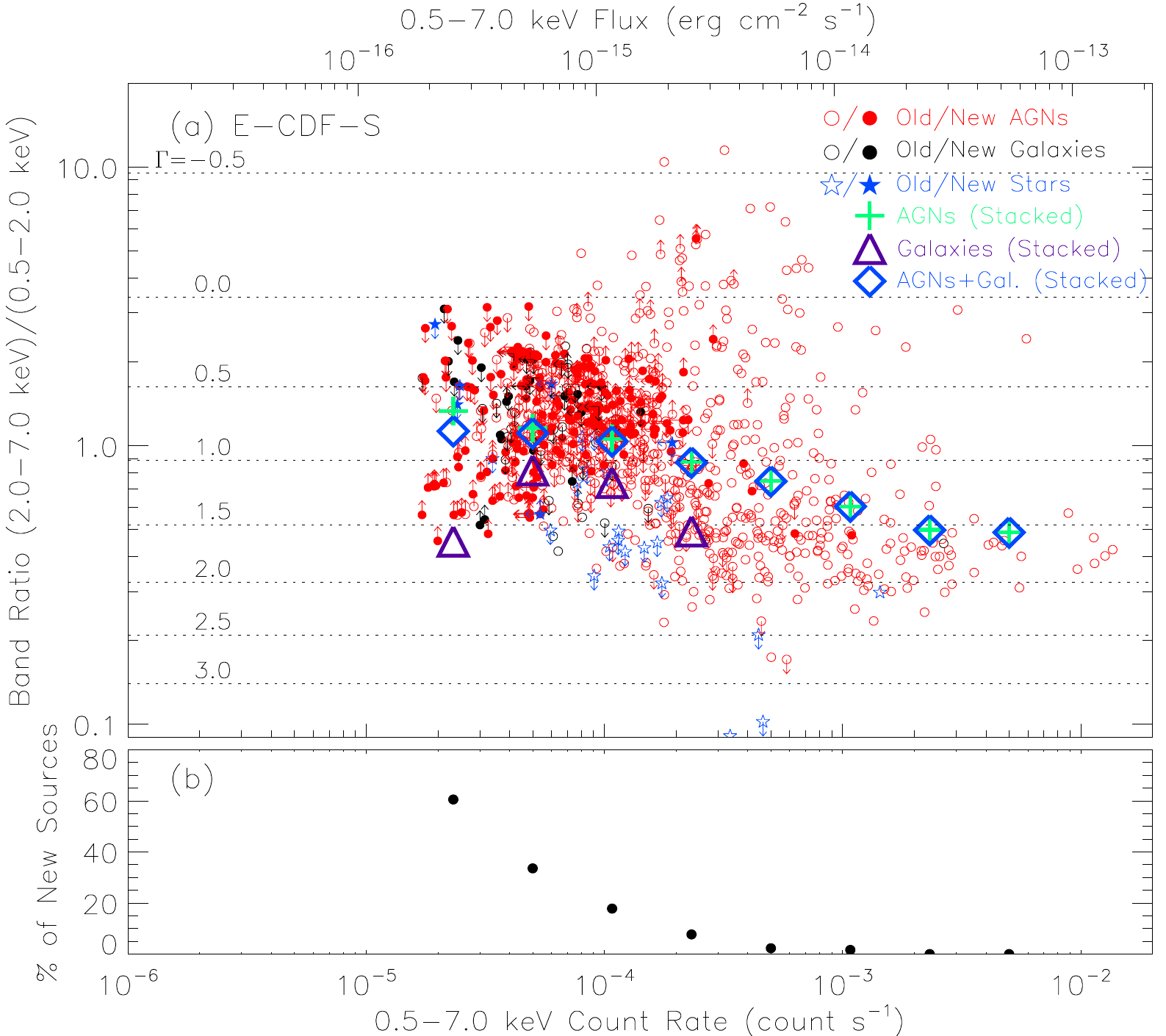}}
\figcaption{(a) Band ratio vs. full-band count rate for the 250~ks \ecdfs\ main-catalog
sources. For reference,
the top $x$-axis displays representative full-band fluxes that are derived
using full-band count rates given an assumed $\Gamma=1.4$ power law.
The meanings of symbols of different types and colors are indicated by the legend.
Arrows indicate limits.
Sources with only full-band detections are not plotted;
there are only 57 (57/1003=5.7\%) such sources, the exclusion of which would not 
affect our results significantly.
Large crosses, triangles, and diamonds denote average/stacked band ratios as a function of full-band count rate that are
derived in bins of $\Delta {\rm log(Count\hspace{0.1cm} Rate)}=0.6$,
for AGNs, galaxies, and both AGNs and galaxies, respectively.
Horizontal dotted lines indicate the band ratios that correspond to
given effective photon indexes.
(b) Fraction of new sources as a function of full-band count rate for the 250~ks \ecdfs\ main-catalog sources,
computed in bins of $\Delta {\rm log(Count\hspace{0.1cm} Rate)}=0.6$.
\label{fig:ecdfs-bratio-new}}
\end{figure*}

\begin{figure}
\centerline{\includegraphics[width=8.5cm]{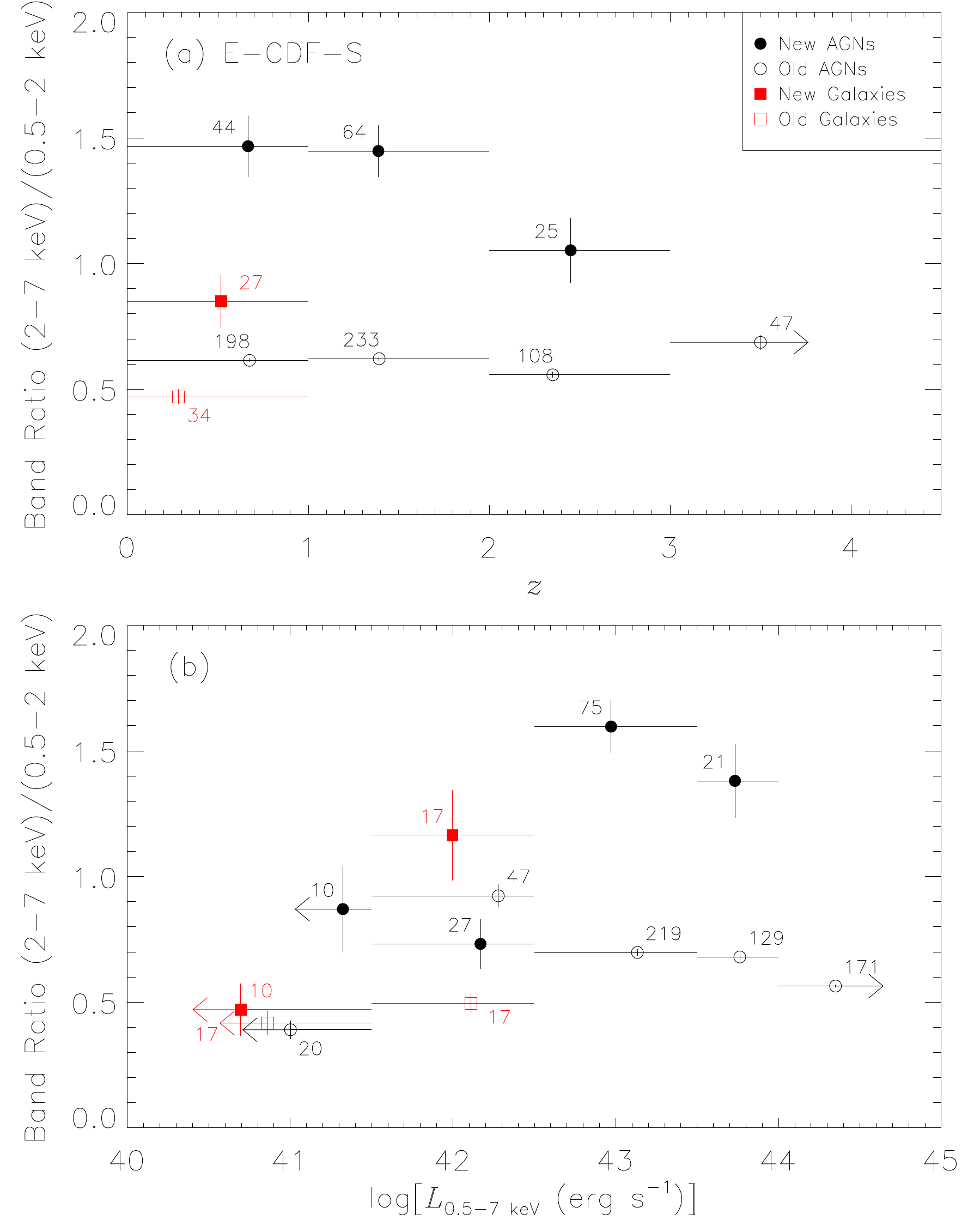}}
\figcaption{Average/stacked band ratios in bins of
(a) redshift ($0<z<1$, $1\le z<2$, $2\le z<3$, and $z\ge 3$) and
(b) absorption-corrected, rest-frame \mbox{0.5--7~keV} luminosity [$\log(L_{\rm X})<41.5$,
$41.5\le \log(L_{\rm X})<42.5$,
$42.5\le \log(L_{\rm X})<43.5$,
$43.5\le \log(L_{\rm X})<44.0$, and
$\log(L_{\rm X})\ge 44.0$] for the new and old 250~ks \ecdfs\ main-catalog sources.
The meanings of symbols are indicated by the legend.
In each bin, the median redshift or \xray\ luminosity is used for plotting;
the number of stacked sources is annotated.
\label{fig:ecdfs-plotstack}}
\end{figure}

Figure~\ref{fig:ecdfs-bratio-new}(a) displays the band ratio 
as a function of full-band count
rate for the new sources (indicated as filled symbols) 
and the old sources (indicated as open symbols),
with the large crosses, triangles, and diamonds
representing the average (i.e., stacked) band ratios
for all AGNs, all galaxies, and all sources 
(counting both AGNs and galaxies), respectively.
The overall average band ratio is, as expected, dominated by AGNs,
which has a rising and then leveling-off shape
toward low full-band count rates
(down to $2\times 10^{-5}$--$3\times 10^{-5}$ count~s$^{-1}$)
that is in general agreement with that seen
in Figure~\ref{fig:cdfn-bratio-new}(a) for the 2~Ms \cdfn.
Figure~\ref{fig:ecdfs-bratio-new}(b) presents 
the fraction of new sources as a function of
full-band count rate for the sources in the main catalog.
From full-band count rates of $2.3\times 10^{-3}$ count~s$^{-1}$
to $2.3\times 10^{-5}$ count~s$^{-1}$,
the fraction of new sources rises monotonically  
from 0\% to $\approx 60\%$.

Figure~\ref{fig:ecdfs-plotstack} presents the average band ratio
in bins of adopted redshift and \xray\ luminosity for the new AGNs, 
old AGNs, new galaxies, and old galaxies, respectively.
A couple of observations can be made, e.g.: 
(1) the new AGNs have larger band ratios than the old AGNs 
no matter which bin of redshift or \xray\ luminosity is considered,
with the only exception of the second lowest luminosity bin, 
reflecting the rise of obscured AGNs toward faint fluxes 
(e.g., Bauer \etal 2004; Lehmer \etal 2012); and
(2) the new galaxies have larger band ratios than the old galaxies
no matter which bin of redshift or \xray\ luminosity is considered
(but note the relatively limited source statistics here).

\begin{figure*}
\centerline{\includegraphics[width=17cm]{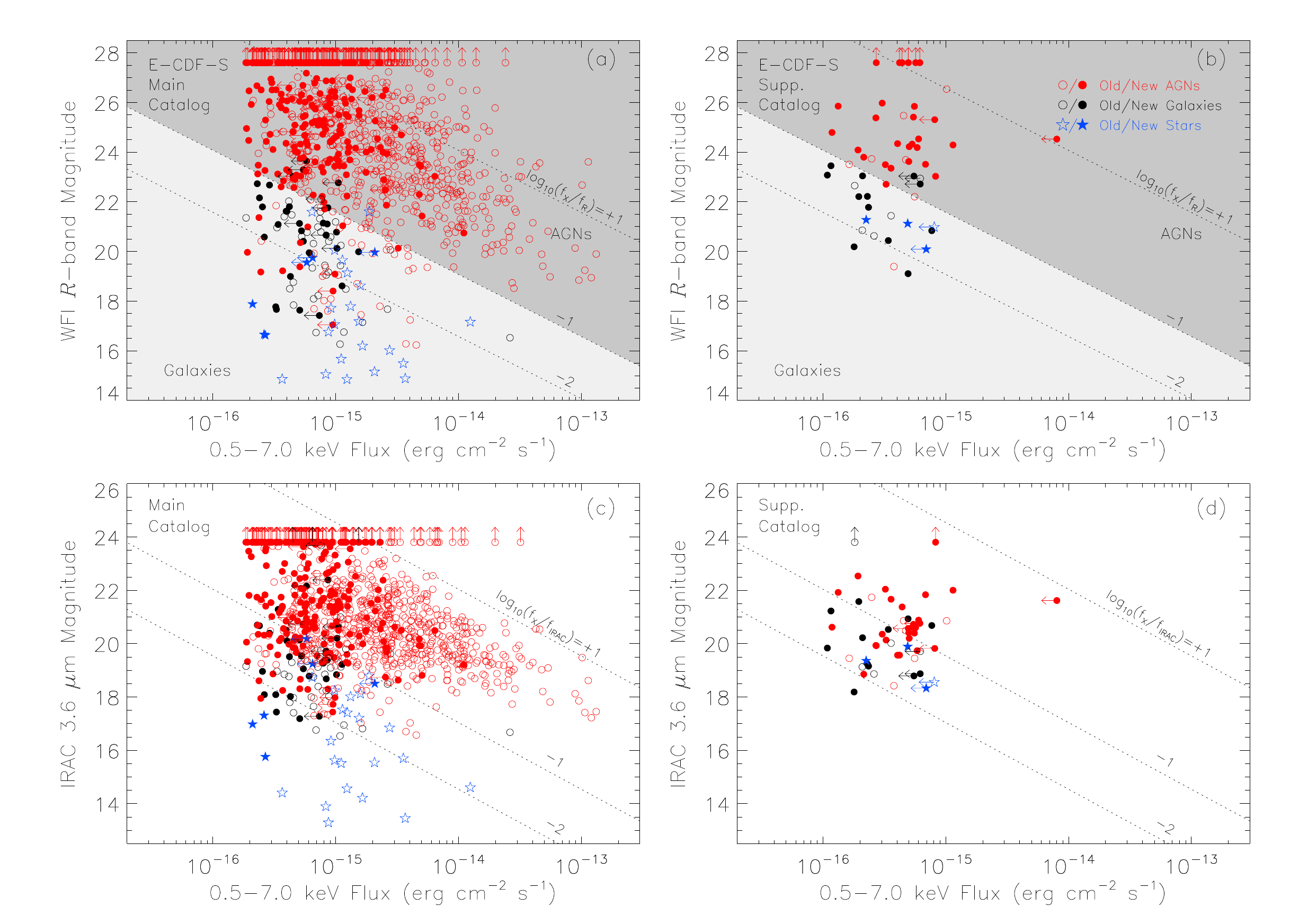}}
\figcaption{(Top) WFI $R$-band magnitude vs. full-band flux for
(a) the 250~ks \ecdfs\ main-catalog sources and (b) the supplementary-catalog sources.
(Bottom) IRAC 3.6~$\mu$m magnitude vs. full-band flux for
(c) the 250~ks \ecdfs\ main-catalog sources and (d) the supplementary-catalog sources.
The meanings of symbols of different types and colors are indicated by the legend.
Arrows denote limits.
In panels (a) and (b),
diagonal dotted lines indicate constant full-band-to-$R$ flux ratios, and shaded areas represent 
approximate flux ratios for AGNs (dark gray) and galaxies (light gray).
In panels (c) and (d),
diagonal dotted lines indicate constant full-band-to-IRAC-3.6~$\mu$m flux ratios.
\label{fig:ecdfs-fox}}
\end{figure*}

Figure~\ref{fig:ecdfs-fox}(a) presents the WFI $R$-band magnitude versus
the full-band flux for the new sources (indicated as filled symbols) 
and old sources (indicated as open symbols), as well as
the approximate flux ratios for AGNs and galaxies,
where the sources are color-coded with red for AGNs, black for galaxies,
and blue for stars, respectively. 
As a comparison, Figure~\ref{fig:ecdfs-fox}(c) presents the IRAC 3.6~$\mu$m 
magnitude versus the full-band flux
for the new sources and old sources.
Overall, a total of 909 (90.6\%) of the sources in the main catalog
are likely AGNs, the vast majority of which lie in the region expected for
relatively luminous AGNs that have $\log (f_{\rm X}/f_{\rm R})>-1$
(i.e., dark gray areas in Fig.~\ref{fig:ecdfs-fox}a);
among these 909 AGNs, 238 (26.2\%) are new.
A total of 67 (6.7\%) of the sources in the main catalog are likely galaxies,
and by selection all of them (excluding several with upper limits on full-band fluxes) 
lie in the region expected for
normal galaxies, starburst galaxies, and low-luminosity AGNs
that have $\log (f_{\rm X}/f_{\rm R})\le -1$ (i.e., light gray areas 
in Fig.~\ref{fig:ecdfs-fox}a);
among these 67 sources, 31 (46.3\%) are new.
Only 27 (2.7\%) of the sources in the main catalog are likely stars,
with all but one having low \hbox{X-ray}-to-optical flux ratios;
among these 27 stars, 6 are new.
Among the new sources, normal and starburst galaxies
total a fraction of 11.3\%,
as opposed to 4.9\% if the old sources are considered, which
is expected due to galaxies having a steeper number-count slope
than AGNs (e.g., Bauer \etal 2004; Lehmer \etal 2012).

\begin{figure}
\centerline{\includegraphics[width=8.5cm]{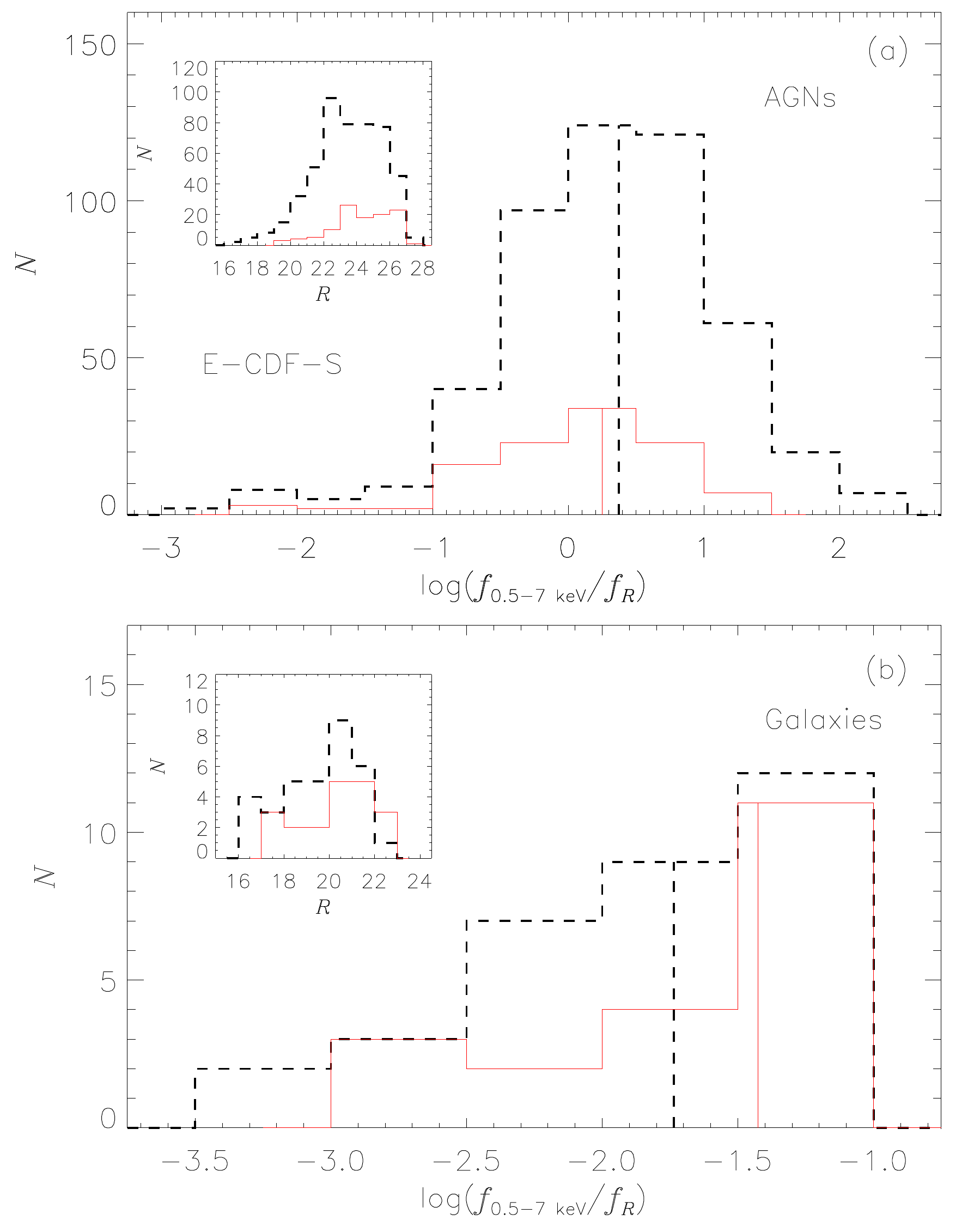}}
\figcaption{Histograms of \xray-to-optical ($R$ band) flux ratio for (a) the new 250~ks \ecdfs\ main-catalog AGNs (solid histogram)
and old AGNs (dashed histogram) and (b) new galaxies (solid histogram) and
old galaxies (dashed histogram), with median flux ratios denoted by vertical lines.
The insets display the histograms of $R$-band magnitude for new sources
(solid histograms) and old sources (dashed histograms).
Only sources with both full-band and $R$-band detections are plotted.
\label{fig:ecdfs-x-to-R}}
\end{figure}

Figure~\ref{fig:ecdfs-x-to-R} presents the
histograms of \xray-to-optical flux ratio for the new AGNs, old AGNs, 
new galaxies, and old galaxies, respectively.
It is apparent that 
(1) there is no significant difference between
the \xray-to-optical flux ratio distributions for the new and old AGNs; and
(2) there is some slight difference between
the \xray-to-optical flux ratio distributions for the new and old galaxies,
with the former having slightly larger \xray-to-optical flux ratios.

\subsection{Supplementary Near-Infrared Bright \chandra\ Source Catalog}\label{sec:ecdfs-supp}

\subsubsection{Supplementary Catalog Production}

Among the 431 (i.e., $1434-1003=431$) candidate-list \hbox{X-ray} sources that
do not satisfy the main-catalog source-selection criterion of $P<0.002$,
271 are of moderate significance with $0.002\le P<0.1$.
In order to retrieve genuine \xray\ sources from this sample of 271 sources,
we create a supplementary catalog that consists of
the subset of these sources having bright near-infrared counterparts,
using again the prior-based source-searching method.
We match these 271 \chandra\ sources with the $K_s\le 22.3$~mag sources 
in the TENIS WIRCam $K_s$-band catalog 
using a matching radius of $1\farcs2$.
A total of 56 near-infrared bright \hbox{X-ray} sources are identified this way,
with $\approx 5.9$ false matches expected 
(i.e., a false-match rate of 10.5\%).
Our supplementary catalog includes
6 L05 main-catalog sources that are not recovered in our main catalog
and 7 L05 supplementary optically bright sources,
thus resulting in a total of $56-6-7=43$ new supplementary-catalog sources that are not present in either of the L05 catalogs.
A point worth noting is that 
the vast majority (28 out of 33; 84.8\%) of the L05 supplementary optically
bright ($R<23$~mag) sources are included either in our main catalog (21 sources) 
or supplementary catalog (the aforementioned 7 sources).

Our 56-source supplementary catalog is
presented in Table~\ref{tab:ecdfs-supp}, in the same format as Table~\ref{tab:ecdfs-main}
(see Section~\ref{sec:ecdfs-maincat} for the details of each column).
A source-detection criterion of $P<0.1$ is adopted for photometry-related calculations
for the supplementary-catalog sources;
and the multiwavelength identification-related columns
(i.e., Columns~18--22) are set to the TENIS WIRCam $K_s$-band matching results.

\begin{table*}
%\tabletypesize{\scriptsize}
%\tablewidth{0pt}
\caption{250~ks \ecdfs\ Supplementary Near-Infrared Bright {\it Chandra} Source Catalog}
\begin{tabular}{lllcccccccccc}\hline\hline
No. & $\alpha_{2000}$ & $\delta_{2000}$ & $\log P$ & {\sc wavdetect} & Pos Err & Off-axis & FB & FB Upp Err & FB Low Err & SB & SB Upp Err & SB Low Err \\
(1) & (2) & (3) & (4) & (5) & (6) & (7) & (8) & (9) & (10) & (11) & (12) & (13) \\ \hline
1 & 03 31 12.66 & $-$27 40 50.6 &     $-$2.3 &  $-$7 &  1.2 &   8.62 &    12.0 &   7.0 &   5.7 &    10.9 &  $-$1.0 &  $-$1.0 \\
2 & 03 31 21.98 & $-$28 00 55.2 &     $-$2.4 &  $-$7 &  1.2 &   7.18 &     \phantom{0}8.4 &   6.3 &   5.0 &     \phantom{0}9.2 &  $-$1.0 &  $-$1.0 \\
3 & 03 31 22.67 & $-$27 35 48.0 &     $-$2.7 &  $-$5 &  1.0 &   8.64 &    16.7 &   7.9 &   6.6 &     \phantom{0}8.0 &   \phantom{0}5.3 &   \phantom{0}4.0 \\
4 & 03 31 24.28 & $-$27 57 52.0 &     $-$1.8 &  $-$5 &  1.3 &   5.70 &     \phantom{0}4.5 &   4.3 &   2.8 &     \phantom{0}7.8 &  $-$1.0 &  $-$1.0 \\
5 & 03 31 24.51 & $-$27 51 49.6 &     $-$1.1 &  $-$5 &  1.3 &   7.80 &     \phantom{0}8.3 &   7.1 &   5.8 &    13.4 &  $-$1.0 &  $-$1.0 \\ \hline
\end{tabular}
The full table contains 97~columns of information for the 56 \xray\ sources.\\
(This table is available in its entirety in a machine-readable form in the online journal. A portion is shown here for guidance regarding its form and
content.)
\label{tab:ecdfs-supp}
\end{table*}

\subsubsection{Properties of Supplementary-Catalog Sources}

Figure~\ref{fig:ecdfs-pos}(b) displays the spatial distribution of the 56
supplementary-catalog sources, with the 43 new
sources denoted as filled symbols;
and Figure~\ref{fig:ecdfs-pos}(d) presents the histograms of off-axis angles
for different source types for the supplementary-catalog sources.
Figures~\ref{fig:ecdfs-fox}(b) and (d) present
the WFI $R$-band magnitude and
the SIMPLE IRAC 3.6~$\mu$m magnitude 
versus the full-band flux for the supplementary-catalog sources, 
respectively.
Among the 56 supplementary-catalog sources,
35 (62.5\%), 17 (30.4\%), and 4 (7.1\%) are likely
AGNs, galaxies, and stars, respectively.
A total of 47 (90.4\%) of the 52 non-star sources have either
\zs's or \zp's, ranging from 0.128 to 2.437 with a median
redshift of 0.838.

\subsection{Completeness and Reliability Analysis}\label{sec:ecdfs-comp}

Following Section~\ref{sec:cdfn-comp},
we produce a set of 9 simulated \mbox{ACIS-I} observations that
closely mimic the real \ecdfs\ observations,
obtain a simulated merged 250~ks \ecdfs\ event file,
construct images for the three standard bands,
run {\sc wavdetect} (sigthresh=$10^{-5}$) to produce a candidate-list catalog,
and make use of AE to extract photometry (including $P$ values) for the candidate-list sources.

\begin{figure*}
\centerline{
\includegraphics[scale=0.64]{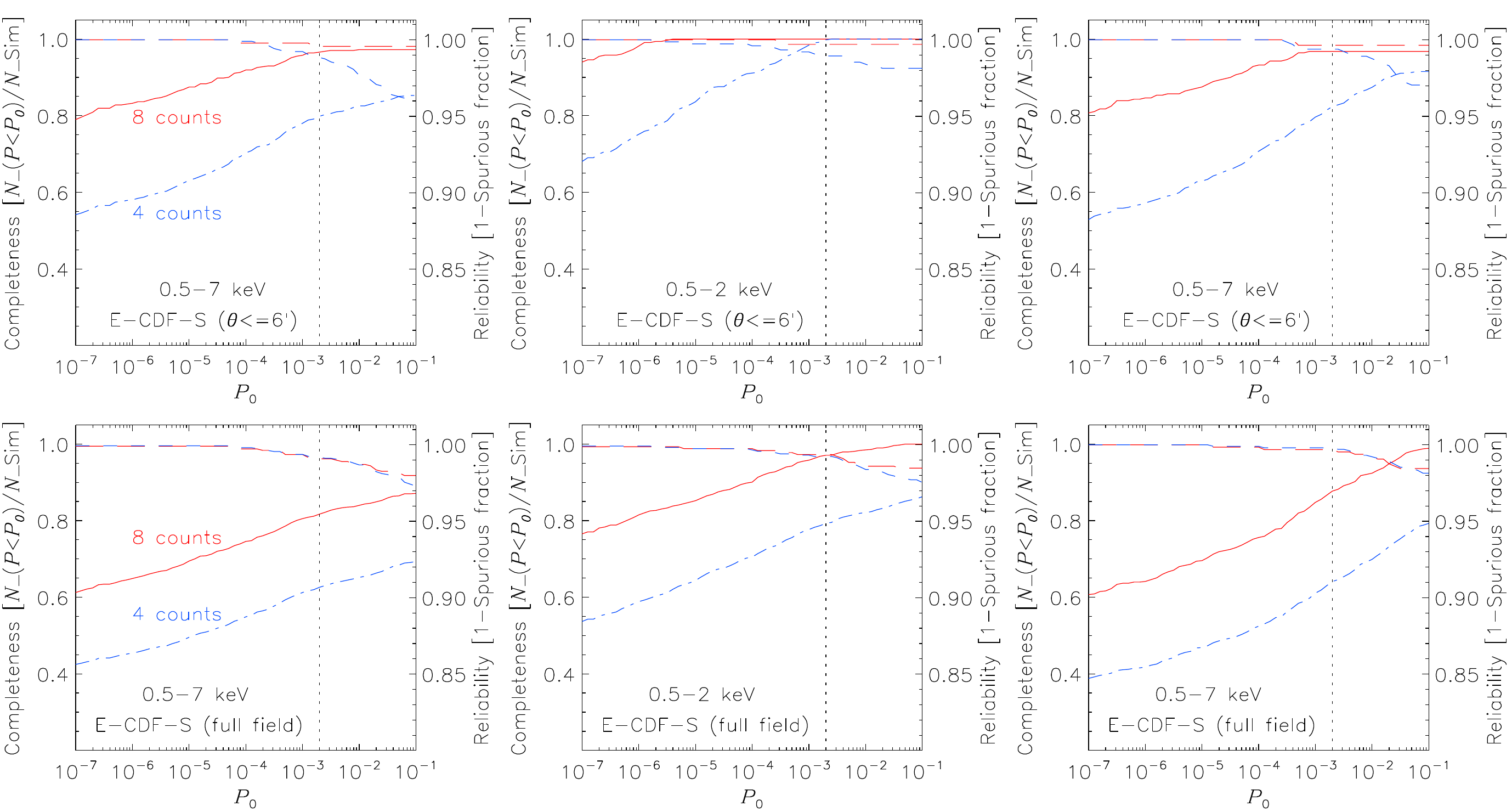}
}
\figcaption{(Top) The $\theta\le 6\arcmin$ case in the 250~ks \ecdfs:
completeness (solid and dashed-dot curves; left $y$-axis) and reliability
(long dashed and short dashed curves; right $y$-axis)
as a function of $P_0$ ($P<P_0$ as the source-detection criterion)
for the simulations in the full, soft, and hard bands,
for sources with $\ge 8$~counts (red solid and long dashed curves)
and $\ge 4$~counts (blue dashed-dot and short dashed curves), respectively.
The vertical dotted lines denote our adopted main-catalog source-detection
threshold of $P_0=0.002$.
(Bottom) Same as top panels, but for the case of the full \ecdfs\ field.\label{fig:ecdfs-comp-rel}
}
\end{figure*}

Figure~\ref{fig:ecdfs-comp-rel} displays
the completeness and reliability as a function of the AE-computed
binomial no-source probability within the $\theta\le 6\arcmin$ regions
and the entire \ecdfs\ field, 
for the simulations in the full, soft, and hard bands,
for sources with at least 8 and 4 counts, respectively.
The case of 4 counts is close to our on-axis (i.e., $\theta\lsim3\arcmin$) 
source-detection limit in the full and hard bands.
It seems clear from Fig.~\ref{fig:ecdfs-comp-rel} that
(1) in all panels, as expected,
each completeness curve goes up
and each reliability curve goes down toward large $P$ threshold
values, and the completeness level for the case of 8~counts is
higher than that for the case of 4~counts; and
(2) the completeness level for the case of either 8~counts
or 4~counts within the $\theta\le 6\arcmin$ regions
is higher than the corresponding completeness level
in the entire \ecdfs\ field.
At our adopted main-catalog $P$ threshold of 0.002,
the completeness levels within the $\theta\le 6\arcmin$ regions are
96.6\% and 79.9\% (full band), 100.0\% and 99.8\% (soft band),
and 96.8\% and 82.4\% (hard band) for sources with $\ge 8$ and $\ge 4$~counts, respectively.
The completeness levels for the entire \ecdfs\ field are
81.7\% and 62.6\% (full band), 97.1\% and 79.3\% (soft band),
and 87.7\% and 64.1\% (hard band) for sources with $\ge 8$ and $\ge 4$~counts, respectively.
At our adopted main-catalog $P$ threshold of 0.002,
the reliability level ranges from 98.8\% to 99.8\% for all panels; and
we estimate that, in the main catalog (i.e., the entire \ecdfs\ field),
there are
about 8, 4, and 2 false detections with $\ge 8$~counts in the full,
soft, and hard bands, and
about 8, 6, and 3 false detections with $\ge 4$~counts in the full,
soft, and hard bands, respectively.

\begin{figure}
\centerline{
\includegraphics[scale=0.55]{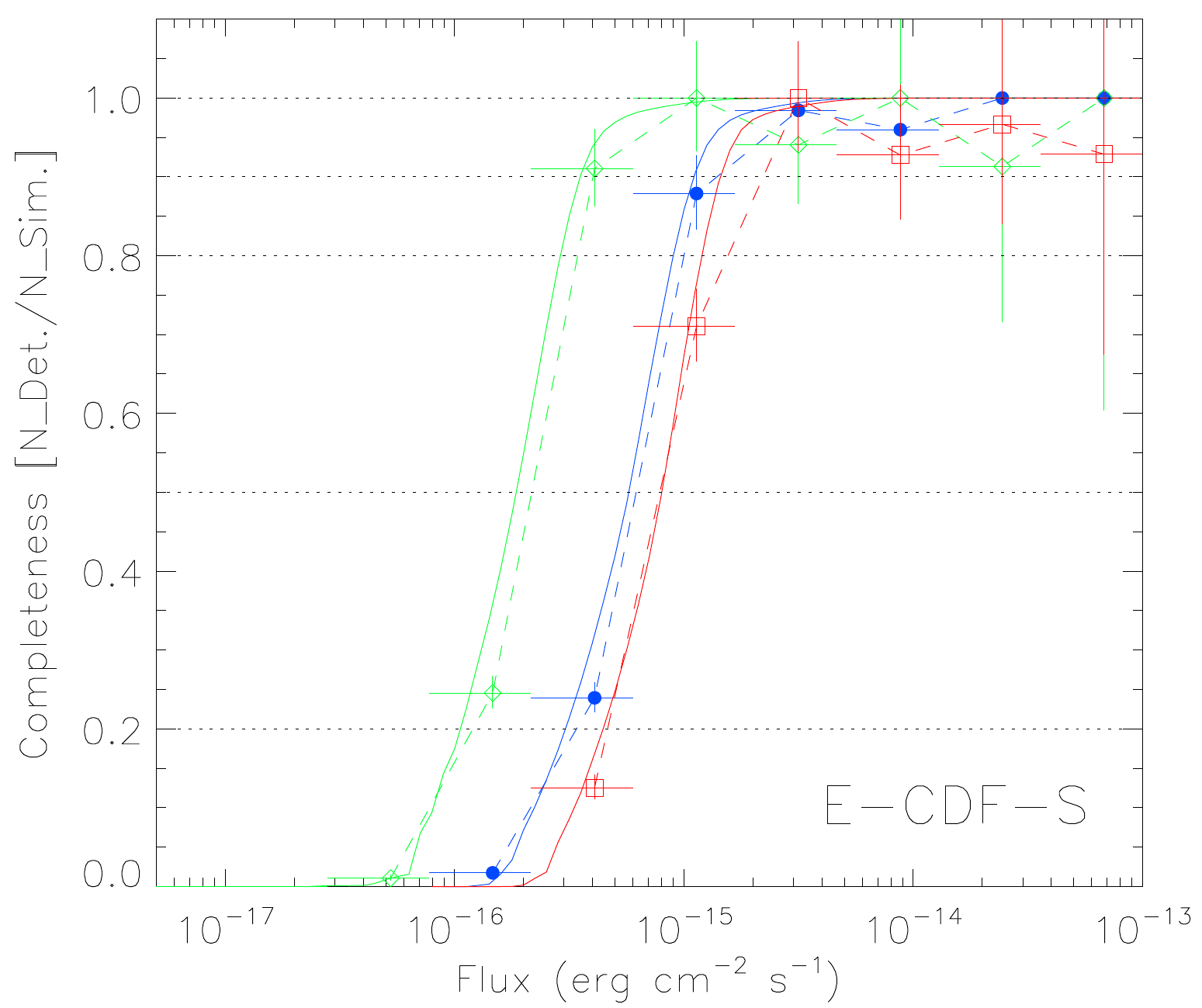}
}
\figcaption{Completeness as a function of flux
given the 250~ks \ecdfs\ main-catalog $P<0.002$ criterion
for the simulations in the full (blue filled circles), soft (green open diamonds), and hard (red open squares) bands,
overlaid with the corresponding sky coverage curves (solid curves)
that are normalized to the maximum sky coverage (see
Fig.~\ref{fig:ecdfs-senhist}).
The dashed lines make connections between the corresponding adjacent cross points.
The horizontal dotted lines denote five completeness levels.
\label{fig:ecdfs-completeness-flux}}
\end{figure}

Figure~\ref{fig:ecdfs-completeness-flux} presents 
the completeness as a function of flux
given the main-catalog $P<0.002$ criterion
for the \hbox{full-,} \hbox{soft-,} and hard-band simulations.
The three curves of completeness versus flux that are derived from the simulations
(dashed lines) approximately track
the normalized sky coverage curves that are
derived from the real \ecdfs\ observations (solid curves).
Table~\ref{tab:ecdfs-completeness} presents
the flux limits corresponding to four specific completeness
levels in the full, soft, and hard bands, which are denoted as horizontal dotted lines in Fig.~\ref{fig:ecdfs-completeness-flux}.

\begin{table}[ht]
\caption{250~ks \ecdfs\ Flux Limit and Completeness}
\centering
\begin{tabular}{lccc}\hline\hline
Completeness & $f_{\rm 0.5-7\ keV}$ & $f_{\rm 0.5-2\ keV}$ & $f_{\rm 2-7\ keV}$ \\
(\%) & (\flux) & (\flux) & (\flux) \\ \hline
90 & $1.1\times 10^{-15}$ & $3.5\times 10^{-16}$ & $1.4\times 10^{-15}$ \\
80 & $9.0\times 10^{-16}$ & $2.9\times 10^{-16}$ & $1.2\times 10^{-15}$ \\
50 & $5.7\times 10^{-16}$ & $1.9\times 10^{-16}$ & $8.0\times 10^{-16}$ \\
20 & $3.0\times 10^{-16}$ & $1.1\times 10^{-16}$ & $4.4\times 10^{-16}$ \\\hline
\end{tabular}
\label{tab:ecdfs-completeness}
\end{table}

\subsection{Background and Sensitivity Analysis}\label{sec:ecdfs-bkg}

\subsubsection{Background Map Creation}\label{sec:ecdfs-bmap}

\begin{figure}
\centerline{\includegraphics[width=8.5cm]{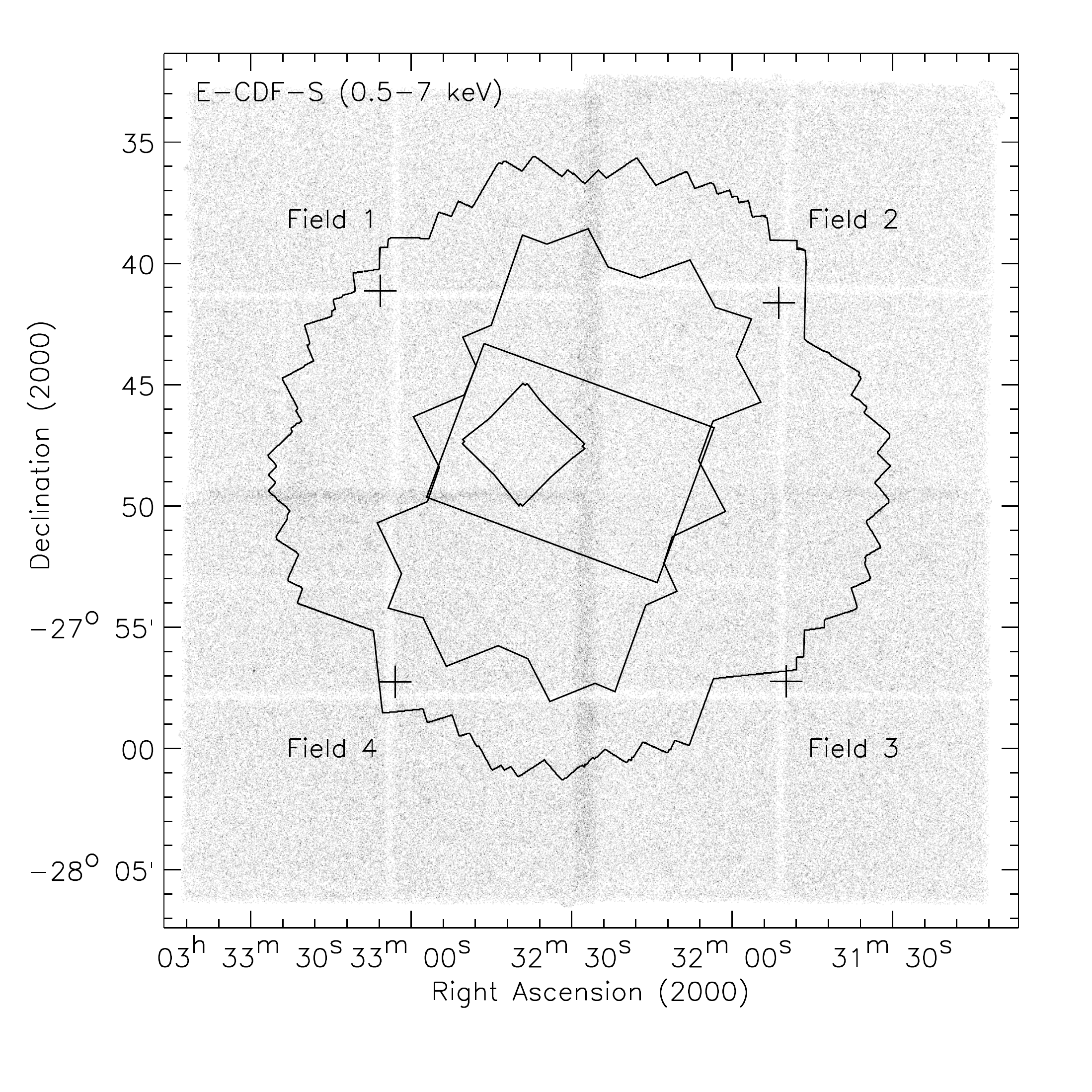}}
\figcaption{Full-band (0.5--7.0 keV) background map of the 250~ks \ecdfs\ rendered using linear gray scales.
The higher background between fields is due to the larger effective exposure caused by overlapping observations.
The regions and the plus signs have the same meanings as those in Fig.~\ref{fig:ecdfs-fb-img}.
\label{fig:ecdfs-fb-bkg}}
\end{figure}

We follow Section~\ref{sec:cdfn-bmap} 
to create background maps for the three standard-band images.
Table~\ref{tab:ecdfs-bkg} summarizes the background properties
including the mean background, total background,
and count ratio between background counts and detected source counts
for the three standard bands.
97.1\%, 99.1\%, and 98.0\% of the pixels 
have zero background counts in the background maps
for the full, soft, and hard bands, respectively.
The values in Table~\ref{tab:ecdfs-bkg} 
are systematically slightly lower than those reported in Table~7 of L05, 
mainly due to the facts  
that we adopt a smaller upper energy bound of 7~keV than 
the value of 8~keV
adopted in L05 and that
we adopt a more stringent approach for data filtering
(see Section~\ref{sec:ecdfs-obs}).
Figure~\ref{fig:ecdfs-fb-bkg} displays the full-band background map.

\begin{figure}
\centerline{\includegraphics[width=8.5cm]{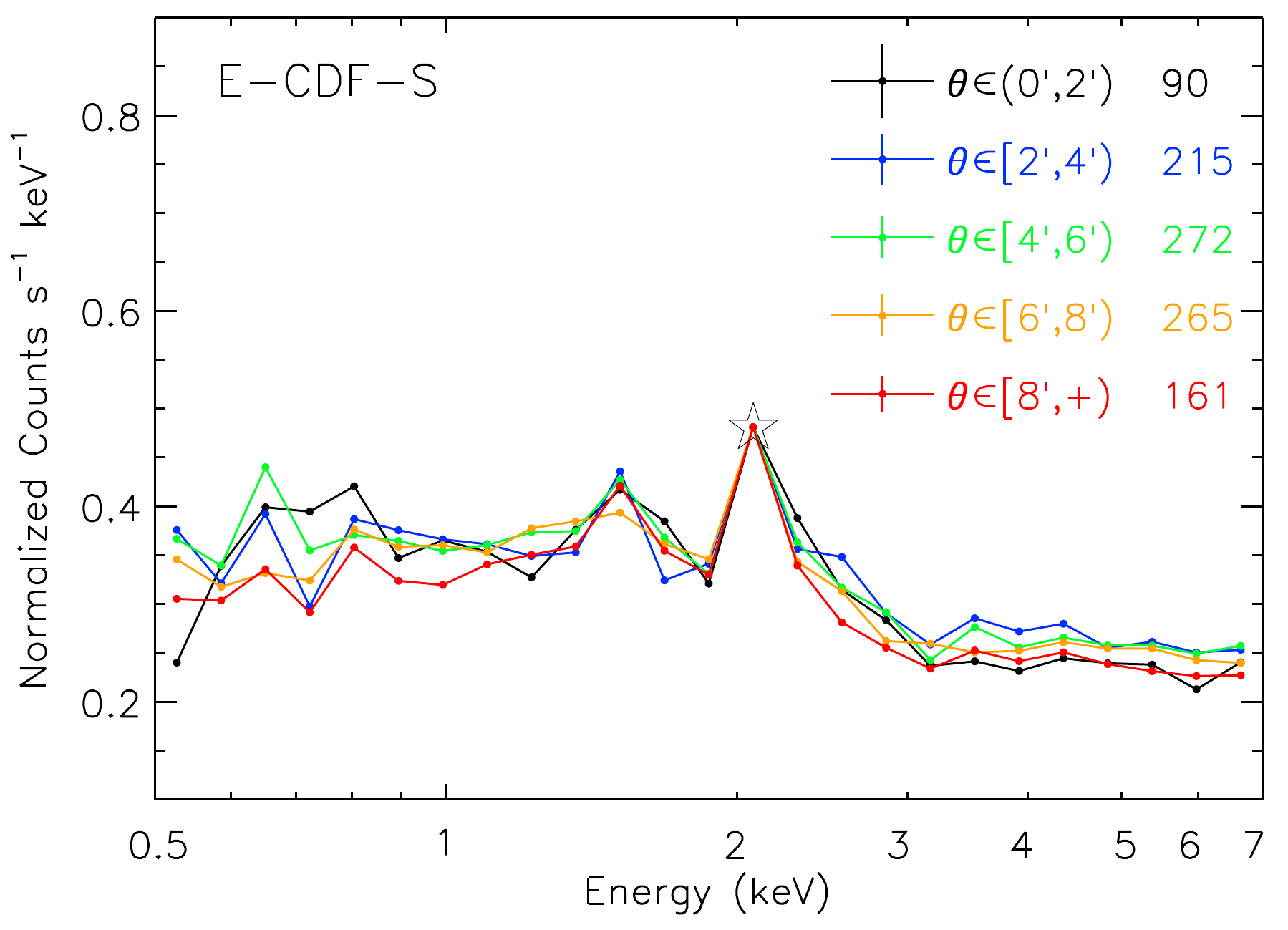}}
\figcaption{Mean background spectra for the 250~ks \ecdfs\ main-catalog sources calculated
in various bins of off-axis angle. The spectra are normalized to have the same value at an
energy slightly above 2~keV, which is indicated by a large 5-pointed star.
For clarity, errors on individual spectral data points are not plotted;
the typical spectral error value and number of sources in each bin of off-axis angle
are annotated in the top-right corner.
\label{fig:ecdfs-bkg-spec}}
\end{figure}

Figure~\ref{fig:ecdfs-bkg-spec} presents the mean \chandra\ background spectra that are calculated
for the 1003 main-catalog sources in various bins of off-axis angle,
using the individual background spectra extracted in Section~\ref{sec:ecdfs-img-cand}.
We find that
(1) the shapes of the mean \chandra\ background spectra remain largely the same across the
entire \ecdfs\ field given the uncertainties, in particular,
as far as the $\gsim 1$~keV parts of the spectra are concerned (with $\lsim 10\%$ variations 
between the shapes);
(2) for the $\lsim 1$~keV parts of the mean background spectra, shape variations seem slightly
more apparent (up to $\approx 20\%$); and
(3) compared to the shapes of the mean \cdfn\ background spectra shown in Fig.~\ref{fig:cdfn-bkg-spec},
the \ecdfs\ background spectra have very similar shapes at $\gsim 1$~keV, but 
seem to level off slightly at $\lsim 1$~keV, 
probably due to cosmic variance and/or variations of \chandra\ instrument status.

\begin{table*}
%\tabletypesize{\small}
%\tablewidth{0pt}
\caption{250~ks \ecdfs: Background Parameters}
\centering
\begin{tabular}{lcccc}\hline\hline
Band (keV) & Mean Background & Mean Background & Total Background$^{\rm c}$ & Count Ratio$^{\rm d}$ \\
 & (count pixel$^{-1}$)$^{\rm a}$ & (count Ms$^{-1}$ pixel$^{-1}$)$^{\rm b}$ & (10$^4$ counts) & (Background/Source) \\ \hline 
Full (0.5--7.0) & 0.031 & 0.160 & 52.4 & \phantom{0}6.5 \\
Soft (0.5--2.0) & 0.009 & 0.048 & 15.7 & \phantom{0}3.2 \\
Hard (2--7) & 0.022 & 0.109 & 36.8 & 12.2 \\\hline
\end{tabular}
\\$^{\rm a}$ The mean numbers of background counts per pixel.
\\$^{\rm b}$ The mean numbers of background counts per pixel
divided by the mean effective exposures.
\\$^{\rm c}$ The total numbers of background counts in the background maps.
\\$^{\rm d}$ Ratio between the total number of background
counts and the total number of detected source counts in the main catalog.
\label{tab:ecdfs-bkg}
\end{table*}

\subsubsection{Sensitivity Map Creation}\label{sec:ecdfs-smap}

We follow Section~\ref{sec:cdfn-smap}
to create sensitivity maps in the three standard bands 
for the main catalog (i.e., using $P<0.002$)
to assess the sensitivity as a function of position 
across the entire field.
We find that there are 12, 11, and 22 main-catalog sources
in the three standard bands that lie typically $\lsim10\%$
below the corresponding derived
sensitivity limits, respectively,
which is likely due to background fluctuations and/or their real $\Gamma$ values
differing significantly from the assumed $\Gamma=1.4$.

\begin{figure}
\centerline{\includegraphics[width=8.5cm]{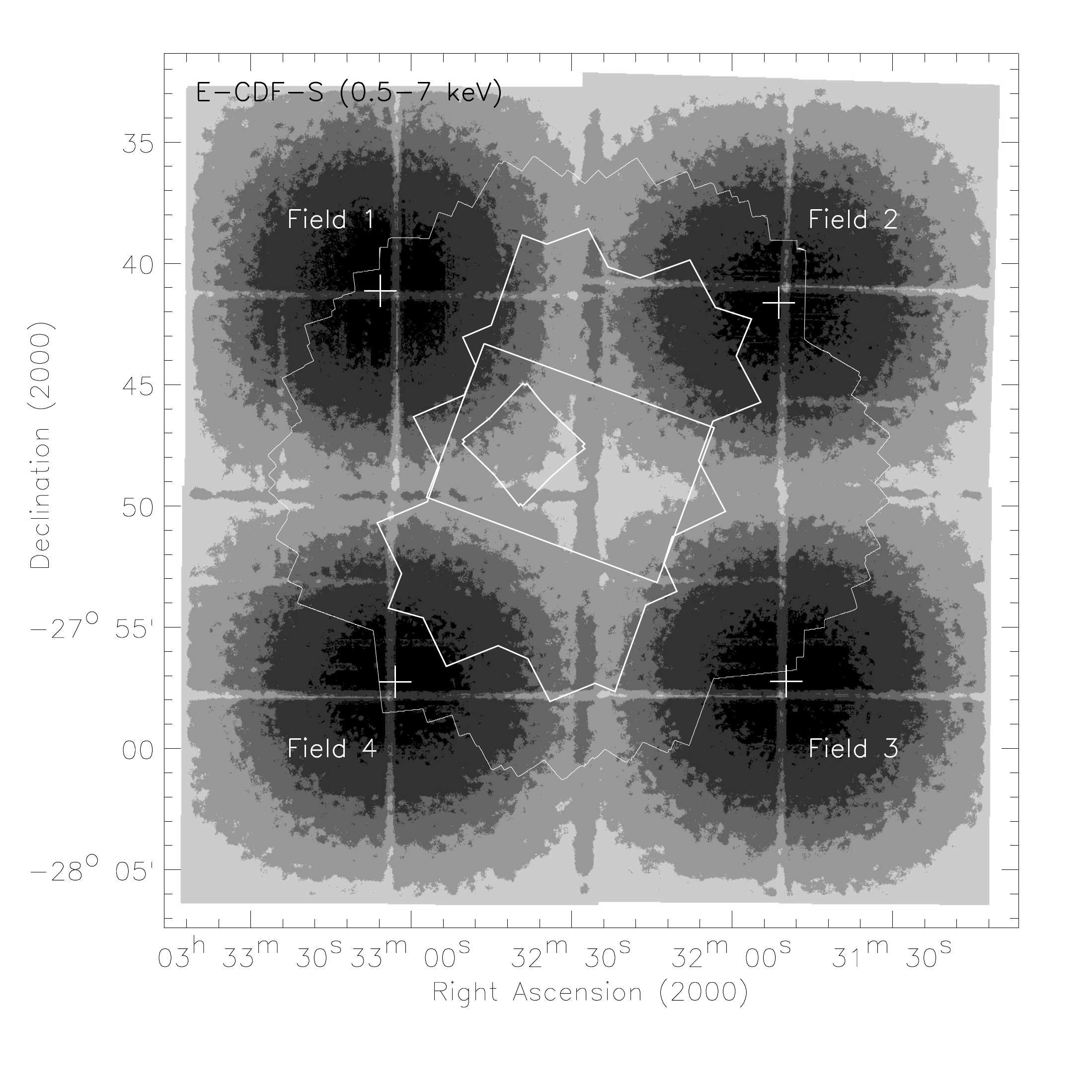}}
\figcaption{Full-band (0.5--7.0 keV) sensitivity map for the 250~ks \ecdfs\ main catalog.
The gray-scale levels, ranging from black to light gray, denote areas with flux limits
of $<2.0\times 10^{-16}$, \hbox{$2.0\times 10^{-16}$} to \hbox{$4.0\times 10^{-16}$}, \hbox{$4.0\times 10^{-16}$} to
\hbox{$6.0\times10^{-16}$}, \hbox{$6.0\times 10^{-16}$} to $10^{-15}$, and $>10^{-15}$ \flux, respectively.
The regions and the plus signs have the same meanings as those in Fig.~\ref{fig:ecdfs-fb-img}.
\label{fig:ecdfs-fb-senimg}}
\end{figure}

\begin{figure}
\centerline{\includegraphics[width=8.5cm]{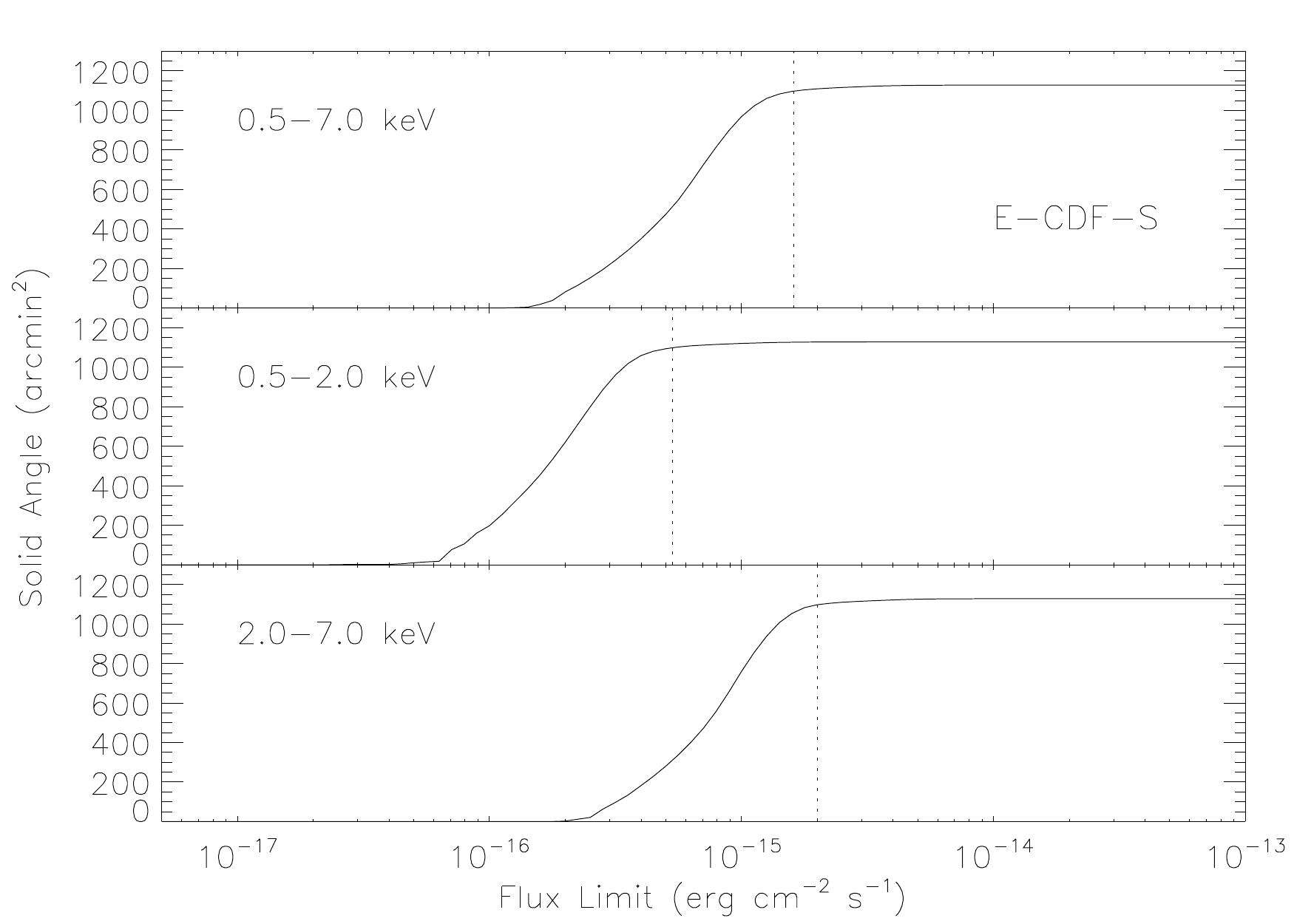}}
\figcaption{Survey solid angle as a function of flux limit in the full, soft, and hard bands
for the 250~ks \ecdfs\ main catalog.
The vertical dotted lines indicate the median fluxes of the main-catalog sources detected in the three bands.
\label{fig:ecdfs-senhist}}
\end{figure}

Figure~\ref{fig:ecdfs-fb-senimg} displays the full-band 
sensitivity map for the main catalog, and 
Figure~\ref{fig:ecdfs-senhist} presents plots of survey solid angle
versus flux limit in the three standard bands given $P<0.002$.
It is clear that higher sensitivities are reached 
at smaller off-axis angles and thus within smaller
survey solid angles.
The central $\approx$1~arcmin$^2$ areas at the four aim points
have mean sensitivity limits of
$\approx 2.0\times 10^{-16}$, $7.6\times 10^{-17}$, and $3.0\times 10^{-16}$ \flux\
for the full, soft, and hard bands, respectively,
which represent a factor of \mbox{$\approx 1.5$--2.0} improvement 
over those of L05, 
due to the facts that we adopt a sensitive two-stage source-detection 
procedure and that L05 adopted a different methodology 
for sensitivity calculations.

\section{Summary}\label{sec:summary}

We present the improved \chandra\ point-source catalogs,
associated data products, and basic analyses of detected 
\hbox{X-ray} sources for the
2~Ms \cdfn\ and 250~ks \ecdfs, 
implementing a number of improvements
in the \chandra\ source cataloging methodology listed in 
Table~\ref{tab:impro}.
In particular,
the combination of sophisticated and accurate \xray\ photometry extraction
as well as the sensitive and reliable two-stage source-detection approach
enables probing fainter and more obscured sources
with high confidence in their validity
than the previous A03 \cdfn\ and L05 \ecdfs\ catalogs, 
without new \chandra\ observational investment.
As such, the improved catalogs allow 
better characterization of all the \mbox{$\approx 1800$}
\cdfn\ and \ecdfs\ sources including the
$\approx 500$ newly-detected ones, thereby
superseding the A03 and L05 catalogs.
In addition to the point-source catalogs, 
we also make other associated data products publicly available, including the final event files, 
raw images, effective exposure maps, background maps, sensitivity maps, and solid-angle vs.
flux-limit curves for the 2~Ms \cdfn\ and 250~ks \ecdfs.$^{\ref{ft:datalink}}$
Below we summarize the most significant results for the \cdfn\
and \ecdfs, respectively.

For the 2~Ms \cdfn, the key results are as follows.

\begin{enumerate}
\item
The entire \cdfn\ is made up of 20 individual observations, 
which have a total effective exposure of 1.896~Ms and cover 
a total solid angle of 447.5~arcmin$^{2}$.

\item
The \cdfn\ main catalog consists of 683 sources that
are detected by running {\sc wavdetect} at a false-positive probability
threshold of $10^{-5}$ and meet our binomial-probability source-selection
criterion of \mbox{$P<0.004$};
such an approach is devised to maximize the number of reliable sources detected.
These 683 sources are detected in up to three standard \mbox{X-ray}
bands, i.e., 0.5--7.0~keV (full band), \mbox{0.5--2.0~keV} (soft band),
and 2--7~keV (hard band).
670 (98.1\%) of these 683 sources have multiwavelength counterparts, and
638 (95.2\% of 670) have either spectroscopic or 
photometric redshifts.

\item
The \cdfn\ supplementary catalog contains 72 sources that
are detected by running {\sc wavdetect} at a false-positive probability
threshold of $10^{-5}$ and meet the requirements of having
$0.004<P<0.1$ and having bright ($K_s<22.9$) near-infrared counterparts.
69 (95.8\%) of these 72 sources have 
either spectroscopic or photometric redshifts.

\item
\mbox{X-ray} source positions for the \cdfn\ main
and supplementary catalogs
are determined utilizing centroid and matched-filter techniques.
The absolute astrometry of \xray\ source positions
is locked to that of the \goodsn\ WIRCam $K_s$-band catalog
and the median positional offset of the 
\xray-$K_s$-band matches is 0\farcs28.
The median \xray\ positional uncertainties at the $\approx 68\%$ confidence level
are $0\farcs47$ and $0\farcs80$
for the main and supplementary catalogs, respectively.

\item
Basic analyses of the X-ray and multiwavelength properties of the \cdfn\
sources indicate that 
86.5\%, 11.0\%, and 2.5\% of the main-catalog sources
are likely AGNs, galaxies, and stars, respectively.
In the central $\theta\le 3\arcmin$ area of the 2~Ms \cdfn,
the observed main-catalog AGN and galaxy source densities reach
$12400_{-1300}^{+1400}$~deg$^{-2}$ and
$4200_{-700}^{+900}$~deg$^{-2}$, respectively.
47.2\%, 52.8\%, and 0.0\% of the supplementary-catalog sources
are likely AGNs, galaxies, and stars, respectively.

\item A total of 196 \cdfn\ main-catalog sources are new
and are generally fainter and more obscured,
compared to the A03 main-catalog sources.
Among the 196 new main-catalog sources,
78.6\% are likely AGNs and 19.9\% are likely normal
and starburst galaxies (with the remaining 1.5\% being likely stars), 
which reflects the rise of normal and starburst
galaxies at these very low flux levels.
Indeed, galaxies become the numerically dominant source
population that emerges
at luminosities less than $\approx 10^{41.5}$ erg~s$^{-1}$,
according to our source-classification results.

\item Simulations demonstrate that
our \cdfn\ main catalog is highly reliable 
(\hbox{$\lsim 5$}, 4, and 3 false detections are expected 
in the full, soft, and hard bands, respectively) 
and is reasonably complete 
(e.g., in the central $\theta\le 6\arcmin$ area,
the completeness levels are $\gsim 82\%$, 95\%, and 68\%
for sources with $\ge 8$~counts 
in the full, soft, and hard bands, respectively).

\item
The \cdfn\ mean background is 0.167, 0.055, and 0.108
count~Ms$^{-1}$~pixel$^{-1}$ for the full, soft, 
and hard bands, respectively;
91.7\%, 97.1\%, and 94.2\% of the pixels 
have zero background counts in the background maps
for the full, soft, and hard bands, respectively.

\item
The 2~Ms \cdfn\ achieves
on-axis flux limits of
\hbox{$\approx 3.5\times 10^{-17}$}, $1.2\times 10^{-17}$, and $5.9\times 10^{-17}$ \flux\
for the full, soft, and hard bands, respectively,
a factor of $\approx 2$ improvement over those of A03,
due to the facts that we adopt a sensitive two-stage source-detection
procedure and that A03 adopted a different methodology
for sensitivity calculations.

\end{enumerate}

For the 250~ks \ecdfs, the key results are as follows.

\begin{enumerate}
\item
The entire \ecdfs\ is made up of 9 individual observations, which have
a depth of $\approx 250$~ks 
and cover a total solid angle of 1128.6~arcmin$^{2}$.

\item
The \ecdfs\ main catalog consists of 1003 sources that
are detected by running {\sc wavdetect} at a false-positive probability
threshold of $10^{-5}$ and meet our binomial-probability source-selection
criterion of \mbox{$P<0.002$};
such an approach is devised to maximize the number of reliable sources detected.
These 1003 sources are detected in up to three standard \mbox{X-ray}
bands, i.e., 0.5--7.0~keV (full band), 0.5--2.0~keV (soft band),
and 2--7~keV (hard band).
958 (95.5\%) of these 1003 sources have multiwavelength counterparts, and
810 (84.6\% of 958) have either spectroscopic or 
photometric redshifts.

\item
The \ecdfs\ supplementary catalog contains 56 sources that
are detected by running {\sc wavdetect} at a false-positive probability
threshold of $10^{-5}$ and meet the requirements of having
$0.002<P<0.1$ and having bright ($K_s<22.3$) near-infrared counterparts.
51 (91.1\%) of these 56 sources have 
either spectroscopic or photometric redshifts.

\item
\mbox{X-ray} source positions for the \ecdfs\ main
and supplementary catalogs
are determined utilizing centroid and matched-filter techniques.
The absolute astrometry of \xray\ source positions
is locked to that of the TENIS WIRCam $K_s$-band catalog 
and the median positional offset of the 
\xray-$K_s$-band matches is 0\farcs38.
The median \xray\ positional uncertainties at the $\approx 68\%$ confidence level
are $0\farcs63$ and $1\farcs20$
for the main and supplementary catalogs, respectively.

\item
Basic analyses of the X-ray and multiwavelength properties of the \ecdfs\
sources indicate that 
90.6\%, 6.7\%, and 2.7\% of the main-catalog sources
are likely AGNs, galaxies, and stars, respectively.
In the areas within respective off-axis angles of $3\arcmin$
of the four \ecdfs\ aim points, 
the mean observed main-catalog AGN and galaxy source densities reach
$5200_{-800}^{+1000}$~deg$^{-2}$ and
$500_{-200}^{+400}$~deg$^{-2}$, respectively.
62.5\%, 30.4\%, and 7.1\% of the supplementary-catalog sources
are likely AGNs, galaxies, and stars, respectively.

\item A total of 275 \ecdfs\ main-catalog sources are new
and are generally fainter and more obscured,
compared to the L05 main-catalog sources.
Among the 275 new main-catalog sources,
86.5\% are likely AGNs and 11.3\% are likely normal
and starburst galaxies (with the remaining 2.2\% being likely stars), which 
reflects the rise of normal and starburst
galaxies when probing fainter fluxes.

\item Simulations demonstrate that
our \ecdfs\ main catalog is highly reliable 
(\hbox{$\lsim 8$}, 6, and 3 false detections are expected 
in the full, soft, and hard bands, respectively)
and is reasonably complete 
(e.g., in the central $\theta\le 6\arcmin$ areas,
the completeness levels are $\gsim 79\%$, 99\%, and 82\%
for sources with $\ge 4$~counts 
in the full, soft, and hard bands, respectively).

\item
The \ecdfs\ mean background is 0.160, 0.048, and 0.109
count~Ms$^{-1}$~pixel$^{-1}$ for the full, soft, 
and hard bands, respectively;
97.1\%, 99.1\%, and 98.0\% of the pixels 
have zero background counts in the background maps
for the full, soft, and hard bands, respectively.

\item
The 250~ks \ecdfs\ achieves
on-axis (i.e., near the four aim points) flux limits of
\hbox{$\approx 2.0\times 10^{-16}$}, $7.6\times 10^{-17}$, and $3.0\times 10^{-16}$ \flux\
for the full, soft, and hard bands, respectively,
a factor of \mbox{$\approx 1.5$--2.0} improvement over those of L05,
due to the facts that we adopt a sensitive two-stage source-detection
procedure and that L05 adopted a different methodology
for sensitivity calculations.

\end{enumerate}

\vspace{0.5cm}

We thank the referees for their helpful feedback that improved this work.
YQX acknowledges support from
the National Thousand Young Talents program (KJ2030220004),
the 973 Program (2015CB857004),
the USTC startup funding (ZC9850290195),
the National Natural Science Foundation of China (NSFC-11473026, 11421303),
the Strategic Priority Research Program ``The Emergence of Cosmological Structures''
of the Chinese Academy of Sciences (XDB09000000),
and the Fundamental Research Funds for the Central Universities (WK3440000001).
BL, WNB, and GY acknowledge support from
\chandra\ \xray\ Center grants AR3-14015X and GO4-15130A, and
\chandra\ ACIS team contract SV4-74018.
DMA acknowledges support from the
Science and Technology Facilities Council through grant code ST/I505656/1.
FEB acknowledges support from CONICYT-Chile (Basal-CATA PFB-06/2007, FONDECYT 1141218, ``EMBIGGEN'' Anillo ACT1101), and the Ministry of Economy, Development, and Tourism's Millennium Science Initiative through grant IC120009, awarded to the Millennium Institute of Astrophysics, MAS.
The Guaranteed Time Observations (GTO) for the \cdfn\ included
here were selected by the ACIS Instrument Principal
Investigator, Gordon P. Garmire, currently of the Huntingdon
Institute for \xray\ Astronomy, LLC, which is under contract
to the Smithsonian Astrophysical Observatory; Contract SV2-82024.

\end{document}